\documentclass[11pt,a4paper,fleqn]{article}
\pdfoutput=1
\usepackage{graphicx}
\usepackage[usenames,dvipsnames]{xcolor} 
\usepackage{amssymb,amsmath,mathtools,mathrsfs} 
\usepackage{epsfig,subfigure} 
\usepackage{booktabs,longtable} 
\usepackage{exscale,relsize} 
\usepackage[normalem]{ulem} 
\usepackage{enumerate} 
\usepackage{jcappub}
\usepackage{aastex}
\usepackage{cancel}
\usepackage{caption}
\usepackage[noabbrev]{cleveref}
\usepackage{hyperref}
\usepackage{float}
\usepackage{appendix}
\usepackage{tabularx}
\usepackage[utf8]{inputenc}
\usepackage{xspace}

\newcommand{\calH}{{\cal H}} 
\newcommand{\p}{\prime} 
\newcommand{\pp}{{\prime\prime}} 
\newcommand{\aM}{{\alpha_M}} 
\newcommand{\aB}{{\alpha_B}} 
\newcommand{\aK}{{\alpha_K}} 
\newcommand{\Meff}{M_{\mathrm{eff}}}
\newcommand{\Mc}{{\cal M}_c}
\newcommand{\tauo}{\tau_{\rm obs}}

\newcommand{\proptoomega}{\ensuremath{\propto \Omega_{\rm DE}}\xspace}
\newcommand{\proptoscale}{\ensuremath{\propto a}\xspace}
\newcommand{\cosmosis}{\textsc{CosmoSIS}\xspace} 
\newcommand{\hiclass}{\texttt{hi\_class}\xspace}
\newcommand{\emcee}{\textsc{emcee}\xspace}

\begin{document}
\title{Constraining Scalar-Tensor Modified Gravity with Gravitational Waves and Large Scale Structure Surveys}

\author[a]{Tessa Baker}
\affiliation[a]{Queen Mary University of London, Mile End Road, London E1 4NS, UK}
\emailAdd{t.baker@qmul.ac.uk}
\author[b,c]{and Ian Harrison}
\affiliation[b]{Jodrell Bank Centre for Astrophysics, Department of Physics \& Astronomy, The University of Manchester, Manchester M13 9PL, UK}
\affiliation[c]{Department of Physics, University of Oxford, Denys Wilkinson Building, Keble Road, Oxford, OX1 3RH, UK}
\emailAdd{ian.harrison-2@manchester.ac.uk}

\abstract{The first multi-messenger gravitational wave event has had a transformative effect on the space of modified gravity models. In this paper we study the enhanced tests of gravity that are possible with a future set of gravitational wave standard siren events. We perform MCMC constraint forecasts for parameters in Horndeski scalar-tensor theories. In particular, we focus on the complementarity of gravitational waves with electromagnetic large-scale structure data from galaxy surveys. We find that the addition of fifty low redshift ($z \lesssim 0.2$) standard sirens from the advanced LIGO network offers only a modest improvement (a factor 1.1 -- 1.3, where 1.0 is no improvement) over existing constraints from electromagnetic observations of large-scale structures. In contrast, high redshift (up to $z \sim 10$) standard sirens from the future LISA satellite will improve constraints on the time evolution of the Planck mass in Horndeski theories by a factor $\sim 5$. By simulating different scenarios, we find this improvement to be robust to marginalisation over unknown merger inclination angles and to variation between three plausible models for the merger source population.
}


\maketitle
\flushbottom

\section{Introduction}
\label{sec:intro}

\noindent The first direct detection of gravitational waves (GWs) from a binary neutron star, event GW170817 \cite{LIGO2017,PhysRevLett.119.161101}, has had a deep impact on the model space of gravity theories proposed to modify General Relativity (GR). The observation of a gamma-ray counterpart arriving 1.7 seconds after the GW merger signal bounded the fractional relative difference in propagation speeds of GWs ($c_T$) and light ($c$) to be less than $10^{-15}$ \cite{Baker2017, Zuma2017, Creminelli2017,Sakstein2017,Boran_2018, Mastrogiovanni2020_jointtests} at redshift zero.\footnote{\noindent The bound here assumes simultaneous emission of GWs and gamma-ray photons at the source. A delay of up to $\sim 100$ seconds between their emission events is possible in some GRB models (though most emission models predict much shorter delays), which weakens the above bound to $10^{-13}$.} In turn, this result maps onto cosmological modified gravity (hereafter, MG) theories that can predict $c_T \neq c$. A small number of theories are incapable of satisfying $c_T = c$ for all redshifts, and are now effectively ruled out, e.g. subject to some fairly mild assumptions\footnote{For example, this result assumes that GW170817 does not occur at a `special' redshift, where the GW speed has been finely tuned to temporarily equal $c$. Further subtleties of the result are described in \cite{2019arXiv190509687I}.}, the quartic and quintic Galileon theories fall into this class. A larger set of theories are capable of satisfying $c_T=c$ for some ranges of their internal parameters, e.g. Generalised Proca theories, TeVeS \cite{Tasinato_2014,2014JCAP...05..015H,2019PhRvD.100j4013S}. Another set of theories do not affect the speed of propagation of GWs, and hence are unconstrained by event GW170817 (for example, $f(R)$ gravity).

The fact that plenty of theories `survived' this constraint is hardly surprising: in terms of cosmological distance measurements, GW170817 represents a single data point at very low redshift ($z\simeq 0.01$). Further tests of gravity will become possible when a larger sample of GW events are available over a wider redshift range. The purpose of this paper is to explore the improvement in constraints which may be obtained from these future tests, and in particular, how they interact with tests of gravity made using electromagnetic (hereafter, EM) observables. 

It is quite natural that EM and GW observations should be highly complementary. GWs probe the spin-2 (tensor) perturbations of an extended gravity theory, whilst traditional cosmological probes --- e.g. the Cosmic Microwave Background (CMB) and measurements of galaxy clustering --- are sensitive to the spin-0 (scalar) perturbations, and coupling between tensor and scalar perturbations is strongly suppressed at leading order. We will see in this paper that these features result in complementary sensitivities to parameters that describe generic deviations from GR. More pragmatically, the very different nature of these probes mitigates against significant systematic errors which may occur in one or the other individually.

Much recent attention has focussed on the prospects of using standard sirens~\cite{Schutz:1986gp,2005ApJ...629...15H} --- GW sources for which a redshift is known, through identification. of a host galaxy or other EM counterpart event --- to provide an independent measurement of the Hubble constant \cite{Chen_2018,Feeney_2018,Feeney_2019,Lagos_2019,2019arXiv190806050G,2019PhRvD.100j3523M,Farr_2019}. In addition, probabilistically assigning `dark sirens' to members of a galaxy catalogue allows them to be used for Hubble constant measurements even without a confirmed EM counterpart \cite{2019ApJ...876L...7S,2020arXiv200614961P}. These measurements can potentially arbitrate on the current tension between low- and high-redshift measurements of the parameter \cite{freedman2017cosmology,Verde_2019}. Given existing works on the subject, we will not pursue this line of enquiry here. Instead we will assume that by the next generation of GW detectors come online, this tension has been understood or resolved. Whilst we do allow the value of $H_0$ to vary in our analyses, our primary focus here is on testing extensions of the $\Lambda$CDM model that incorporate deviations from GR. We make the choice of including CMB data but not standard candle data in our constraints, as this allows us a consistent picture within which to simulate our catalogues of GW sources.

As we will see shortly, the key effect we study here accumulates with increasing propagation distance of a gravitational wave. As such, high-redshift massive black hole binaries (MBHs) observed by the LISA mission --- detected up to redshifts of potentially $z\sim10$ or higher --- are excellent candidates for constraining this aspect of modified gravity theories. These LISA binary mergers we consider are those at the centre of galaxies with mass $10^4$ -- $10^8 \, M_{\odot}$. The statistics of the massive black hole population are still highly uncertain, so we will adopt three commonly-used models that span the range of likely formation scenarios, and study their impact on our results. The gravitational waveform for a compact object binary is famously degenerate in the chirp mass and the redshift of the source, meaning identifying EM counterparts or host galaxies for which a redshift may be obtained remains crucial. Of course, this strongly restricts the number of GW sources which can be used for distinguishing cosmologies based on distance-redshift relations --- we discuss this issue in \cref{sec:gw_simulations}.

The potential of LISA to constrain deviations of the GW luminosity distance (hereafter, $d_{GW}$ --- we will introduce this formally in \cref{sec:theory}) from its expected value in GR has been previously studied in \cite{Belgacem_2018_gwlum, Belgacem_2018_modified, Belgacem2019, Barausse2020}. The authors of \cite{Belgacem_2018_modified} developed a very general parameterisation of changes to the effective luminosity distance, encapsulating a broad range of theories including Horndeski, DHOST and non-local gravity \cite{Crisostomi_2016,Langlois_2016,Ben_Achour_2016,Achour_2016,Maggiore_2014,2018JCAP...03..002B}. This parameterisation, along with bigravity oscillation effects \cite{Max_2017,Jim_nez_2020}, was studied lately in \cite{Belgacem2019}; the authors employed geometric probes of the background expansion rate to constrain $d_{GW}$ in combination with the constant dark energy equation of state parameter $w_0$, forecasting a headline bound of $\sim 4.5\%$ on $w_0$ and $2-5\%$ on modifications to $d_{GW}$. In this paper we adopt a different parameterisation, but we will connect our work to these results later on (see section \ref{sec:results}).

\begin{figure}
  \begin{centering}
    \includegraphics[width=\textwidth]{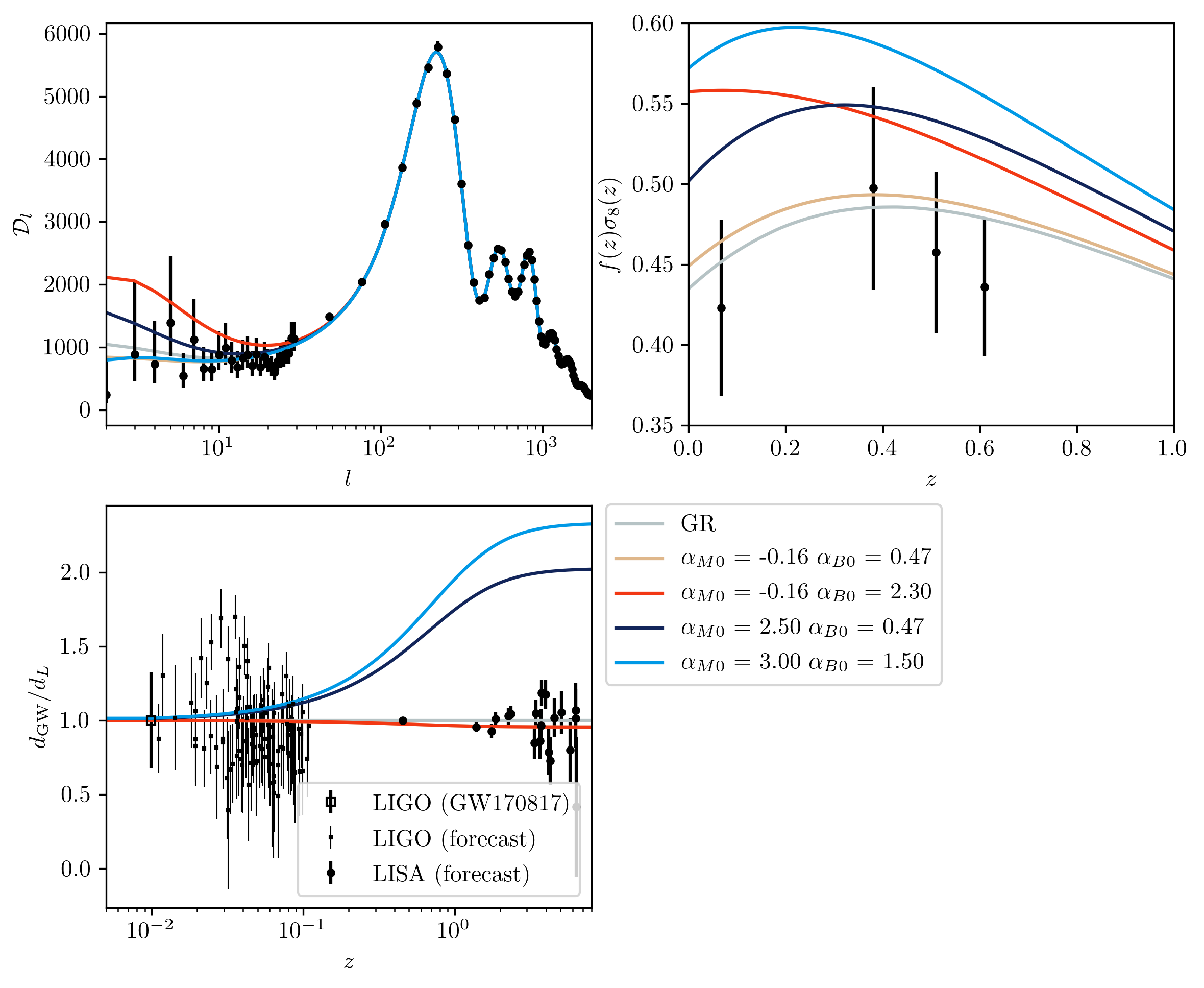}
  \end{centering}
  \caption{Cosmic Microwave Background temperature power spectrum, $D_\ell = \ell(\ell+1)C_\ell^{TT}/2\pi $ $[\mu K]^2$ (\emph{upper left}), Redshift Space Distortion (\emph{upper right}) and gravitational wave luminosity distance (\emph{lower left}) data considered in this work, shown for comparison with a selection of predictions from General Relativity (GR) and Horndeski models with parameterisation $\alpha_{X}\propto \Omega_{\Lambda}$. Theoretical curves are described in further detail in \cref{sec:theory} and data points in \cref{sec:gw_simulations} (LISA) and \cref{sec:experiments} (others).}
  \label{fig:test_models}
\end{figure}

In this paper we also consider the GW luminosity distance as measured by LISA, but we do so in combination with \textit{perturbative} probes of gravity on cosmological scales, as well as near-term data from the Advanced LIGO-VIRGO network. Through the use of carefully designed potential functions, many models of MG and dark energy are able to reproduce a $\Lambda$CDM-like expansion history.\footnote{Of course this represents a high degree of fine-tuning, but this is generic to nearly all current models of cosmological gravity, including $\Lambda$CDM.} As such the primary electromagnetic testing ground for extensions of GR is not their expansion history, but their ability to yield a modified theory of cosmological perturbations that consistently fits the matter power spectrum, redshift space distortions (RSD), weak lensing spectra, and all other observations of large-scale structure. The goal of this paper is put together these two puzzle pieces --- GW and EM probes --- and ask if together they can perform an effective `pincer movement' that tightly bounds MG models on all fronts (see \cite{palmese2020probing} for an analysis sharing the same philosophy). \Cref{fig:test_models} demonstrates the utility of this approach: a number of plausible models extending $\Lambda$CDM (parametrised by $\alpha_M$ and $\alpha_B$ --- see \cref{sec:theory}) are shown, alongside current CMB and RSD data in the top panels. The lower panel indicates the potential of GW standard siren data: in particular, GW observations at high redshift are highly inconsistent with models which are not confidently excluded by the CMB and RSD data alone.

In order to consistently link predictions for GW and EM observables, we necessarily must work within a more restricted set-up than the authors of \cite{Belgacem2019}. We choose to work in the framework of Horndeski theories of gravity \cite{1974IJTP...10..363H, 2009PhRvD..79f4036N,PhysRevD.84.064039}, which constitute the family of gravity models that couple one new scalar degree of freedom to the gravitational metric, and have at most second-order derivatives in their equations of motion (a feature connected to the energetic stability of the theory the via Ostrogradsky instability \cite{Ostrogradsky:1850fid, Woodard2007}). The Horndeski family has been widely adopted as a primary target for the next generation of galaxy surveys that aim to test models of dark energy and modified gravity; for example, it was recently selected as a priority for tests of gravity by the DESC consortium of the Vera Rubin Observatory \cite{2019arXiv190509687I}. As such, it is an ideal framework in which to study the complementarity of EM and GW data (see \cite{Frusciante_2020} for a useful review). Constraints from current EM data on the parameters in Horndeski theories are already available \cite{Bellini_2016,Kreisch_2018,2019PhRvD..99j3502N, Spurio_Mancini_2019}, and here we closely follow their approaches alongside the addition of the GW data. For future combinations of a `Stage IV' CMB experiment, Vera Rubin Observatory data and intensity mapping data from the Square Kilometer Array, \cite{Alonso_2017} provide forecasts Fisher matrix forecasts, finding expected improvements of a factor $\sim 5$ over current experiments, whilst \cite{Reischke_2018, Spurio_Mancini_2018} considered stage IV CMB, tomographic galaxy clustering and cosmic shear.

The structure of this paper proceeds as follows: in \cref{sec:theory} we present the framework of Horndeski scalar-tensor gravity and explore its effects on GW propagation. In \cref{sec:gw_simulations} we simulate catalogues of LISA detections and their measurement errors. \Cref{sec:experiments} describes the mock LIGO catalogues and other data we will use in our analysis, and \cref{sec:forecasting} our likelihood and sampling methods. Our results are presented in \cref{sec:results} and we discuss their wider implications in \cref{sec:conclusions}.

\section{Gravitational Waves in Scalar-Tensor Theories}
\label{sec:theory}

\subsection{Horndeski Gravity}
\label{sub:theory_horndeski}
Horndeski gravity~\cite{1974IJTP...10..363H, 2009PhRvD..79f4036N,PhysRevD.84.064039} is a broad category of scalar-tensor theories that encompasses well-known models such as quintessence, k-essence, KGB, f(R) gravity, DGP gravity, the Galileon family, and many others. Each term in the Lagrangian of Horndeski gravity is parameterised by a coefficient function of the scalar degree of freedom, $\phi$, and its kinetic term, commonly denoted by $X = -\nabla^\mu\phi\nabla_\mu\phi/2$. Individual models correspond to setting these coefficients to particular functional forms. It is in this sense that Horndeski gravity is a `parent' theory; an expression derived in terms of the general Horndeski coefficients can be specialised to any of its `offspring' theories by substituting in the corresponding functional forms. We choose to work with the Horndeski Lagrangian here because it provides a concrete, yet flexible, system in which to connect EM and GW constraints on a common set of parameters. Its parameters have been widely tested using electromagnetic observables \cite{Bellini_2016,Kreisch_2018,2019PhRvD..99j3502N, Spurio_Mancini_2019}. We note in passing though, that the general effects we study here --- predominantly modifications to the GW luminosity distance --- are also present in theories outside of the Horndeski family, for example non-local gravity, bigravity and generalised Proca theories. In fact, \cite{Lagos_2018} showed that the key parameter $\alpha_M$ we introduce in eq.~(\ref{aM}) below has corresponding equivalents in the most general vector-tensor and tensor-tensor gravity theories. Hence our study can be adapted to other families of gravity theories with minor modifications.

In the aftermath of GW170817, some of the terms in the original Horndeski action are no longer viable (subject to some mild assumptions, as laid out in section 3.3 of \cite{2019arXiv190803430B}), because they predict GWs propagating at speeds significantly different from $c$. Here we will assume these constraints hold at all redshifts (see \cite{Bonilla2020} for forecasts without this assumption). Hence we will show here only the remaining \textit{reduced} Horndeski action, consistent with luminal GWs \cite{heisenberg2020horndeski}:
\begin{equation}
S=\int d^4x \sqrt{-g}\left[\frac{1}{2}M_{\mathrm{eff}}^2(\phi)R+K(\phi,X)-G_3(\phi, X)\square\phi\right]+{S}_m\left(g_{\mu\nu}, \psi_m\right)
\label{rHD_action}
\end{equation}
where $\psi_m$ are matter fields minimally coupled to the metric $g_{\mu\nu}$, and $S_m$ denotes the matter action. Note that the conformal coupling to the Ricci scalar, $M_{\mathrm{eff}}^2$, is a function of $\phi$ \textit{only}. GR is recovered in the limit $K=G_3=0$, $M_{\mathrm{eff}}^2=M_P^2$, where $M_P$ is the canonical Planck mass. Using the equivalence of action \ref{rHD_action} to KGB gravity with non-minimal matter coupling, \cite{Deffayet:2010qz} showed that this action has luminal GWs around arbitrary spacetime backgrounds.

A non-constant $M_{\mathrm{eff}}$ function has the interpretation of an evolving effective Planck mass. The rate of evolution of this effective Planck mass is encapsulated in the widely adopted `property function' first introduced in \cite{Bellini:2014fua}:
\begin{equation}
\label{aM}
\alpha_M(z)\equiv \frac{\mathrm{d} \ln \left(M_{\mathrm{eff}}/M_P\right)^2}{\mathrm{d}\ln a}
\end{equation}
where we suppress the argument of $M_{\rm eff}$ for clarity.
The function $\alpha_M(z)$, along with two others, parameterises the dynamics of linear cosmological perturbations resulting from eq.~(\ref{rHD_action}) when expanded about a $\Lambda$CDM-like cosmological background.\footnote{By this we mean a cosmology for which the only non-standard addition is an effective fluid dark energy equation of state, usually encapsulated by the standard CPL $\{w_0, w_a\}$ parameters.} The other two property functions describing the perturbative dynamics are defined by \cite{Bellini:2014fua}:
\begin{align}
\aK(z) &= \frac{1}{H^2\Meff^2}\left\{2X \left(K_X + 2X K_{XX} - 2G_{3\phi} - 2XG_{3\phi X}\right) + 12\dot{\bar\phi}XH \left(G_{3X} + XG_{3XX}\right)\right\} \label{aKdef}\\
\aB(z) &=\frac{2\dot{\bar\phi}}{H\Meff^2}  \left(XG_{3X} - M_{{\rm eff},\phi} \right) \label{aBdef}
\end{align}
where dots indicate derivatives with respect to (physical) time, $\bar\phi$ indicates the homogeneous component of the Horndeski scalar, and subscripts indicate derivatives, e.g. $G_{3X}\equiv \partial G_3(\phi,X)/\partial X$. The functions defined by eqs.~(\ref{aKdef}) and (\ref{aBdef}) are often referred to as the \textit{kineticity parameter} and \textit{braiding parameter}, respectively. This is slightly misleading terminology since all three $\alpha_i$ are functions of time (not constant parameters); but the expression is common enough that we will adopt it hereafter. The names stem from attempts to describe physically their role in a gravity theory: the kineticity parameter, in loose terms, parameterises the kinetic energy of the Horndeski scalar field, and affects the sound speed of its perturbations. The braiding parameter quantifies mixing between the kinetic terms of the scalar field and the metric, and causes the scalar field perturbations to cluster.

In the original Horndeski formulation, two more parameters are important. One, $\alpha_T$, describes deviations of the GW propagation speed as $c_T^2=c^2[1+\alpha_T(z)]$; as described above, observations of GW170817 imply this is effectively zero in the form of eq.~(\ref{rHD_action}). The other, $\alpha_H(z)$, is known as the Beyond Horndeski parameter, and allows for generalisations of the original Horndeski Lagrangians under disformal transformations \cite{Langlois_2016,Ben_Achour_2016,Achour_2016}. However, non-zero values of $\alpha_H(z)$ have been found to destroy the energetic stability of GWs \cite{Creminelli_2018}, permitting them to rapidly decay into perturbations of the Horndeski scalar field \textit{if} the sound speed of the perturbations is equal to unity, $c_s^2=1$. On these theoretical grounds, we have set $\alpha_H=0$ in this work. More recently, further investigations have indicated that the stability of perturbations to the scalar in the presence of GWs may imply a further bound linking two of the property functions as $|\alpha_B +\alpha_M|\lesssim 10^{-2}$ \cite{Creminelli_2019, Noller2020} in some subsectors of the Horndeski family. We will explore this in \cref{subsec:creminell_prior}. 

There is an important subtlety regarding bounds on $c_T$, and by implication, the Horndeski parameter $\alpha_T(z)$. The authors of \cite{PhysRevLett.121.221101} point out that in a low-energy effective field theory (EFT) of dark energy, $c_T$ must return to unity at some energy scale in order to permit a Lorentz-invariant UV completion of the EFT. They argue that this energy scale \textit{could} be as low as $\left(M_P H_0^2\right)^{1/3}$, which is $\sim 260$ Hz when expressed as an equivalent frequency. This is dangerously close to the frequency range event GW170817 swept through before merger \cite{PhysRevLett.119.161101}, raising the possibility that the constraints reported above were obtained in a regime when modified gravity effects are already suppressed. This motivates further tests of the speed of GW propagation at lower frequencies with LISA standard sirens. However, as this is not the focus of our present work, we will use Lagrangian \ref{rHD_action} which has $\alpha_T=0$ at all times by construction.

So it is that we are left with three remaining property functions $\{\aM, \aB, \aK\}$ which span the general space of scalar-tensor theories consistent with initial LIGO-Virgo GW detection. Note that these parameters describe the \textit{perturbative} dynamics of a theory; the background cosmological expansion rate must be independently specified via the standard $\Lambda$CDM parameters and dark energy equation of state, $w(z)$, which we take here to be $-1$; see \cite{Belgacem_2018_modified, Belgacem2019} for a study including a modified expansion history.

\subsection{GW Propagation in Horndeski Gravity}
\label{sec:theory_gw_horndeski}
The behaviour of cosmologically propagating GWs in reduced Horndeski gravity is given by studying the equation of motion for tensor perturbations on an FRW background metric. Numerous authors have provided derivations (for example \cite{Bellini:2014fua, Lagos_2018}), so we will simply present the resulting equation here:
\begin{equation}
\label{hprop}
h^\pp_{A}+\left[2+\alpha_M(z)\right]\calH h_{A}^\p+k^2h_{A}=0
\end{equation}
where $h$ is a linear tensor perturbation of the metric, $A$ indicates either of the two $+, \,\times$ GW polarisations, primes denote derivatives with respect to conformal time, and ${\cal H}=a^\p/a$ is the conformal Hubble factor. In more general scenarios, a massive graviton, non-luminal GWs, and a non-zero source of effective anisotropic stress can all result in further modifications to this equation (see \cite{2018PhRvD..97j4037N, Lagos_2018, 2018JCAP...08..030A} ). However, all of these effects require more exotic kinds of deviations from GR than are possible in the reduced Horndeski scalar-tensor class.

In this paper we will assume that some form of screening is an essential feature of viable extensions of GR. Screening is the generic name for a set of properties found commonly in modified gravity theories, which act to suppress deviations from GR in certain environments --- extensive reviews can be found in \cite{Joyce_2015, 2019arXiv190803430B}. Generally these environments equate to those with densities substantially above the cosmic mean density (such as galaxy interiors), providing a mechanism by which the stringent local tests of gravity can be automatically respected \cite{Will_2014}. The result of this assumption is that we take the generation of GWs during a binary merger to occur as in GR; we focus instead on modifications to these waveforms that take place during the propagation of the GW towards Earth.

On these grounds, we can write the solution of eq.~(\ref{hprop}) as $h_{MG,A} = {\cal C}\, h_{GR,A}$, where $h_{GR,A}$ is the GR solution for the amplitude. The GW amplitude changes over cosmological distances and timescales, such that $|{\cal C}^\prime/C|\ll |k|$, where $k$ is the GW wavenumber (even for LISA sources, GW wavelengths are a tiny fraction of a parsec); this justifies the use of the WKB approximation. One can then solve eq.~(\ref{hprop}) for ${\cal C}$ \cite{2018PhRvD..97j4037N}: 
\begin{eqnarray}
{\cal C}&=&{\rm exp}\left[\frac{1}{2}\int_0^x dx\; \alpha_M(x)\right]\equiv \frac{M_{\rm eff}(x)}{M_{\rm eff}(x=0)}
\label{Cfac}
\end{eqnarray}
where $x=\ln a$ and the second equality employs eq.~(\ref{aM}). The two equivalent expressions above allow one to choose between specifying the evolution of the property functions, or the values of the  effective Planck mass at source and observer (see section \ref{sec:screening} for some subtleties of this point).

To lowest PN order, the plus and cross polarisation amplitudes for a circular, inspiralling binary in GR are given by \cite{Nissanke_2009}:
\begin{eqnarray}
h_+(\tau_{\rm obs})&=&h_0 \;\frac{\left(1+\cos^2\iota\right)}{2}  \;\cos[2\Phi(\tau_{\rm obs})], \quad\quad
h_\times(\tau_{\rm obs})=h_0 \;\left[-\cos\iota\right] \;\sin[2\Phi(\tau_{\rm obs})]  \label{hamp2}\\
h_0&=&\frac{4}{d_L}\left(\frac{G_N\Mc}{c^2}\right)^{5/3}\left(\frac{\pi f_{gw}^{\rm obs}}{c}\right)^{2/3} \label{hamp1}
\end{eqnarray}
where $f_{gw}^{\rm obs}$ and $\tau_{\rm obs}$ are the GW frequency and time to coalescence measured in the observer's frame. $\Mc$ is the redshifted chirp mass of the binary, \mbox{$\Mc = (1+z)\mu^{3/5} m^{2/5}$}, where \mbox{$m=m_1+m_2$} is the total binary mass, and \mbox{$\mu=m_1 m_2/m$} is the reduced mass. $\iota$ is the inclination angle between the rotation axis of the binary and the observer's line of sight, i.e. \mbox{$\iota=0^\circ$} corresponds to a face-on system. The luminosity distance $d_L$ is given by:
\begin{eqnarray}
d_L&=&c (1+z)\int_0^z\,\frac{d z^\prime}{H(z^\prime)},
\label{eq:dLexpr}
\end{eqnarray}
and the phase factor $\Phi$ is (where $\Phi_0$ is the value at coalescence, an integration constant):
\begin{eqnarray}
\Phi(\tau_{\rm obs})&=&\Phi_0-\left(\frac{5G_N\Mc(z)}{c^3}\right)^{-\frac{5}{8}}\tau_{\rm obs}^{\frac{5}{8}}
\end{eqnarray}
For gravity models represented by eq.~(\ref{rHD_action}), the phase of the waveform is not altered from its GR behaviour. A related property is that the frequency evolution of the waveform is governed by standard relations, and measurement of this allows for accurate determination of the chirp mass $\Mc$ via the relation:
\begin{eqnarray}
\frac{df_{gw}}{d\tau_{\rm obs}}&=&\frac{96}{5}\pi^{\frac{8}{3}}\left(\frac{G_N\Mc}{c^3}\right)^{5/3}f_{gw}^{11/3}
\label{eqn:fevolve}
\end{eqnarray}
The two remaining unknowns governing the waveform amplitude are then the luminosity distance $d_L$, and the inclination $\iota$. We will consider the effects on any degeneracy between these two parameters in \cref{subsec:dl_error}.

Under Horndeski extensions of GR, eq.~(\ref{hamp1}) is multiplied by the factor $\cal C$ in eq.~(\ref{Cfac}). This modification is absorbed into a redefinition of the luminosity distance appearing in the GW amplitudes \cite{2018PhRvD..97j4037N,Belgacem2019, Lagos_2019, Dalang_paper2, mastrogiovanni2020probing,dagostino2019}. The resulting effective GW luminosity distance, $d_{GW}$, is given by:
\begin{eqnarray}
 {d_{GW}}&=&{\cal C}^{-1}\,{d_L}={d_L}\,{\rm exp}\left[-\frac{1}{2}\int_0^x dx\; \alpha_M(x)\right]
 \label{eq:dgw_ratio}
\end{eqnarray}
Note that via eq.~(\ref{Cfac}), the ratio $d_{GW}/d_L$ can be expressed in terms of the effective Planck masses at the source and observer; \cite{Dalang_2019} proved the generality of this result for general background spacetimes.

Let us suppose that we have a electromagnetically-confirmed redshift for a GW event. The apparent slip between the luminosity distance inferred from the GW amplitude and the one implied by the source redshift (via the standard formula of eq.~\ref{eq:dgw_ratio}) is a potentially detectable signature of modified gravity effects. Whilst for a single event the uncertainties on $d_{GW}$ inferred from the waveform amplitude are likely to be large, a set of tens or more events could offer constraints on $\aM(z)$ that are entirely independent of existing EM probes. Furthermore, as \cref{fig:test_models} indicates, the cumulative nature of the distance slip favours LISA GW sources, which should be detectable out to redshifts five and beyond (see \cite{dagostino2019} for an extension to the Einstein Telescope).

In reality, only a fraction of LISA GW events are expected to have detectable EM counterparts or identifiable host galaxies from which spectroscopic redshifts can be identified. Exactly how many is highly uncertain at present, and depends sensitively on the population model and gas environments of binary MBHs. 
We discuss the models we use for this in \cref{sub:em_counterparts}.

\subsection{Large-Scale Structure in Horndeski Gravity}
\label{sec:theory_lss_horndeski}
The effective luminosity distances of GW sources are directly sensitive to the property function $\aM$ alone. This is distinctly different to the behaviour of scalar cosmological perturbations in Horndeski gravity, where all three of the property functions $\left\{\aM, \aB, \aK\right\}$ feature in the linearised field equations. Implementation of these modified field equations in an Einstein-Boltzmann solver is available in the {\tt hi$\_$class} code \cite{Blas_2011,2017JCAP...08..019Z, Bellini_2020}, and in equivalent schemes \cite{Hu_2014,Raveri_2014,Bellini_2018}. We will not reproduce the full system of equations solved by these codes here. As an indicative example, the Fourier-space Poisson equation corresponding to the Lagrangian eq.~(\ref{rHD_action}) in conformal Newtonian gauge is \cite{Bellini:2014fua}:
\begin{eqnarray}
\label{eqn:poisson}
&&-2\frac{k^2}{a^2}\Phi=\frac{8\pi G_N\rho_{m}}{M_{\rm eff}^2}\Delta-3\aB H\dot{\Phi}-H^2\Psi(\aK+3\aB)-\aK H^2\delta\dot{\varphi} -\aB H\delta\varphi\left(\frac{k^2}{a^2}-3\dot{H}\right)\nonumber\\
\end{eqnarray}
For convenience the scalar field perturbation appearing in these equations has been normalised as \mbox{$\delta\varphi = -\delta\phi/\dot{\bar\phi}$}, where $\phi$ is the scalar field in the Lagrangian of eq.~(\ref{rHD_action}).   

The role of $\aK$ in this equation requires comment. It always appears multiplied by a term of order $\sim H^2$ and a metric potential or scalar field perturbation. These are the terms that are suppressed in the widely-adopted quasistatic limit $|k^2\gg H^2|$, i.e. a regime of subhorizon scales on which spatial derivatives dominate over time derivatives of variables.\footnote{We note that large values of $\aK$ can lower the effective sound horizon of the scalar field \cite{Bellini:2014fua, Sawicki_2015} to be smaller than the standard cosmological horizon. However, the effects of $\aK$ on cosmological perturbations are only significant above the scalar sound horizon, so unless $\aK$ takes very large values, it still evades constraint.} The quasistatic approximation applies to the scales probed by galaxy surveys and GW experiments; it has been validated for specific gravity models \cite{PhysRevD.89.023521} and is expected to hold for generic members of the Horndeski class \cite{PhysRevD.87.104015, perenon2020improvements, linder2020limited}. As a consequence of their accompanying factors, all terms in eq.~(\ref{eqn:poisson}) that contain $\aK$ are neglected when the quasistatic limit is taken. This feature is repeated in the other components of the linearised gravitational field equations, though we do not show them explicitly here. 

The result of the quasistatic approximation is that $\aK$ completely drops out of the perturbation equations in the subhorizon limit, rendering it largely unconstrained by the data we consider in this work. Hence we will focus on constraining only the property functions $\aB$ and $\aM$ from here on. Some physical insight into the irrelevance of $\aK$ can be gleaned by realising that in models representable as perfect fluids, e.g. pure quintessence models, $\aK$ is the only non-zero Horndeski parameter. In this type of model the modifications to the field equations depend on the factor $(w-1)$, rendering them very small if the dark energy equation of state is close to $-1$. Indeed, in this work we will assume the background expansion rate of the universe is close to that of the standard $\Lambda$CDM model, rendering the $\aK$ parameter largely ineffective. 

\Cref{fig:test_models} shows some examples of Horndeski models with several values of \mbox{$\alpha_M(z=0)$} and \mbox{$\aB(z=0)$}. The upper left panel shows predictions for the CMB TT power spectrum overlaid with \emph{Planck} 2015 data. The upper right panel shows the scale-independent growth rate, \mbox{$f=d\ln D/d\ln a$} (where $D(a) = \delta_m(a)/\delta_m^{ini}$ is the usual growth factor) multiplied by the power spectrum amplitude normalisation parameter, $\sigma_8$. Overlaid on this plot are four data points from current Redshift Space Distortion surveys (details in \cref{sub:rsd}).

The measurements shown in the upper right panel of \cref{fig:test_models} are obtained under the assumption that the growth rate $f(z)$ is independent of scale, as occurs in $\Lambda$CDM. In practice, this means that data from all $k$-bins of the redshift space galaxy power spectrum are fitted using a single (redshift-dependent) value of $f$, up to some cut-off $k_{NL}$ where the model becomes invalid due to non-linearities. In contrast, scale-independence of the growth rate does not generally hold in modified gravity models. If the scale-dependence is strong, then it becomes questionable whether measurements such as those in \cref{fig:test_models} can be used at all without a careful reanalysis of the original power spectrum data.

Fortunately, the authors of \cite{2019PhRvD..99j3502N} showed that in a large region of Horndeski parameter space, the growth rate becomes virtually scale-independent below a certain transition scale, $k_t$, such that scale-independence holds well for $k_t<k<k_{NL}$. Therefore we can use the RSD measurements detailed in section \ref{sub:rsd} provided we apply them at a fiducial wavenumber in this regime. We choose $k_{\rm fid}=0.05$h Mpc$^{-1}$, and stay within the range of $\{\aB, \aM\}$ values explored by \cite{2019PhRvD..99j3502N}.

\subsection{Ansatzes for the $\alpha_i(z)$}
\label{sec:ansatz}

For a given model of modified gravity, the functional forms $K(\phi, X)$, $G_3(\phi, X)$ and $\Meff^2(\phi)$ in eq.~(\ref{rHD_action}) are known explicitly, and the corresponding $\alpha_i$ functions could be worked out via eqs.~(\ref{aM}-\ref{aBdef}). However, this is not the intended use case of parameterised formalisms for testing gravity. Instead, we generally wish to test an entire space of models by constraining the $\alpha_i$ parameters directly. The operational downside of such generality is that without a specific Lagrangian, we don't know the evolution of the homogeneous scalar field component $\dot{\bar\phi}$. We therefore do not have access to the precise redshift-dependence of the $\alpha_i$ parameters.

The most common approach to make an \textit{ansatz} for the time evolution of the $\alpha_i$, i.e. we assume a particular functional form that itself contains a handful of parameters. This us allows to translate unknown functional degrees of freedom into unknown constant parameters. The two forms below are the most commonly adopted ansatzes for the Horndeski alpha parameters:
\begin{align}
  \quad\quad  \alpha_i(a) &= \alpha_{i0} \Omega_\Lambda(a) &
    \alpha_i(a) &= \alpha_{i0} a
    \label{ansatzes}
\end{align}
We will explore both of these options below; in either case, our key MG parameters of interest are then $\{\alpha_{M0}, \alpha_{B0}\}$. For the parameter ranges explored in this paper, our functional forms lie comfortably within the `safe' zone represented by figure 2 of \cite{mastrogiovanni2020probing}, and hence produce a monotonically increasing GW luminosity distance.

The two ansatz forms in eq.~(\ref{ansatzes}) are smooth, monotonically increasing functions that reduce to zero at high redshifts and approach order unity at low redshifts. These represent generic properties that we expect viable deviations from GR to have, and hence motivates their use here. We stress, however, that these forms are not derived from fundamental theory; they should be considered as approximations, at best, which could allow our parameterised framework to `catch' the signatures of underlying new physics. 

We note that these ansatz forms have been criticised in \cite{Linder_2016, Linder_2017} for not accurately representing the evolution of some mainstream gravity models, e.g. versions of $f(R)$ gravity. However, \cite{2017PhRvD..96f3516G} have argued that observables are not sensitive to the precise behaviour of the MG functions on short timescales, and that smooth ansatzes described by two constant parameters are the preferred model (as quantified by the Bayesian Information Criterion) for $86\%$ of the theory space they consider. An example of this would be the form $\alpha_i =\alpha_{i0} a^p$, which was considered in \cite{Kreisch_2018, 2019PhRvD..99j3502N} in addition to eqs.~(\ref{ansatzes}). However, those works found the power-law indices to be essentially unconstrainable with current EM data. For this reason, we choose to focus on single-parameter forms in our work. Developing new ansatzes or model-specific treatments is beyond the scope of the present paper; we will adopt eqs.~(\ref{ansatzes}) because they are the standard in the literature, and therefore offer the best opportunity for comparison of our results to existing work.

There is a further set of restrictions imposed upon the $\alpha$ functions, regardless of the ansatz used. These ensure that two types of pathologies are not present in the Horndeski model under consideration. These are i) ghost instabilities, when the kinetic term of perturbed fields becomes negative, and ii) gradient instabilities, when the sound speed of the perturbations becomes negative, leading to their exponential growth on a (potentially short) timescale related to the cut-off of the effective theory. The following conditions are necessary for their avoidance, and are automatically checked in the \hiclass software we will use in our analyses:
\begin{align}
Q_{\text{S}} & =\frac{2\Meff^{2}D}{(2-\alpha_{\textrm{B}})^{2}}>0\,,\qquad\qquad Q_{\text{T}}  =\frac{\Meff^{2}}{8}>0\label{eq:tensorstab} \\ 
c_{\text{s}}^{2} & =-\frac{\left(2-\alpha_{\textrm{B}}\right)\left[\dot{H}-\frac{1}{2}H^{2}\alpha_{\textrm{B}}-H^{2}\alpha_{\textrm{M}})\right]-H\dot{\alpha}_{\textrm{B}}+\left({\rho}_{\textrm{m}}+{p}_{\textrm{m}}\right)/\Meff^2}{H^{2}D}>0\label{eq:scalarstab} 
\end{align}
where $Q_{\text{S}}$ and $Q_{\text{T}}$ are coefficients of the kinetic terms of scalar and tensor perturbations, and
\begin{align}
 D&\equiv\alpha_{\textrm{K}}+\frac{3}{2}\alpha_{\textrm{B}}^{2}\,.
\end{align}
We see at a glance from eq.~(\ref{eq:tensorstab}) that the value $\alpha_B = 2$ could cause problems within our calculations. Fortunately, the data will be sufficiently constraining to keep our likelihood contours well below this boundary.

\subsection{Can the Planck Mass be Screened?}
\label{sec:screening}
Drawing on concerns first expressed in \cite{2018JCAP...08..030A}, the authors of \cite{Dalang_2019} have recently suggested that in some models, the requirement of screening at a GW source eliminates any detectable modification of its GW luminosity distance. In other words, the assumptions that permit the GW generation process to happen as per GR result in GW propagation being also virtually identical to GR. Conversely, if the GW source is unscreened then the GW propagation can be significantly modified, \textit{but} the stringent Solar System constraints on gravity then apply \cite{Burrage2020}.

To understand when this restriction applies, we must be aware that under modifications of GR, generally two different gravitational couplings become relevant. One of these, $G_N(t)$ (where we have denoted a possible time-dependence) describes the gravitational coupling between two matter sources, whilst the other, $G_{gw}(t)$ is the coupling that appears in the quadratic self-interaction term of the action for gravitational waves. GR has the property \mbox{$G_N = G_{gw} =$} constant, but modified gravity theories can generically break this equivalence. 

$G_{gw}(t)$ is related to the effective Planck mass appearing in eq.~(\ref{rHD_action}) as $8\pi G_{gw}(t)= \Meff^{-2}(t)$, so it is this  coupling that alters the effective luminosity distance of GW sources via eqs.~(\ref{aM}) and (\ref{hprop}). Conversely, Solar System constraints such as those from Lunar Laser Ranging experiments bound the evolution of $G_N(t)$. Binary pulsar constraints, such as those from the spin-down rate of the Hulse-Taylor pulsar, in principle are sensitive to the time variation of both gravitational couplings. However, the authors of \cite{2019arXiv191010580W} calculate that the effects of a plausible variation in $G_{gw}$ on the period decay of a local binary system are many orders of magnitude suppressed with respect to the effects of a change in $G_N$. As a result, $\dot{G}_N/G_N$ is constrained to $\sim 10^{-3}$, whilst $G_{gw}$ is effectively unconstrained by observations of the binary pulsar orbital decay rate.

The statements put forwards in \cite{Dalang_2019} apply to theories in which $G_N(t) \simeq G_{gw}(t)$. For example, in scalar-tensor theories that screen via the Vainshtein mechanism, one can show that in the screened limit (well below the Vainshtein radius) the gravitational coupling in the Poisson equation is equivalent to $8 \pi M_{\rm eff}^2$. In this case, the effective Newton's constants controlling Solar System dynamics (say) and cosmological GW propagation \textit{are} effectively the same. If we insist that $8 \pi M_{\rm eff}^2$ matches Lunar Laser Ranging measurements of Newton's constant at all redshifts, then $\alpha_M=0$ and GW propagation is unmodified.

In the chameleon screening mechanism, a general argument presents that the conformal factor relating the Einstein and Jordan frame --- which also controls the density-dependence of the scalar potential --- is not able to evolve rapidly over cosmological distances and times, in order to preserve successful screening \cite{PhysRevLett.109.241301}. Since this conformal factor is related to $M_{\rm eff}$ in eq.~(\ref{rHD_action}), one generically expects that $M_{\rm eff}$ must evolve slowly too, constraining the potential impact on GW luminosity distances to be small. However, the specifics here must be worked out in a particular model.

In this paper we will maintain an agnostic approach to the underlying mechanisms by which GR is recovered near a GW source. Although the above arguments indicate that GW luminosity distances are not a good test of Vainshtein and most chameleon-screened theories (though see \cite{koyama2020testing}), we cannot rule out the possibility of new screening mechanisms, or new properties of existing ones, being discovered in future. Indeed, recent works by \cite{Creminelli_2019} have found that GW predictions of scalar degrees of freedom (here described in an EFT framework) depend in a subtle way on the presence of matter sources and their screening. Furthermore, as we mentioned in the introduction, a modification of luminosity distances is a general property of gravity theories beyond the scalar-tensor sector. As such, the ability of LISA data to constrain these properties remains key to agnostic tests of gravitational laws on cosmological distance scales.

\section{Simulated LISA Data}
\label{sec:gw_simulations}
\Cref{fig:test_models} demonstrates the ability of LISA observations of massive ($10^4-10^8 \, M_{\odot}$) binary merger standard siren measurements of luminosity distances to usefully constrain Horndeski modified gravity models, in addition to the cosmography constraints given by \cite{Tamanini2016, wang2019}. Here, we detail our procedure for simulating catalogues of LIGO standard siren events, following closely the prescription of \cite{Tamanini2016}. In particular, we discuss our choices of merger progenitor populations, the signal-to-noise ratios (SNR) at which they may be detected by LISA, detection of electromagnetic counterparts, and how uncertainty on merger inclination angle $\iota$ propagates into the uncertainty on the standard siren luminosity distance.

\subsection{GW Merger Populations}
\label{subsec:gw_populations}
A major source of uncertainty in the modelling of massive black hole merger events is the production mechanism and mass of the `seeds' from which they grow. One key scenario considered is that of `light' seeds, which are formed from the collapse of population III stars at redshifts $z\simeq 15-20$, and have masses up to a few hundred solar masses. An alternative to this is `heavy seed' models, in which the inflow of cold gas onto protogalactic nuclei leads to seed black holes of masses up to $~10^5$ M$_\odot$, at similar redshifts.

The different seed mass scenarios are then further modulated by the delay between the merger of two galaxies and the merger of their central black holes. After a galaxy merger, the two MBHs sink rapidly to the minimum of the new galactic potential by dynamical friction and form a bound system \cite{Begelman:1980vb}. The binary is then subject to the famous `last parsec problem': unless there is either a substantial amount of gas in the new galactic centre or frequent three-body interactions, it is likely that the binary cannot lose sufficient energy to merge within a Hubble time \cite{2003AIPC..686..201M, Vasiliev_2015}.
\begin{figure}
    \centering
    \includegraphics[width=0.6\textwidth]{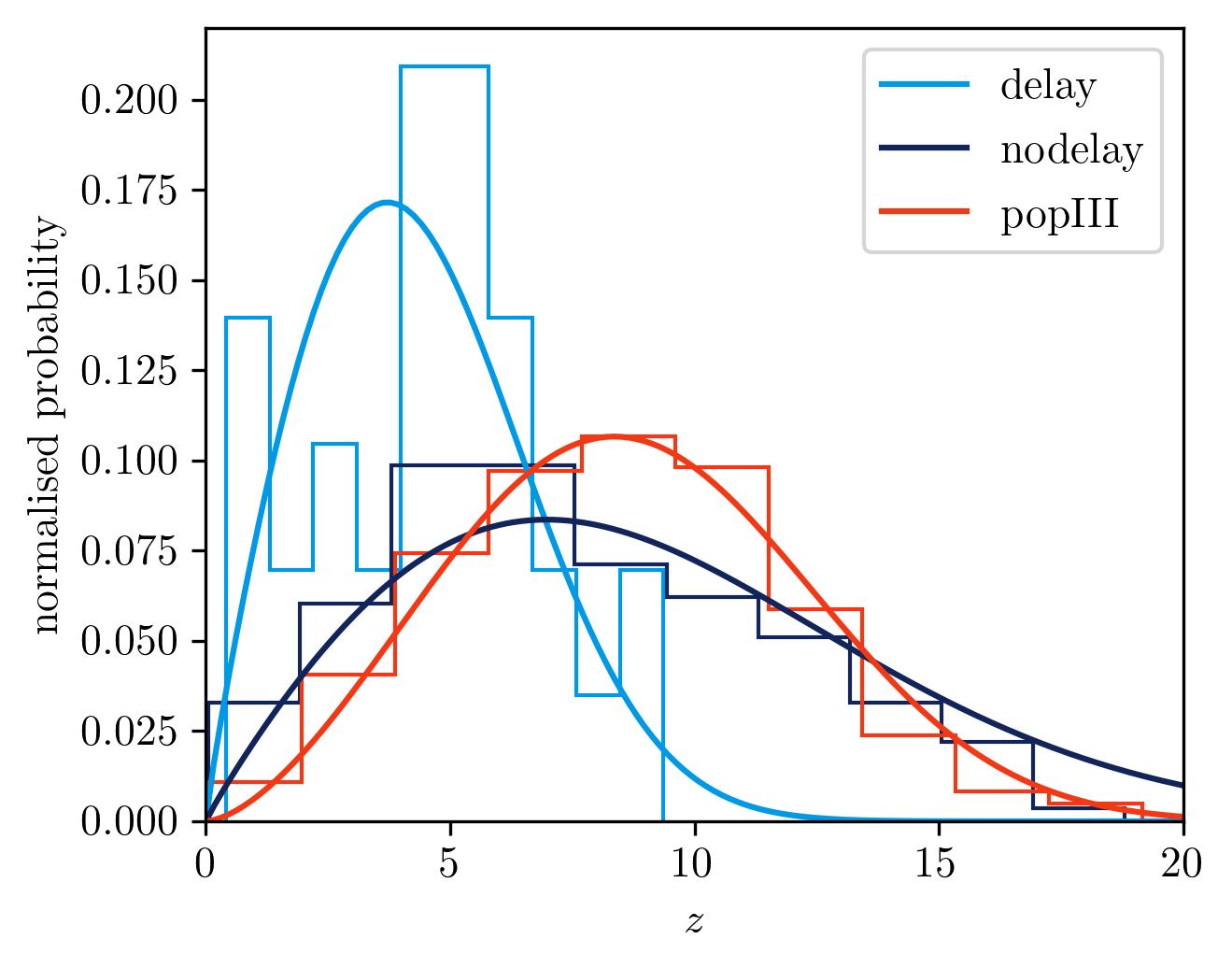}
  \caption{Discrete and normalised redshift probability distributions for the three MBH binary population models discussed in \cref{subsec:gw_populations}. These are sampled to generate our mock input catalogues of merger events.}
  \label{fig:pops-nz}
\end{figure}
With these considerations in mind, \cite{Barausse_2012,2016PhRvD..93b4003K,Tamanini2016} studied three models for MBH source populations, which we will use here. These models are characterised by the mass and redshift distributions of the mergers. For the redshift distributions we will assume a form of:
\begin{equation}
    \label{eqn:nz_gw}
    n(z) =  z^\alpha \exp{\left[-(z/z_0)^\beta\right]}.
\end{equation}
for each of the three populations described below.
\begin{itemize}
     \item \textbf{Population III:} Constraints using this model are labelled `popIII' in our plots. In this model, the light seed mechanism dominates, producing a MBH population with a substantial tail to lower masses, reaching down to $\sim 10^4 \,M_\odot$. Delay between host galaxy and MBH mergers is included. However, for light-seed models the authors of \cite{2016PhRvD..93b4003K} found the delay between the mergers has a small (order unity) effect on LISA event rates, so the case for these progenitors without a delay is not considered separately here. 
     
      The red curve of \cref{fig:pops-nz} shows the redshift distribution of merger events in this model. To obtain this, we discretize the merger rates presented in \cite{2016PhRvD..93b4003K} (using the model and code originally devloped in \cite{Barausse_2012}), and fit a smooth model using \cref{eqn:nz_gw} --- this process removes some of the statistical fluctuations resulting from the simulations in \cite{2016PhRvD..93b4003K}. For the popIII case the best-fit parameters are $\lbrace \alpha, \beta, z_0 \rbrace = \lbrace 1.60, 2.75, 10.2 \rbrace$.
    \item \textbf{Delay:} Constraints using this model are labelled `delay' in our plots. In this model, the heavy seed mechanism operates. It is assumed that the last parsec problem is significant, leading to a slow final inspiral and consequent low merger rate. As \cref{fig:pops-nz} shows, this pushes the redshift distribution of merger events to lower redshifts than the other models, with virtually all events occurring at $z\leq 10$. It also results in the delay merger model being the most pessimistic in terms of final event rate (note that the normalisation of the probability distributions in \cref{fig:pops-nz} obscures this fact). We model the redshift distribution for this population using \cref{eqn:nz_gw} with parameters $\lbrace \alpha, \beta, z_0 \rbrace = \lbrace 0.877, 2.39, 5.65 \rbrace$.
    \item \textbf{No Delay:} Constraints using this model are labelled `nodelay' in our plots. Here again the heavy seed mechanism dominates, producing a mass distribution that extends to slightly higher chirp masses than the popIII case. However, unlike the delay model above, this model assumes that sufficient gas and three-body interactions occur such that delays between host galaxy and MBH mergers are negligible. This leads to an increased event rate, and represents the most optimistic model in the set in terms of raw numbers of merger events. The redshift range of events is comparable to that of the popIII model, peaking around $z=8-10$ and extending to $z\sim 20$. We model the redshift distribution for this population using \cref{eqn:nz_gw} with parameters $\lbrace \alpha, \beta, z_0 \rbrace = \lbrace 0.928, 1.89, 10.2 \rbrace$.
\end{itemize}
The population models described above dictate the number of massive black hole binary mergers occurring as a function of redshift and chirp mass \cite{Baibhav_2019}. However, these do not directly translate into distributions for \textit{detected} LISA events. This is consequence of the `bucket' shape of the LISA sensitivity curve and the calculation of SNR, which we will present shortly. 

\subsection{Masses, Redshifts and Inclinations}
\label{sub:mri}
From the merger populations above, we simulate realisations of catalogues which may be generated by the LISA experiment. For a given MBH population model, we find the mean expected number of binary MBH merger events as:
\begin{eqnarray}
\bar{N}&=&T_{\rm obs}\int_0^\infty\,\frac{d^2N}{dz\,dt}\;dz
\end{eqnarray}
where $T_{\rm obs}$ is the duration of observation (four years for the standard LISA mission scenario), and ${d^2N}/{dz\,dt}$ is the number of merger events per year per unit redshift, as measured in the observer's frame. From the curves given in figure 3 of \cite{2016PhRvD..93b4003K} we obtain a raw expected number of events per year $\bar{N}$ for each population model. We then sample a Poisson distribution with the expectation value $\bar{N}$ to obtain the actual number of events, $N$, entering a given realisation of our catalogue.

To assign redshifts to our events, for each model we draw $N$ samples from its respective probability distribution in \cref{fig:pops-nz}. We treat the total redshifted mass distributions in \cite{2016PhRvD..93b4003K} similarly, converting them to smooth, normalised probability distributions and sampling $N$ masses from them, which we assign to our events. We do this randomly, thereby neglecting any galactic evolution properties which correlate redshift and binary MBH mass. The event redshifts can be converted into an effective GW luminosity distance via eq.~(\ref{eq:dLexpr}) and eq.~(\ref{eq:dgw_ratio}); note that the resulting value of $d_{GW}$ will have sensitivity to the time evolution ansatz used for $\alpha_M$ (see \cref{sec:ansatz}).
        
\begin{figure}
  \centering
    \includegraphics[width=0.5\textwidth]{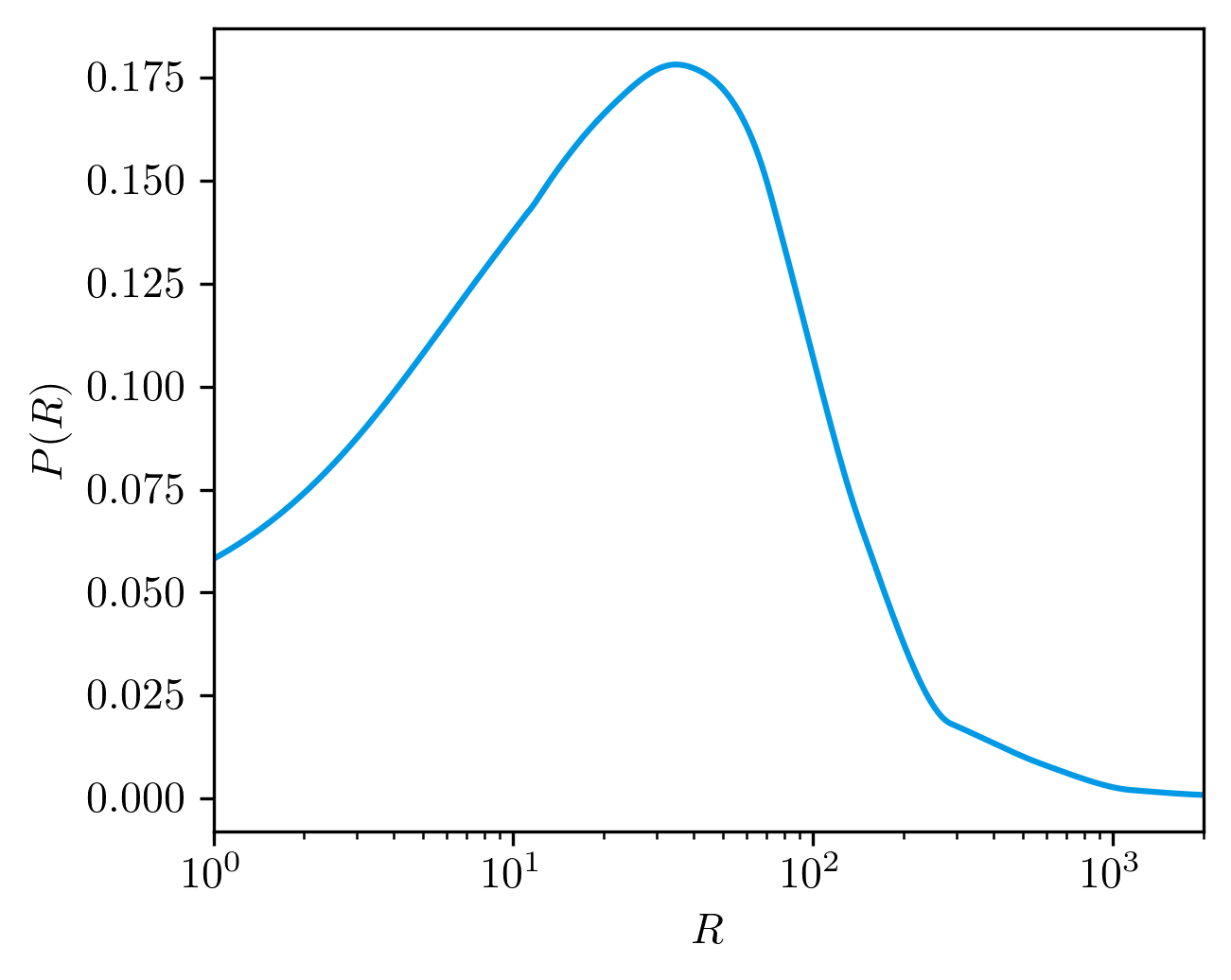}
  \caption{Distribution of the mass ratios of merger events assumed when generating the simulated catalogues for LISA events. Here the mass ratio is defined as $R=m_1/m_2$, where $m_1\geq m_2$.}
  \label{fig:ratio_curve}
\end{figure}

The samples drawn above give the redshifted {\it total} masses of our binary mergers; however, what appears in the waveform amplitude eq.~(\ref{hamp1}) is the redshifted {\it chirp mass}. These two quantities are related as:
\begin{eqnarray}
{\cal M}_c &=& {\cal M}_{\rm Tot} \left[\frac{R}{(1+R)^2}\right]^{\frac{3}{5}}
\end{eqnarray}
where the $ {\cal M}_{\rm Tot} =(1+z)(m_1+m_2)$,  $R=m_1/m_2$ is the ratio of the two binary companion masses, here defined such that $m_1\geq m_2$ always. We use the distribution of MBH mass ratios found by \cite{2012arXiv1208.5251G}, which ranges $1\leq R\leq 10^3$, peaking around $R\sim 30$. We sample from the normalised distribution shown in \cref{fig:ratio_curve} to assign $R$ values to our catalogue of events, and calculate their chirp masses. 

\begin{figure}
    \includegraphics[width=1.0\textwidth]{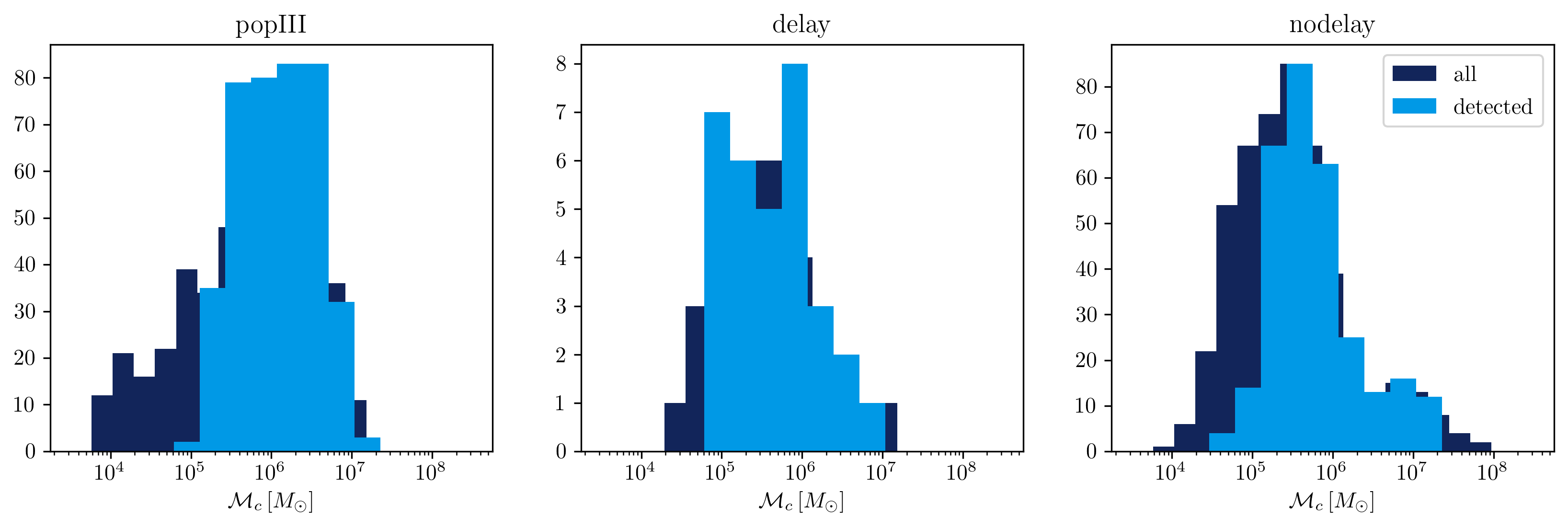}
  \caption{Distribution of redshifted chirp masses of binaries in our LISA event catalogue. Darker bars indicate all events (including those with SNR$<8$, whilst lighter bars show only detected events. Note that the middle plot has a different y-axis scale. Note that where the bars overlap, the darker bars must always be at least as high as the lighter bars.}
  \label{fig:masses}
\end{figure}

\Cref{fig:masses} displays the resulting mass distributions for the three MBH population models, showing both the input catalogue of events and the sub-population of them that are detected, as we will describe next. We see that both the PopIII and No Delay models produce a significant population of low-mass events which fall below the detection threshold.

We assign a characteristic frequency to each binary in our catalogue as:
\begin{eqnarray}
f_{\rm char}&=&\frac{1}{8\pi}\frac{c^3}{G_N \Mc} 
\label{fchar}
\end{eqnarray}
We cut from our catalogue any events whose characteristic frequency falls outside the LISA band ($10^{-4}\,{\rm Hz} < f_{\rm char} < 1\,{\rm Hz}$). Finally, for each of our simulated merger events we draw values of the binary inclination angle $\iota$ from a distribution proportional to $\sin \iota$ \cite{2019PhRvD.100j3523M}.  

\subsection{LISA Noise Model} 
\label{subsec:SNR}
From the population of merger events simulated as described above, we then wish to select events which will be detectable with the LISA mission. We use the analytical form of the LISA sensitivity curve presented in \cite{Noisecurve}, for a `L6A2M4N2' mission scenario (six laser links, 2.5 Gm interferometer arm length, a four-year mission lifetime and well-controlled instrumental noise levels). We note that this analytical form is not sufficiently accurate for high-frequency GW sources that potentially exit the LISA band during their evolution, e.g. early-stage stellar binaries; this does not pose a problem for our MBH sources, which must generally be at lower frequencies to accrue sufficient SNR for detection.

We integrate eq.~(\ref{eqn:fevolve}) over a four-year mission period (redshifted to the corresponding interval in the restframe of the source), using the characteristic frequency of eq.~(\ref{fchar}) as the upper integral value. This yields the initial observed frequency of the source four years prior. Following \cite{2016PhRvL.116w1102S}, we use these initial and final frequencies to estimate the signal-to-noise accumulated in the LISA detectors from both GW polarisations as follows:
\begin{eqnarray}
\rho^2&=&2\int_{f_{\rm init}}^{f_{\rm char}}\frac{h_c(f)^2}{f^2 \left<S_n(f)\right>}\, d{f}
\label{SNRdef1}
\end{eqnarray}
with the strain amplitude given by:
\begin{eqnarray}
h_c&=&\frac{1}{\pi\, d_C}\left(\frac{2G_N}{c^3}\frac{dE}{df}\right)^{1/2},
\label{SNRdef2}
\end{eqnarray}
and
\begin{eqnarray}
\frac{dE}{df}&=&\frac{\pi}{3G_N}\frac{(G_N \Mc)^{5/3}}{1+z}(\pi f)^{-1/3}.
\label{SNRdef3}
\end{eqnarray}
Here $\rho$ is the SNR, $d_C=d_L/(1+z)$ is the comoving distance to the merger and $\left<S_n(f)\right>$ is the one-sided power spectral density (PSD) of the LISA noise, averaged over sky location and polarisations (the quantity $h_c/f$ here can be shown to be equivalent to $\tilde{h}(f)$, the Fourier transform of $h(t_{\rm obs})$, matching eq.~(\ref{SNRdef1}) to the conventional SNR definition found in, for example, \cite{1994PhRvD..49.2658C,Moore_2014}).

We treat events with SNR $\rho>8$ as detected by LISA and include them in our simulated catalogues which go into the cosmological analysis. We remove any events that enter or exit the LISA frequency band during the observation period, since our SNR calculation is not accurate at the band edges. With regards to our SNR estimator, we further note that in reality i) not all GW sources will be observed for a full four years, and ii) our upper frequency limit $f_{\rm char}$ is only a rough proxy for the real evolution of the source towards the final stage of inspiral. In reality eqs.~(\ref{hamp1}) and (\ref{hamp2}) are valid up to at most the frequency of the innermost stable circular orbit (ISCO), which is about a factor four lower than $f_{\rm char}$, and after which more complex dynamics takes over. Given our focus on tests of gravity, in this paper we do not enter into full models of binary coalescence times and evolution close to merger. We have confirmed that our SNR estimates agree within a factor of order unity to the approximate model for SNR given by eq.~(19) of \cite{2019PhRvD.100j3523M}.

\begin{figure}
    \includegraphics[width=\textwidth]{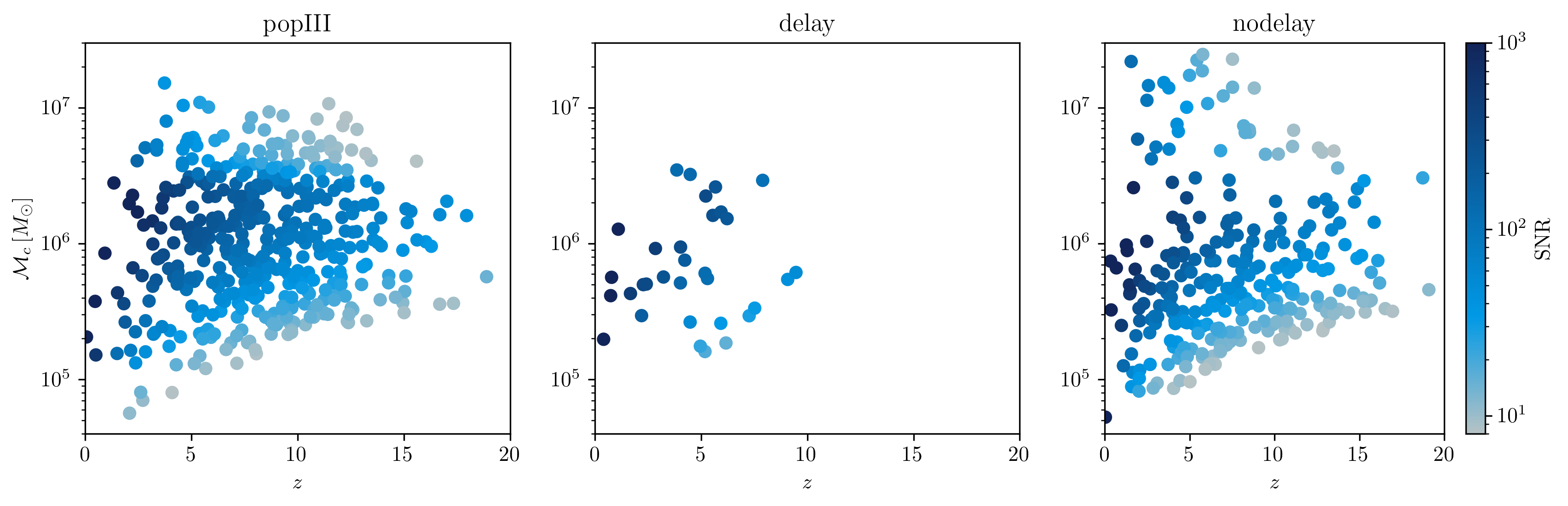}
  \caption{Distributions in mass and redshift of merger events observed during a four-year LISA mission for three MBH population models (see text for description). The colourbar indicates the associated SNR of the detection, see section \ref{subsec:SNR}.}
  \label{fig:SNR_models}
\end{figure}

\Cref{fig:SNR_models} shows the redshift and mass distributions\footnote{These `waterfall' plots are more commonly presented with the redshift and mass axes transposed. Since redshift leverage is key to our tests on gravity, the authors found this version more helpful for the present work.} for LISA detections in each of the three population models, in a single realisation of a simulated catalogue for each. We can see that at redshifts greater than five, only an increasingly narrow mass range centred around ${\cal M}_c\sim 10^6$M$_\odot$ persists. The dropout of low masses here is intuitive: at progressively higher redshifts, a larger chirp mass is needed to exceed the SNR threshold of 8. The dropout of higher-mass binaries is instead caused by the `bucket' shape of the LISA sensitivity curve (see, for example \cite{2017arXiv170200786A}), which rises sharply at $f<10^{-3}$ Hz, selectively penalising the SNR of high-mass systems. The redshifting of GW frequency pushes increasingly lower-mass systems into this zone at high redshifts, leading to the observed tapering of detections. The low event rate of the delay model (middle panel) is particularly noticeable, and to a lesser extent, the extended high-mass range of the nodelay heavy seed model.

\subsection{Luminosity Distance Error Model}
\label{subsec:dl_error}
In order to model the distribution of errors on GW luminosity distances in our simulated catalogue, we adopt a simplified version of the approach from \cite{2019PhRvD.100j3523M}. Our catalogue contains all the components needed to calculate the true plus and cross polarisation amplitudes $h_+, h_\times$ expected for each event via eqs.~(\ref{hamp2}) and (\ref{hamp1}). To create realistic catalogues of observations, we must add to these true $h_+$, $h_\times$ values a simulated measurement error on both polarisations. We use the square root of the LISA power spectral density curve evaluated at the characteristic frequency of the source, denoted $\sigma_h(f) = \sqrt{S_n(f)}$, to simulate these errors as follows.
 
We generate a joint Gaussian likelihood for the measured plus and cross polarisations of each event in luminosity distance and inclination as follows:
\begin{eqnarray}
    \label{eqn:Aplus_Across_inclination}
    P(h_{+}, h_{\times} | D_{L}, \iota, \Mc, \vec{\theta}_{\rm cosmo})\propto N\left[h_+; \frac{G_N\Mc}{d_L(\vec{\theta}_{\rm cosmo})}\,\frac{(1+\cos^2\iota)}{2},\sigma_h^2(f)\right]\nonumber \\ 
    \times N\left[h_{\times}; -\frac{G_N\Mc}{d_L(\vec{\theta}_{\rm cosmo})}\,\cos\iota,\sigma_h^2(f)\right]
\end{eqnarray}
where $N[A;B;\sigma^2]$ indicates a Gaussian distribution for $A$ with mean $B$ and standard deviation $\sigma$ and $\vec{\theta}_{\rm cosmo}$ is the vector of relevant cosmological parameters. Following \cite{2019PhRvD.100j3523M}, we have used here a simplified expression for the amplitude of the waveform, derived in appendix \ref{app:waveform}, and dropped the phase information.

\begin{figure}
  \begin{centering}
    \includegraphics[width=\textwidth]{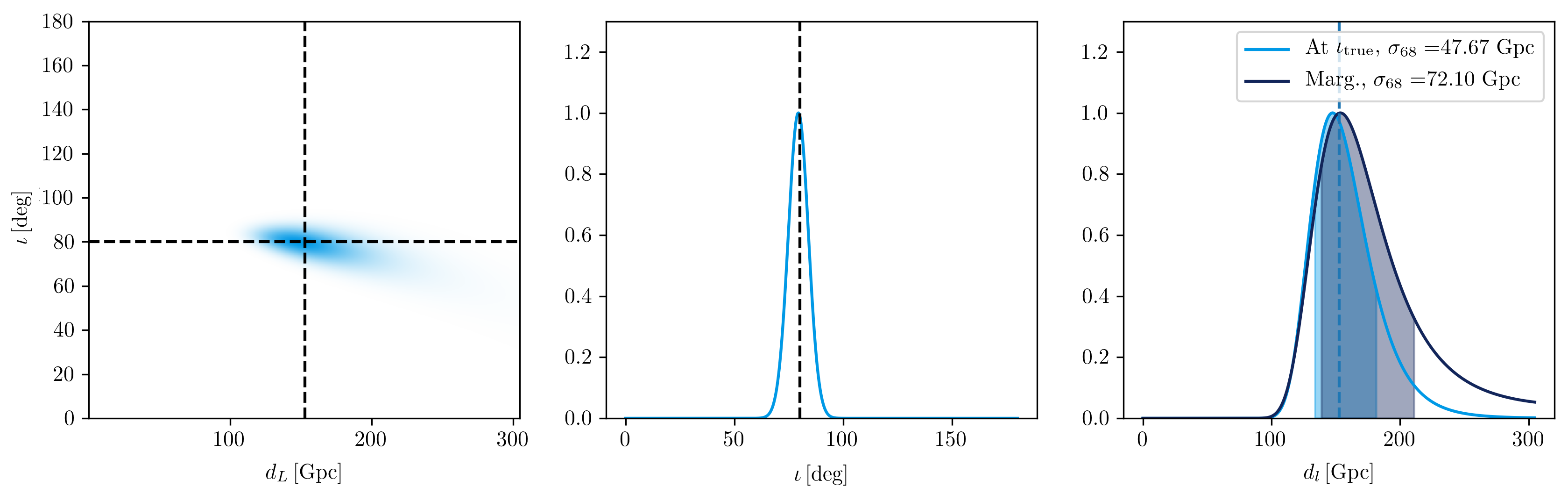}\\
      \includegraphics[width=\textwidth]{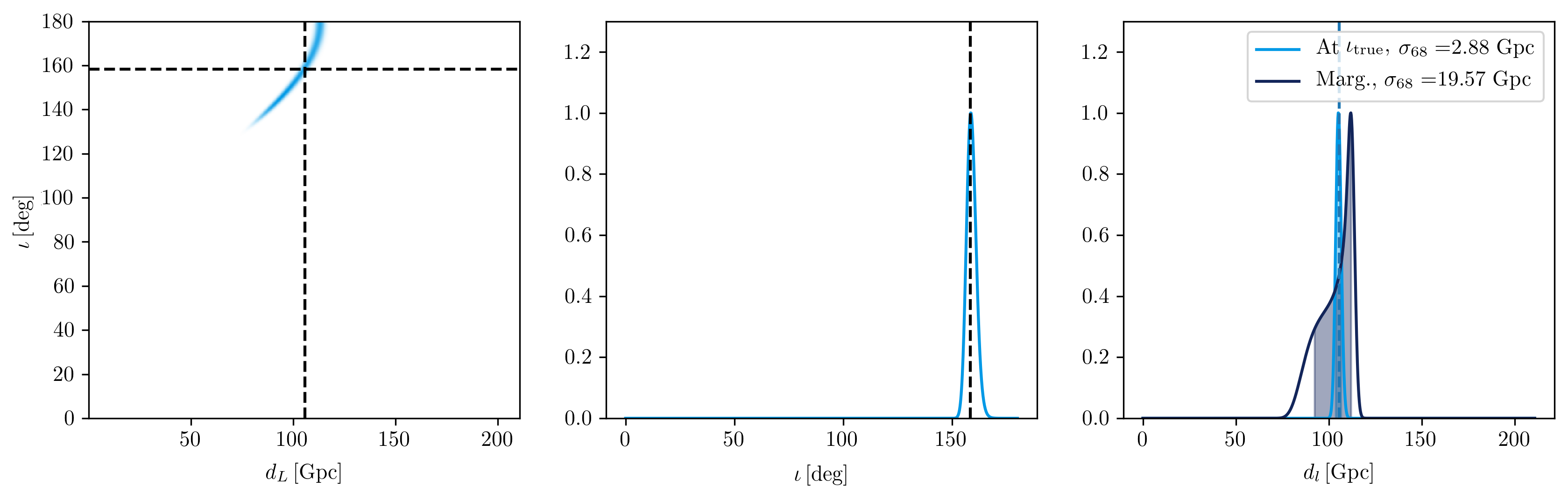}
  \end{centering}
  \caption{Impact of uncertainties in the merger inclination angle $\iota$ on the estimation of the GW luminosity distance $d_L$ for two sources in out simulated catalogues. \emph{Left} panels show the GW likelihood in both parameters, \emph{centre} panels show the conditional likelihood for inclination angle only (at the true value of luminosity distance), \emph{right} panels show the conditional (at the true inclination angle) and marginalised distributions on luminosity distance only.}
  \label{fig:Aamps_likelihood}
\end{figure}

The left-most panels in \cref{fig:Aamps_likelihood} show examples of this likelihood distribution for $d_L$ and $\iota$. We have neglected the comparatively small uncertainty in the measurement of $\Mc$ \cite{2019PhRvD.100j3523M}, assuming it can be measured from the frequency evolution of the waveform (see eq.~(\ref{eqn:fevolve}). We see that for events observed with an orientation close to edge-on ($\iota = 90^\circ$) the likelihood broadens, whilst for events far from edge-on it narrows to a well-defined degeneracy curve.
The rightmost panel of \cref{fig:Aamps_likelihood} shows the corresponding likelihoods for the GW luminosity distance for the two simulated sources. Shown are the likelihood evaluated at the true value of inclination $\iota_{\rm true}$, and the likelihood marginalised over unknown inclination angle. These provide us with two potential error models for the luminosity distance, depending on whether information on the merger inclination angle can be extracted from the electromagnetic counterpart of the merger. We approximate $\sigma_{\iota_{\rm true}}$ and $\sigma_{\iota_{\rm marg.}}$ as the widths of these distributions which contain $68\%$ of the probability. We see that, depending on the true inclination angle, the degeneracy between luminosity distance and inclination angle can broaden the width of the $d_L$ distribution by appreciable amounts. Hence the inclination angle $\iota$ could potentially be an important nuisance parameter for cosmology with GW standard sirens, if it is not well-measured. Electromagnetic follow-up observations of the merger system potentially provide information on the merger inclination, including from optical and radio telescopes \cite{Ravi:2018jyk}. In \cref{sec:results} we will present our results in both the optimistic (unmarginalised) and pessimistic (marginalised in $\iota$) cases, to quantify the impact of this degeneracy.

We must also include in our final estimate of the luminosity distance error contributions from gravitational weak lensing of the GW signals \cite{ezquiaga2020apparent, Mukherjee_2020_multimessenger, Mukherjee_2020_probing}, which (de)magnify the GW signal, and peculiar velocities of the GW sources. We write this as:
\begin{eqnarray}
\label{eqn:sigma_dgw}
\sigma_{d_{GW}}^2&=&\sigma_{\iota}^2 + \sigma_{\rm lens}^2 + \sigma_{v}^2.
\end{eqnarray}
where $\sigma_{\iota} = \lbrace \sigma_{\iota_{\rm marg.}},\sigma_{\iota_{\rm true}}\rbrace$. The weak lensing error $\sigma_{\rm lens}^2$ is given by \cite{2010PhRvD..81l4046H, Tamanini2016}:
\begin{eqnarray}
  \sigma_{\rm lens} = 0.066 \,d_L(z) \left[\frac{(1 - (1 + z)^{-0.25})}{0.25}\right]^{1.8},
\end{eqnarray}
and $\sigma_{v}$ is the error associated with uncertainty on peculiar velocity of the source \cite{Tamanini2016,2019arXiv190908627M}:
\begin{equation}
    \sigma_v = d_L(z) \left[ 1 + \frac{c (1 + z)}{H(z) d_L(z)} \right] \frac{\sqrt{\langle v^2 \rangle}}{c}.
\end{equation}
For the r.m.s. peculiar velocity $\langle v^2 \rangle$ we assume a value of $500\,$km$\,$s$^{-1}$ for all sources.

\subsection{Electromagnetic Counterpart Detections}
\label{sub:em_counterparts}
In order to make use of binary mergers as standard sirens, we also require redshifts from electromagnetic counterparts of the mergers. If a significant fraction of MBH mergers are accompanied by quasar activity, flares or increased variability, then electromagnetic facilities contemporary with LISA may be able to identify candidate host galaxies \cite{2006ApJ...637...27K}, from which spectrographic redshifts can subsequently be obtained. The pre-merger localisation of most LISA sources is expected to be $\sim 10$ deg$^2$, well-matched to the field of view of upcoming surveys such as the Vera Rubin Observatory and the SKA \cite{Tamanini2016}. 

The fraction of LISA sources expected to have counterparts is highly sensitive to the gas environments and formation channels of MBH binaries. We will not simulate these in detail, but instead will use the counterpart fractions predicted by the `optimistic' scenario in table 10 of \cite{Tamanini2016}, which are $F_c=\left\{10\%, 50\%, 5\%\right\}$ for the popIII, delay and nodelay MBH models, respectively. From our raw catalogues of merger events detected by LISA, we remove objects according to a probability $\propto D^2_{l}$, such that the surviving fraction of the original catalogue is $F_c$. This encodes an expectation that the likelihood of a detectable counterpart falls off with the observable brightness of the source.

\begin{figure}
    \includegraphics[width=1.0\textwidth]{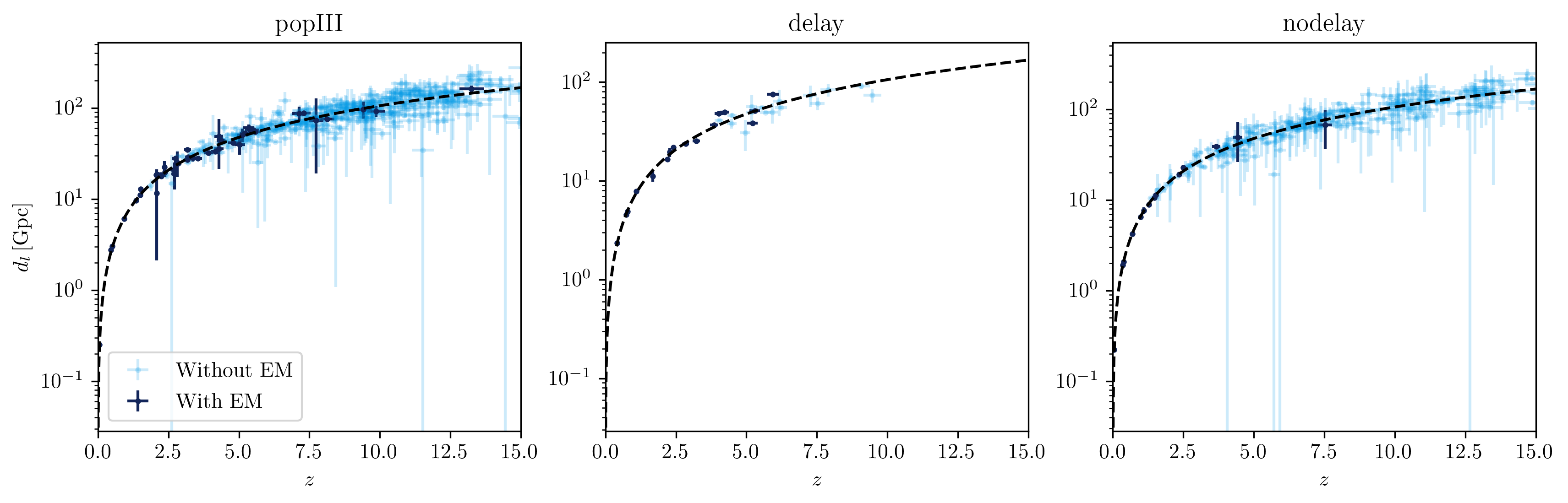}
  \caption{Examples showing our simulated LISA catalogues and their errors (when marginalised over uncertainty in the merger inclination angle) for the three population models considered. Dark blue points correspond to sources with electromagnetic counterparts, which we expect to have a measured redshift and use in our cosmological constraints.}
  \label{fig:pops-catalogues}
\end{figure}

In line with this optimistic redshift scenario, we assume spectroscopic redshifts for all counterparts, with errors $\sigma_z = 0.005(1 + z)$. \Cref{fig:pops-catalogues} shows our final catalogues of LISA standard sirens (dark points), along with their simulated errors on luminosity distance and redshift; note that the redshift errors are very small on the scale of these plots. Lighter points indicate sources that are detectable by LISA but do not have an associated EM counterpart. 

\section{Additional Data}
\label{sec:experiments}
In addition to the gravitational wave data described in section \cref{sec:gw_simulations}, we also include constraints from a set of current cosmological large scale structure data. This is done to provide addition constraints on cosmological parameters, as well as context for the improvement in constraints they offer for the $\alpha_M$ Horndeski parameter. For the $\alpha_B$ Horndeski parameter (which does not affect GW propagation), they are the dominant source of constraints. Our implementation and data sets used closely follow \cite{2019PhRvD..99j3502N}.

We form the joint likelihood over data from Cosmic Microwave Background, Redshift Space Distortion and Baryon Acoustic Oscillation (BAO) data, with the specific data sets described below. All data sets are included in a \cosmosis pipeline, as modules from the \cosmosis standard library.

\subsection{Cosmic Microwave Background}
\label{sub:cmb}
For CMB data we use the \emph{Planck} 2015 likelihoods \citep{2014A&A...571A..15P}. Specifically we use the \texttt{Plik lite} likelihood in temperature TT only at high-$\ell$ ($30 \leq \ell \leq 2508 $), combined with the \texttt{bflike} likelihood in temperature and polarisation TT, EE, BB,TE at low-$\ell$ ($0 \leq \ell \leq 29 $). We also include information from the \texttt{SMICA} CMB $\phi\phi$ lensing likelihood in the multipole range $0 \leq \ell \leq 2048$.

\subsection{Redshift Space Distortions}
\label{sub:rsd}
We include Redshift Space Distortion data, constraining $f(z)\sigma_8(z)$, from the `consensus' results of DR12 of the Baryon Oscillation Spectroscopic Survey (BOSS) \citep{2017MNRAS.470.2617A}, and from the 6dF Galaxy Survey \citep{2012MNRAS.423.3430B}. As mentioned in \cref{sec:theory_gw_horndeski}, it is important when testing modified gravity models that the constraints from this data set are only applied at a `safe' fiducial scale where the growth rate is effectively scale-independent. Following \cite{2019PhRvD..99j3502N}, we use the scale $k_{\rm fid}=0.05$h Mpc$^{-1}$.

\subsection{BAO}
\label{sub:bao}
Our BAO data, constraining the volume averaged distance $D_V(z)$, are from WiggleZ \citep{2014MNRAS.441.3524K}, Sloan Digital Sky Survey Main Galaxy Sample (SDSS MGS) \citep{2015MNRAS.449..835R}, and the `consensus' BOSS-DR12 results \citep{2017MNRAS.470.2617A}.
\\
\\
We first run an MCMC chain using only the CMB+RSD+BAO data sets (i.e. without including simulated GW data) to constrain the cosmological parameters. Details of the priors used and posteriors obtained are given in \cref{app:LSSonly}. From these chains we extract the maximum posterior values for all parameters varied and use them when simulating our gravitational wave catalogues. The cosmological parameter constraints for these LSS-only chains are shown in \cref{app:LSSonly} and are consistent with the previous results (using slightly different data) of \cite{2019PhRvD..99j3502N}.

\subsection{LIGO Data}
\label{sub:ligo_data}
\begin{figure}
    \centering
    \includegraphics[width=0.5\textwidth]{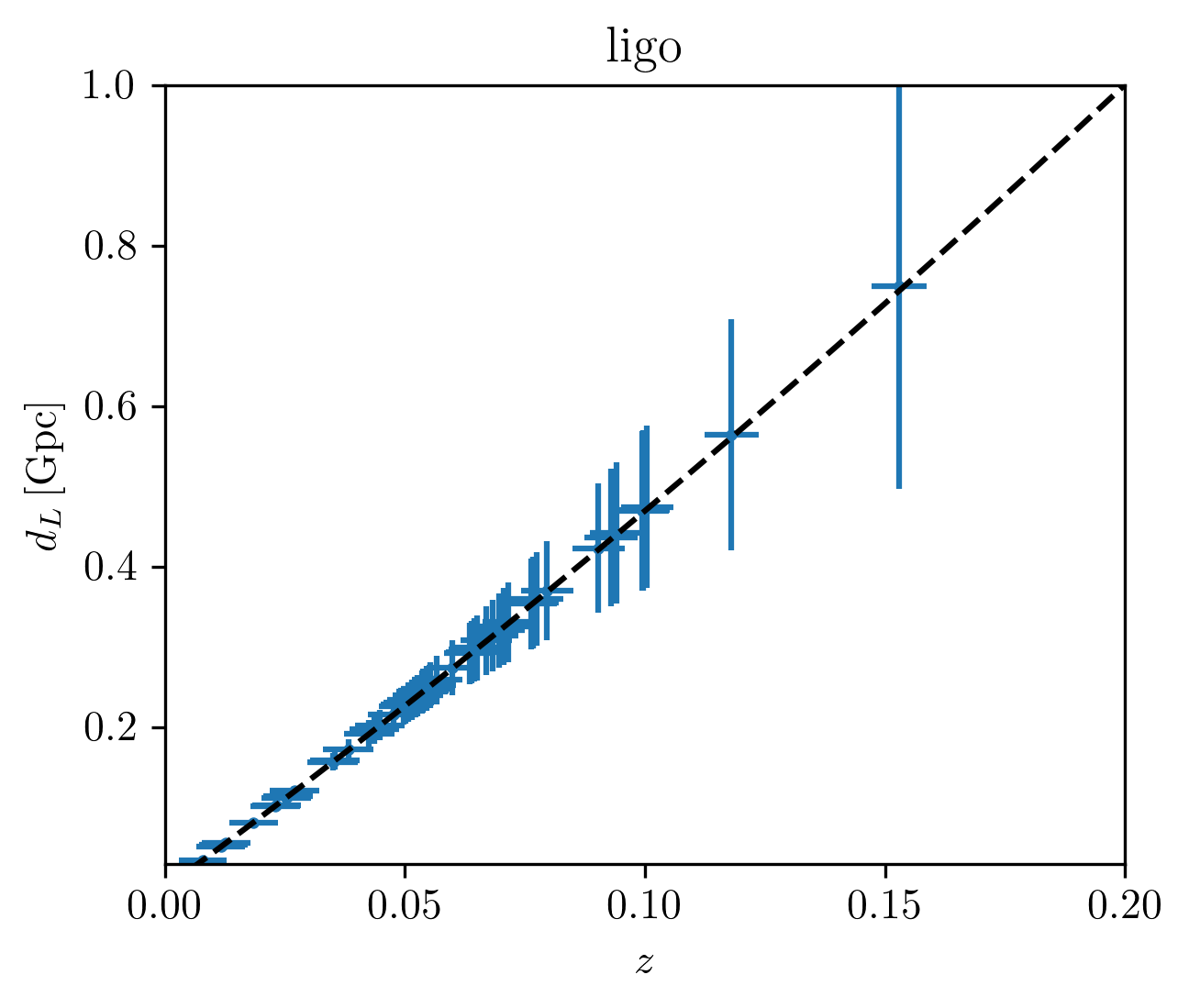}
  \caption{Examples showing our simulated LIGO data and their errors.}
  \label{fig:ligo_cat}
\end{figure}

We also include forecast constraints from standard sirens which may be detectable with the existing Advanced LIGO-VIRGO (advLIGO) detector network. These standard sirens are from a very different merger population (generally binary neutron stars) to the LISA sources we mainly consider, existing at much lower redshifts. Qualitatively, the lower panel of \cref{fig:test_models} shows that such sources are not as useful for constraining Horndeski models as LISA sources, because of their weaker redshift leverage on $d_{GW}/d_L$. However, we include them here to show the improvement in constraints that will be possible in the near future, well before LISA is launched. In addition to the current single detection of a gravitational wave event with a confirmed electromagnetic counterpart --- the binary neutron star merger GW170818 \citep{PhysRevLett.119.161101} --- we also simulate a catalogue of fifty further similar events from the advLIGO-VIRGO network. These events have a redshift distribution given by \cref{eqn:nz_gw} with parameters $\lbrace \alpha, \beta, z_0 \rbrace = \lbrace 3.21, 0.83, 0.008 \rbrace$ which matches well the distribution in figure 2 of \cite{Lagos_2019}. For these events we will assume a simplified model for the error on luminosity distance as follows.
Under rough approximation, the SNR of a GW signal is inversely proportional to the GW luminosity distance of the source (we are ignoring here evolution of the frequency). This enables us to write
\begin{align}
    \rho d_{GW} &= \rho^* d_{GW}^*
    \label{LIGO1}
\end{align}
where $\rho^*$ and $d_L^*$ are fiducial values of the SNR and luminosity distance. Previous works have shown that the one-sigma error on the luminosity distance of LIGO-VIRGO standard sirens scales approximately as \cite{Lagos_2019, mastrogiovanni2020probing, Chen_2018}, 
\begin{align}
    \sigma_{d_{GW}}&\simeq \frac{1.8}{\rho}\, d_{GW}
        \label{LIGO2}
\end{align}
Combining eqs.~(\ref{LIGO1}) and (\ref{LIGO2}) gives
\begin{align}
     \sigma_{d_{GW}}&\sim \frac{1.8}{\rho^* d_{GW}^*}\, d_{GW}^2
        \label{LIGO3}
\end{align}
Since we generate our mock catalogue at the fiducial model very close to $\Lambda$CDM, $d_{GW}$ here is coincides with the normal luminosity distance inferred from EM signals. The maximum distance at which the advLIGO-VIRGO network is expected to detect BNS standard sirens is $\sim 400$ Mpc \cite{Lagos_2019}. The SNR of an event at this distance must be achieve a minimum of $\rho=8$ for detection. Using this to set the fiducial values of $\rho^* = 8$ and $d_{GW}^*=400$ Mpc, eq.~(\ref{LIGO3}) yields:
\begin{align}
     \sigma_{d_{GW}}&\sim \frac{1.8}{8\times 400 \,{\rm Mpc}} \,d_{GW}^2\simeq 5.63\times 10^{-4}\, d_{GW}^2
        \label{LIGO4}
\end{align}
where $\sigma_{d_{GW}}$ and $d_{GW}$ are in units of Mpc. The resulting simulated catalogue is shown in \cref{fig:ligo_cat}. As for LISA, redshift errors are assumed to be spectroscopic with $\sigma_z = 0.005(1 + z)$.

\section{Forecasting Method}
\label{sec:forecasting}
In order to forecast constraints on cosmological and modified gravity parameters, we perform a Monte Carlo evaluation of the parameter posterior probability distributions, given the real and simulated data described in \cref{sec:gw_simulations,sec:experiments}. As for constructing our simulated data, we make use of the \cosmosis \footnote{\url{https://bitbucket.org/joezuntz/cosmosis/wiki/Home}} \cite{Zuntz_2015} modular cosmological parameter estimation code, which allows us to form a pipeline for calculating the relevant observables and for sampling over the parameter space. Before running the chains we set broad priors on the parameters which are considered, as detailed in \cref{tab:model_params}. 

\subsection{Likelihood}
\label{sub:likelihood}
We assume a joint likelihood with the simplifying approximation that each of the data sets (CMB, RSD, BAO, LIGO GW and LISA GW) considered are independent and hence that the likelihood can be factorised into individual Gaussian likelihoods for each data set:
\begin{equation}
    \mathcal{L}(D|\vec{\theta}) = \mathcal{L}(D_{\rm CMB}|\vec{\theta}) \mathcal{L}(D_{\rm RSD}|\vec{\theta})
    \mathcal{L}(D_{\rm BAO}|\vec{\theta})
    \mathcal{L}(D_{\rm GW}|\vec{\theta}),
    \label{eqn:likelihood}
\end{equation}
where $\vec{\theta}$ is a vector containing all of the model parameters, and $D_{\rm X}$ is the data vector for each type of observable (and the GW likelihood contains either LISA or LIGO data as specified). Each part of the likelihood is a simple Gaussian, for the CMB, RSD and BAO parts using the real data vectors and covariance matrices provided in the \cosmosis modules. For the GW parts the data vector is that from our LISA and LIGO simulations at the best fitting cosmology from the LSS-only runs, along with a diagonal covariance matrix with errors given by the $\sigma_{d_{GW}}$ found from eq.~(\ref{eqn:sigma_dgw}) for LISA and eq.~(\ref{LIGO3}) for LIGO. Depending on the LISA chain these errors are either marginalised over or conditioned on $\iota$ using eq.~(\ref{eqn:Aplus_Across_inclination}) --- again, at the best-fitting cosmology (we do not simultaneously fit inclination and cosmology). Note that our method here differs from \cite{Tamanini2016}, who use a Fisher matrix forecast to use study the full information content of the waveform.

For a given set of cosmological parameters $\vec{\theta}$ (including Horndeski parameters) we calculate observables for each data set in \cref{sec:experiments} using the Boltzmann code \hiclass\footnote{Our \cosmosis wrapper for \hiclass can be found at \url{https://github.com/itrharrison/hi_class}.} \cite{2017JCAP...08..019Z,Bellini_2020}. These predicted observables are then compared to the data sets in the likelihood function.

\subsection{Sampling}
\label{sub:sampling}
We use the Markov Chain Monte Carlo ensemble sampler \emcee \cite{2013PASP..125..306F} to sample from this parameter space according to the posterior probability. Chains are run using an ensemble of 32 walkers, for at least $10^5$ total samples, of which 5000 are removed for burn-in. 

\section{Results}
\label{sec:results}
\Cref{tab:model_params_propto_scale} and \cref{tab:model_params_propto_omega} summarise the best-fit values and one dimensional $68\%$ confidence intervals corresponding to the posterior distributions shown in \cref{fig:exps}. We discuss the observed trends in the Horndeski parameters below. The full posteriors in all of our varied parameters (i.e. including the standard cosmological parameters as well as Horndeski parameters) from the existing electromagnetic LSS-only data are shown in \cref{app:LSSonly}, and are highly consistent with the existing results from \cite{2019PhRvD..99j3502N}.

\subsection{Fiducial Parameter Constraints}
\label{subsec:constraints_fiducial}
\Cref{fig:exps} shows the joint constraints on $\aM$ and $\aB$ at $z=0$ for the popIII population model, for the $\propto \Omega_\Lambda$ (left panel) and $\propto a$ (right panel) ansatze respectively (eqs.~\ref{ansatzes}). The contours here compare the relative constraining power offered by the EM data sets described in section \ref{sec:experiments} i) alone (blue), ii) combined with 50 mock LIGO standard siren events as per section \ref{sub:ligo_data} (green), and iii) combined with both LIGO and the LISA events generated in section \ref{sec:gw_simulations} (red). As can be seen, and in the numerical values in \cref{tab:model_params_propto_scale} and \cref{tab:model_params_propto_omega}, the inclusion of LIGO data offers only a modest improvement in the widths of the parameter constraints for both models. In contrast, the inclusion of the LISA data improves the constraints by a factor $\sim 5$ (where a factor 1 would indicate no improvement), similar to the improvement expected from future LSS experiments alone \cite{Alonso_2017}.
\begin{table}
    \centering
    \begin{tabular}{lccc}
        \hline
    Experiment & $\sigma_{\alpha_{B0}}$ & $\sigma_{\alpha_{M0}}$ & $\sigma_{\Xi_0}$\\ 
    \hline
    LSS-only        & 0.56 & 0.32 &  0.16 \\
    LSS+LIGO (forecast) & 0.45 & 0.24 & 0.12 \\
    \hline
    LSS+LISA (pop. III) & 0.19 & 0.06 & 0.03\\
    LSS+LISA (delay) & 0.20 & 0.09 & -\\
    LSS+LISA (no delay) & 0.21 & 0.08 & -\\
    \hline
    \end{tabular}
    \caption{Widths of one dimensional 68\% confidence intervals for Horndeski and phenomenological (see \cref{sub:alt_param}) parameters in the \proptoscale model.}
    \label{tab:model_params_propto_scale}
\end{table}

\begin{table}
    \centering
    \begin{tabular}{lccc}
        \hline
		Experiment & $\sigma_{\alpha_{B0}}$ & $\sigma_{\alpha_{M0}}$ & $\sigma_{\Xi_0}$\\ 
		\hline
        LSS-only        & 0.59 & 0.73 & 0.23\\
        LSS+LIGO (forecast) & 0.60 & 0.68 & 0.21\\
        \hline
        LSS+LISA (pop. III) & 0.49 & 0.11 & 0.03\\
        LSS+LISA (delay) & 0.50 & 0.15 & -\\
        LSS+LISA (no delay) & 0.51 & 0.13 & -\\
        \hline
    \end{tabular}
    \caption{Widths of one dimensional 68\% confidence intervals for Horndeski and phenomenological (see \cref{sub:alt_param}) parameters in the \proptoomega model.}
    \label{tab:model_params_propto_omega}
\end{table}

\subsubsection{Effect of EM Data}
\label{subsec:constraints_lss}
We note that there is a considerable difference in contour shapes for the two ansatzes used in \cref{fig:exps}. The ansatz $\alpha_i\propto a$ represents a deviation from GR that grows more strongly at high redshifts than the $\alpha_i\propto \Omega_{DE}$; for example, the Horndeski functions are roughly twice as large at $z=2$ in the former ansatz. 

In Horndeski models, the Integrated Sach Wolfe ISW part of the CMB TT power spectrum has a known insensitivity to parameter combinations lying along the line $2\alpha_M + \alpha_B =0$, which modify the ISW plateau in `opposite' directions \cite{Kreisch_2018}. It is this which gives rise to the clear degeneracy direction observed in right panel of \cref{fig:exps}; the earlier growth of the $\alpha_i\propto a$ ansatz makes it particularly sensitive to departures from this line, effectively correlating $\alpha_M$ and $\alpha_B$. This degeneracy direction also exists in the $\alpha_i\propto \Omega_{DE}$ constraints (left panel of \cref{fig:exps}) before the addition of RSD data (see figures 1 and 3 of \cite{2019PhRvD..99j3502N}). 

However, the different time evolutions of the two ansatze causes them to respond differently to the inclusion of RSD data. The RSD data disfavour large, positive values of $\alpha_M$, which would significantly enhance the growth rate. Under the $\alpha_i\propto \Omega_{DE}$ ansatz, adding these data shifts the blue contour downwards and off the $2\alpha_M + \alpha_B =0$ axis. However, under the $\alpha_i\propto a$ ansatz, negative values of $\alpha_M$ are not permitted as they lead to the formation of gradient instabilities as described in section \ref{sec:ansatz}. Instead of shifting downwards, the blue contour shrinks along the degeneracy axis (not visible in \cref{fig:exps}, since the blue contours there already include RSD data). This results in the very different shapes of the blue and green contours in \cref{fig:exps}.

Under the $\alpha_i\propto \Omega_{DE}$ ansatz the required absence of gradient instabilities manifests as the shared sloping lower bound of the LSS-only and LSS+LIGO contours, permitting mildly negative values of $\alpha_M$.

\subsubsection{Effect of GW Data}
\label{subsec:constraints_gw}
\begin{figure}
  \begin{centering}
    \includegraphics[width=0.5\textwidth]{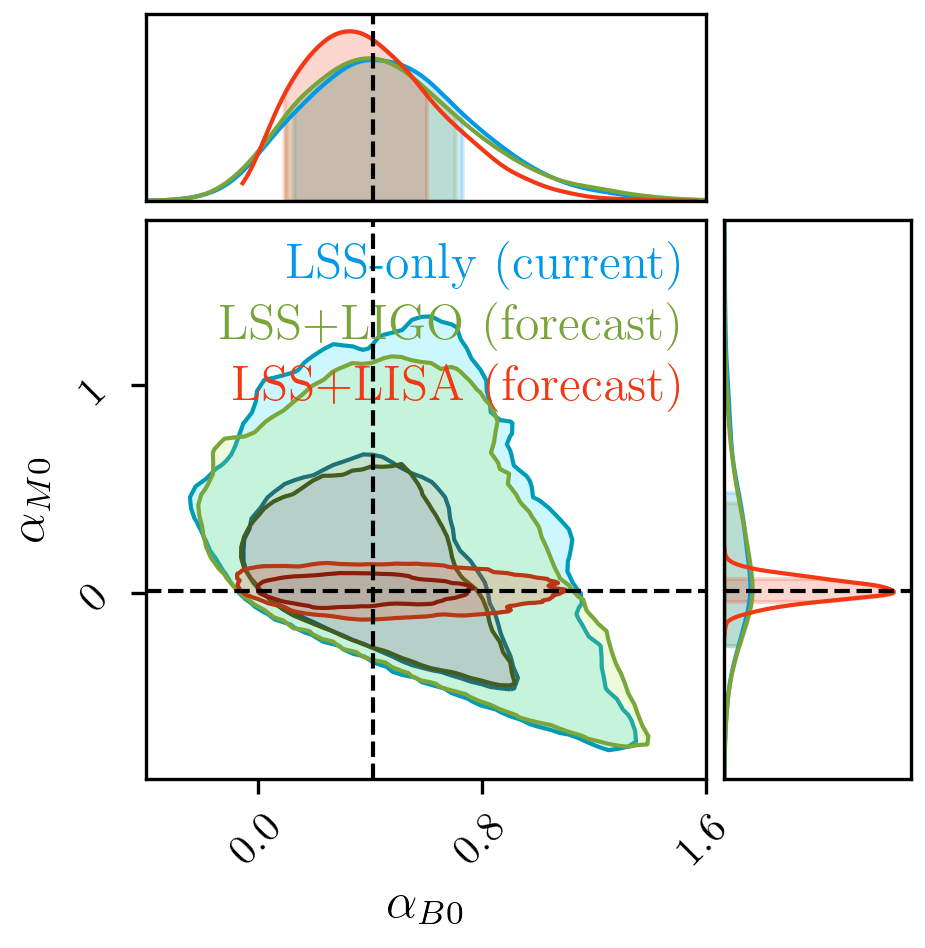}
    \includegraphics[width=0.5\textwidth]{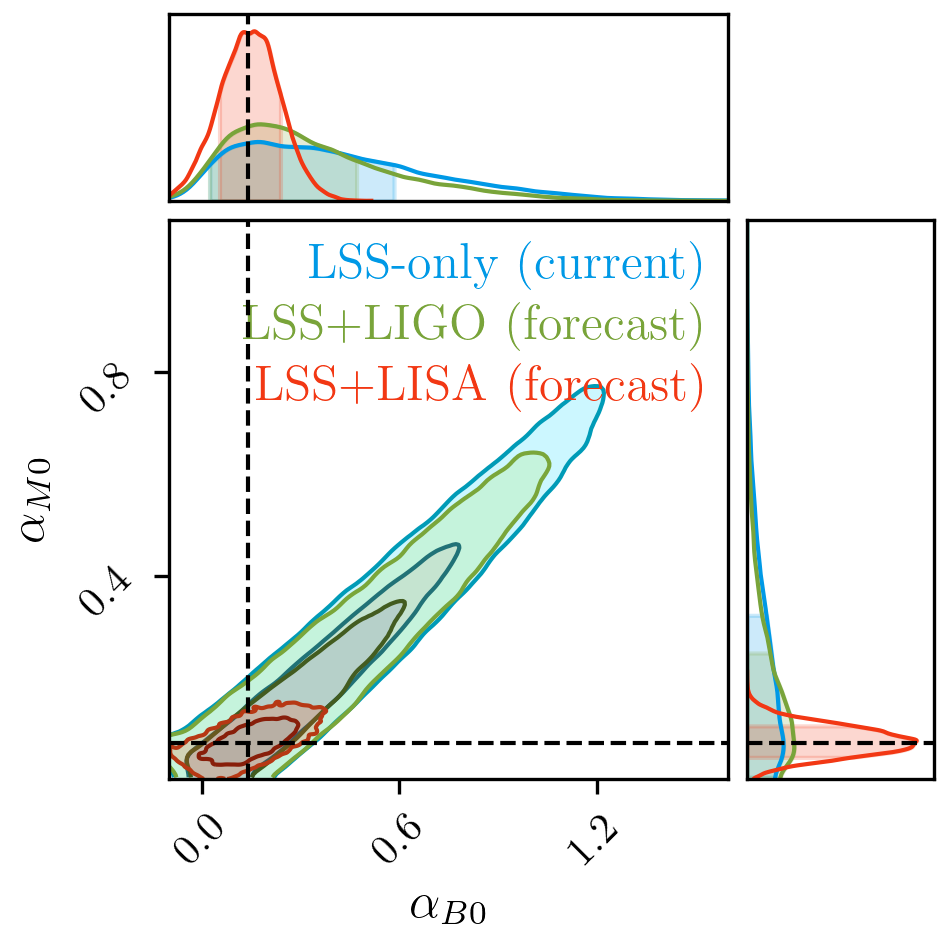}
  \end{centering}
  \caption{Constraints on coefficients appearing in the Horndeski $\alpha_M$ and $\alpha_B$ functions, in both the \proptoomega (\emph{left}) and \proptoscale (\emph{right}) cases. The three sets of contours correspond to constraints from Large Scale Structure data only, those with LSS and forecast LIGO data, and those with forecast LSS and LISA data. LISA constraints are shown for the popIII population model, but are very similar for the other two models.}
  \label{fig:exps}
\end{figure}

As is clear from our expressions in section \ref{sec:theory}, the luminosity distance of GW sources is sensitive to \emph{only} the $\alpha_M$ parameter. For the left panel of \cref{fig:exps}, adding GW data from LISA results in a straightforwards compression of contours in the vertical direction with no change in $\alpha_B$. For the right panel there is a corresponding reduction along the degeneracy axis, for the reasons explained above. In both of these panels, we show the constraints for the popIII population model, see \cref{sub:results_pops} for a discussion of the effects of varying this assumption.

We see that the LIGO events add relatively little constraining power, as they are predominantly at low redshifts where our cumulative deviations from GR are relatively small (lower left panel of \cref{fig:test_models}). In contrast, the LISA data reduces the 68\% confidence interval on $\aM$ by a factor of approximately $5.3$ for the \proptoomega case and $6.6$ for the \proptoscale case.

\subsection{Effect of Merger Population Model}
\label{sub:results_pops}
\begin{figure}
  \begin{centering}
          \includegraphics[width=0.5\textwidth]{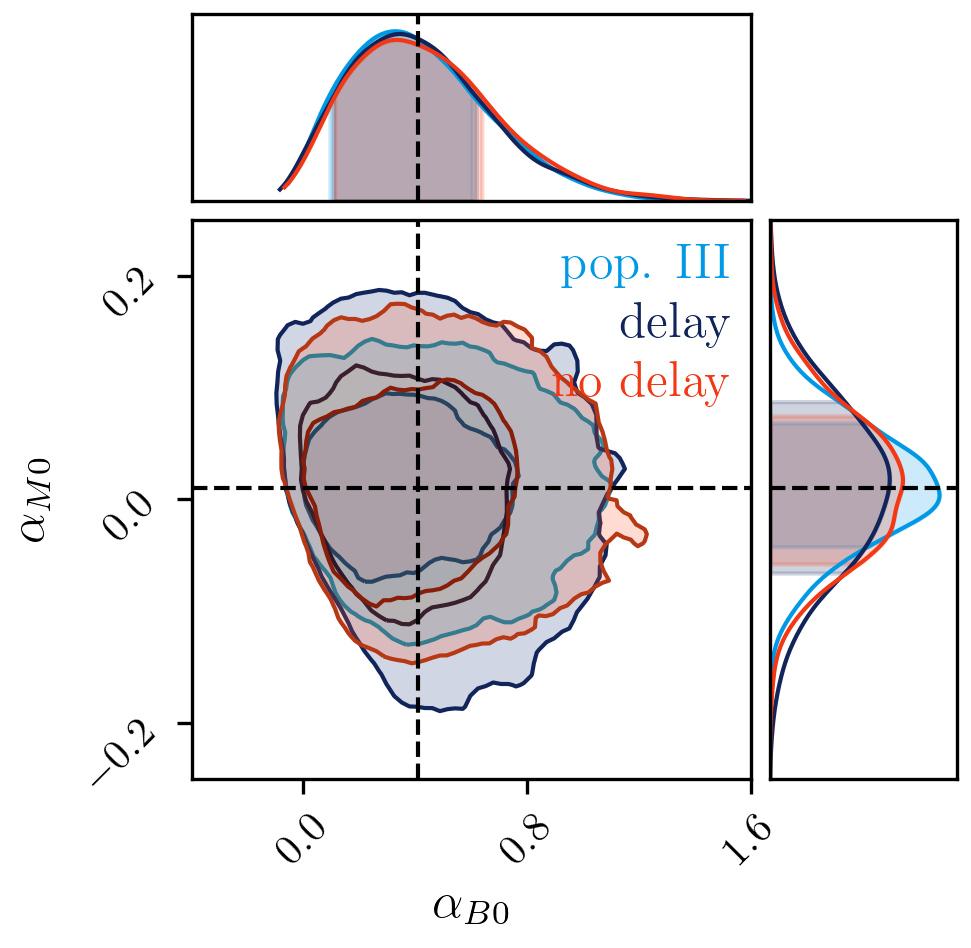}
      \includegraphics[width=0.5\textwidth]{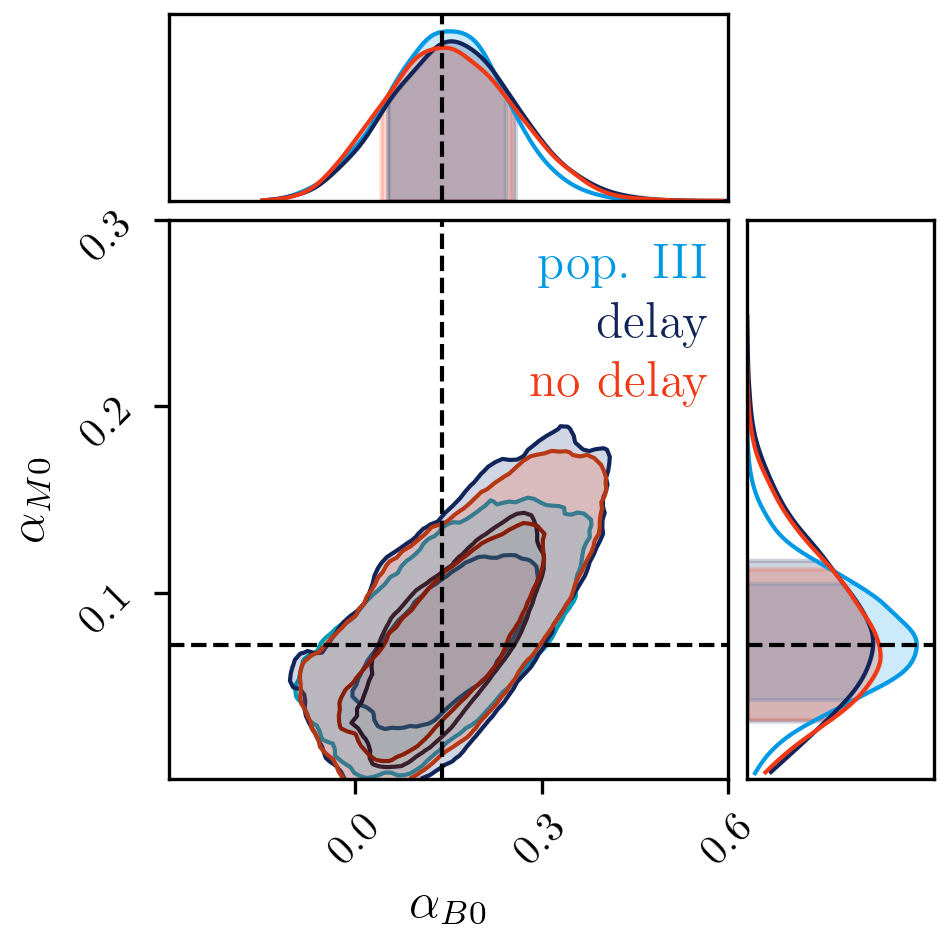}
  \end{centering}
  \caption{Constraints on Horndeski parameters in both the \proptoomega (\emph{left}) and \proptoscale (\emph{right}) cases, showing the effect of the choice of GW progenitor population model.}
  \label{fig:pop_models}
\end{figure}

\Cref{fig:pop_models} shows the effect on the constraints on $\alpha_{M0}$ and $\alpha_{B0}$ of varying the three population models described in \cref{subsec:gw_populations}, for both $\proptoomega$ and $\proptoscale$ ansatze. As can be seen, the forecast constraints on the Horndeski parameters is very robust to the true underlying population model. This can be understood through \cref{fig:pops-catalogues}, which shows that, whilst the three population models differ significantly in their predictions for the total numbers of detectable mergers, many of the additional events occuring in the popIII and nodelay models occur at high redshifts. This means they are very unlikely to have detected EM counterparts, and so do not contribute to our analysis. The delay model offers the weakest constraints still, but only by a very small margin. \Cref{tab:model_params_propto_scale} and \cref{tab:model_params_propto_omega} show the effect of this variation on the inferred 68\% confidence intervals for the Horndeski parameters.

\subsection{Effect of Merger Inclination Uncertainty}
\label{sub:results_inclinations}
\begin{figure}
  \begin{centering}
        \includegraphics[width=0.5\textwidth]{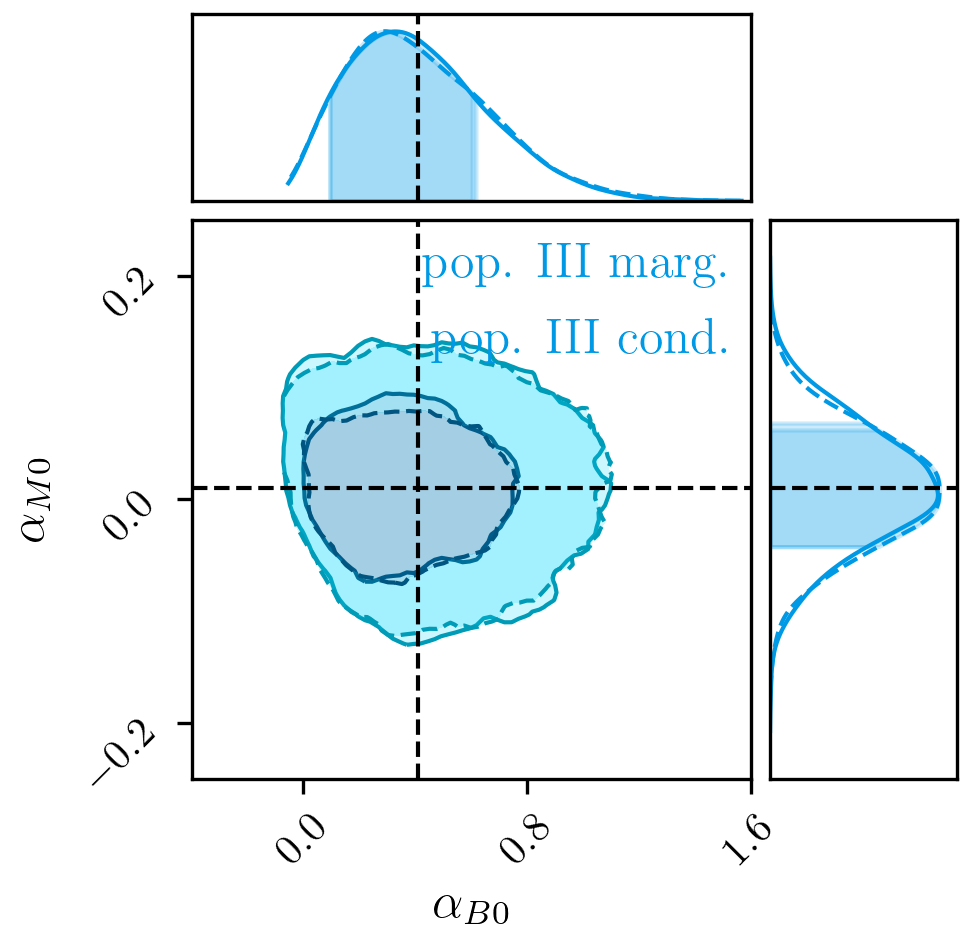}
    \includegraphics[width=0.5\textwidth]{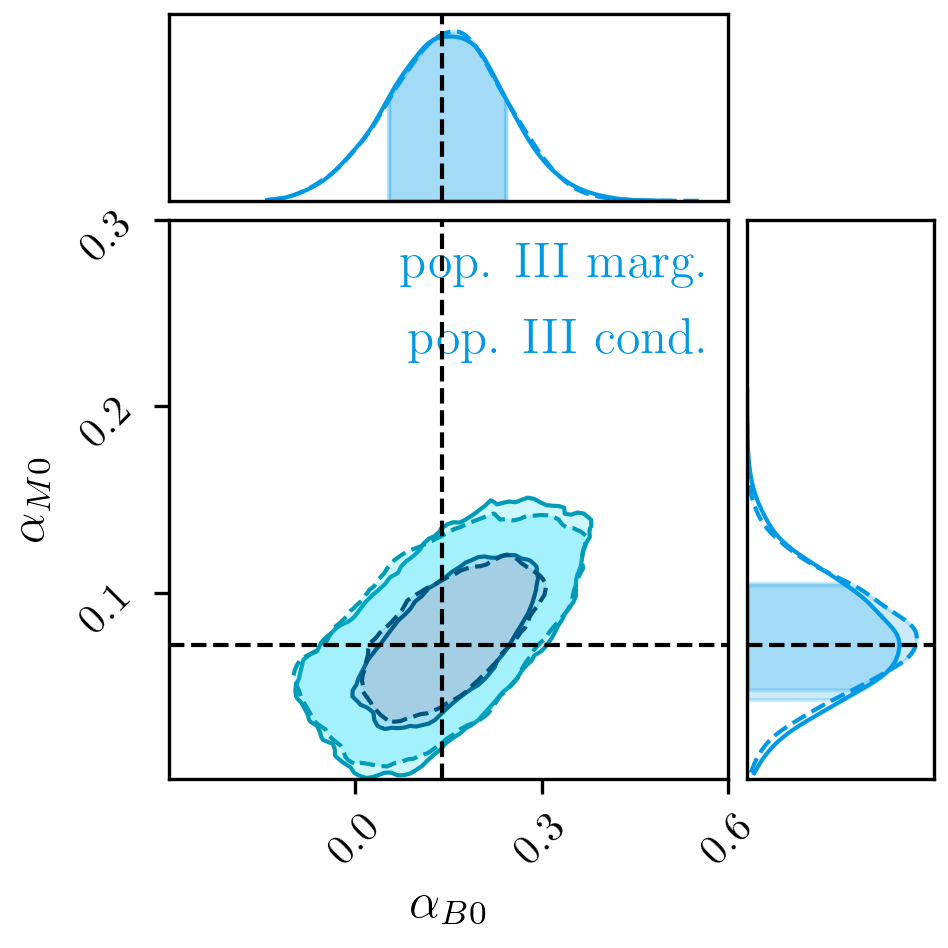}
  \end{centering}
  \caption{Constraints on Horndeski parameters in both the \proptoomega (\emph{left}) and \proptoscale (\emph{right}) cases, showing the effect of the inclusion of uncertainty from marginalisation over unknown GW binary merger inclination angle. The change in contours is negligible.}
  \label{fig:inc_models}
\end{figure}

\Cref{fig:inc_models} shows the effect, within the popIII merger population model, of including the extra uncertainty in luminosity distance measurements from marginalisation over the merger inclination angle $\iota$. Contours labelled `marg.' include the effect of this marginalisation, whilst contours labelled `cond' (or `conditional') do not. As can be seen, the inclusion of this extra uncertainty on a per-source basis has very little effect on the constraining power. The errors on individual source's luminosity distances are typically inflated by up to 50\%; for individual LIGO sources, \cite{mastrogiovanni2020probing} states that uncertainty on $d_L$ with LIGO is 20-40\% for one event. For large populations of sources as considered for our LISA analysis, however, the effect of this increased uncertainty on individual sources is minimised.

\subsection{Effect of the Horndeski Scalar Stability Prior}
\label{subsec:creminell_prior}
In addition to the stability criteria for Horndeski models discussed in \cref{sec:theory}, the authors of \cite{Creminelli2017} propose an additional theoretical prior in the $\{\aM, \aB\}$ parameter space which severely restricts the viable space of models with Vainshtein screening. They consider the stability of nonlinear solutions to cubic Galileon-like theories, by evaluating the kinetic matrix of scalar field perturbations. The requirement that the theory be free from ghost and gradient instabilities (see \cref{sec:ansatz}) translates into a set of positivity requirements on the kinetic matrix, which they argue are generically violated over cosmological scales. In terms of the Horndeski $\alpha_i$ parameters, the stability bound translates into the condition \cite{Noller2020}:
\begin{equation}
    |\aM + \aB| \lesssim 10^{-2}.
    \label{eqn:creminelli}
\end{equation}
For the reasons explained in \cref{sec:screening}, theories with a Vainshtein screening mechanism are unlikely to be significantly constrained by our work here. The authors of \cite{Creminelli_2019} note that without a UV completion of their framework, it is difficult to assess whether the instability has detectable effects on GW signals (though it raises concerns about theoretical viability in general). Hence we have chosen to consider the effects of this bound separately from our forecasts. The effect of this additional theory prior on LSS-only constraints is considered in \cite{Noller2020}, showing the extremely large reduction it places on the remaining space of viable models. 

\begin{figure}
  \begin{centering}
        \includegraphics[width=0.5\textwidth]{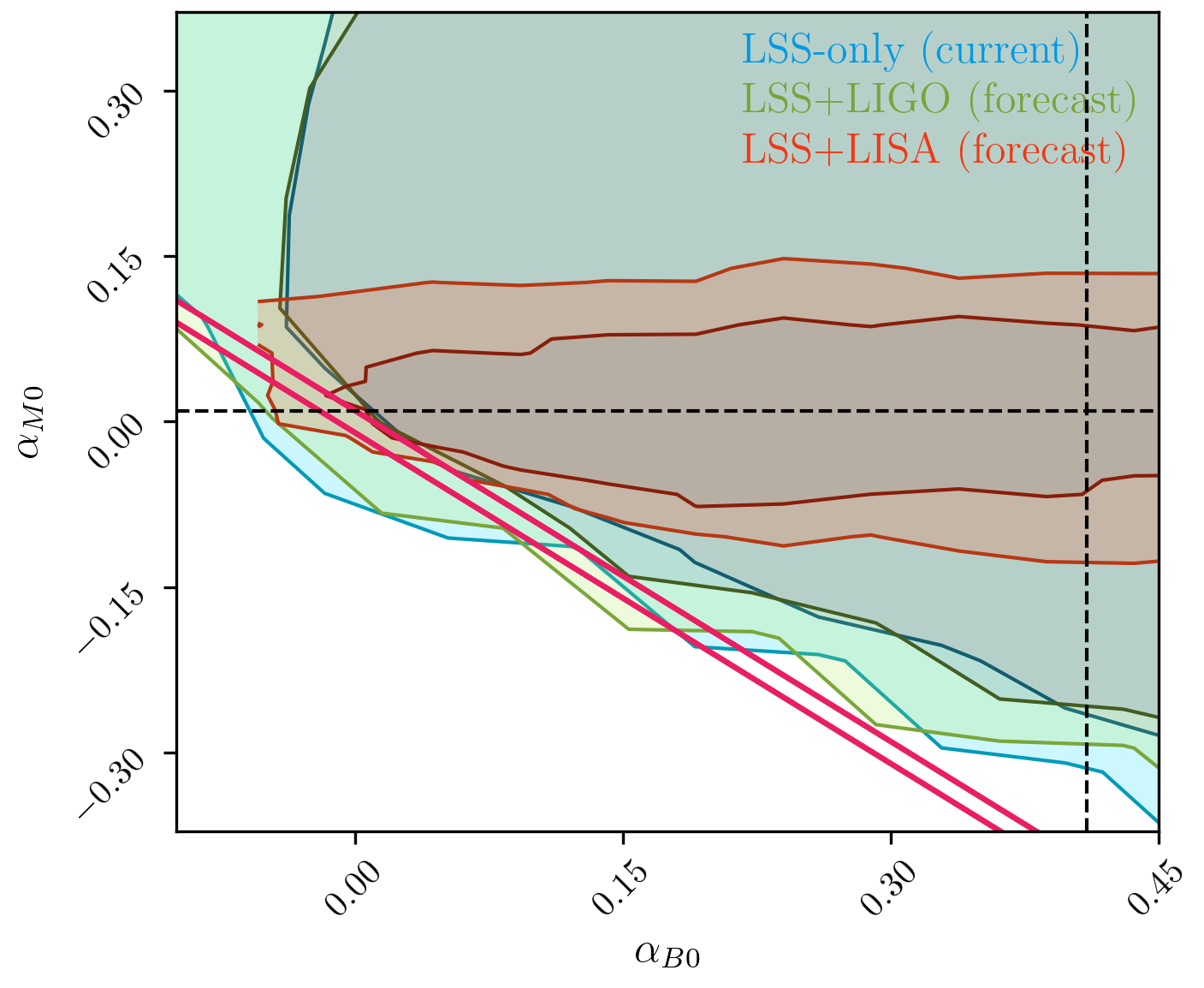}
    \includegraphics[width=0.47\textwidth]{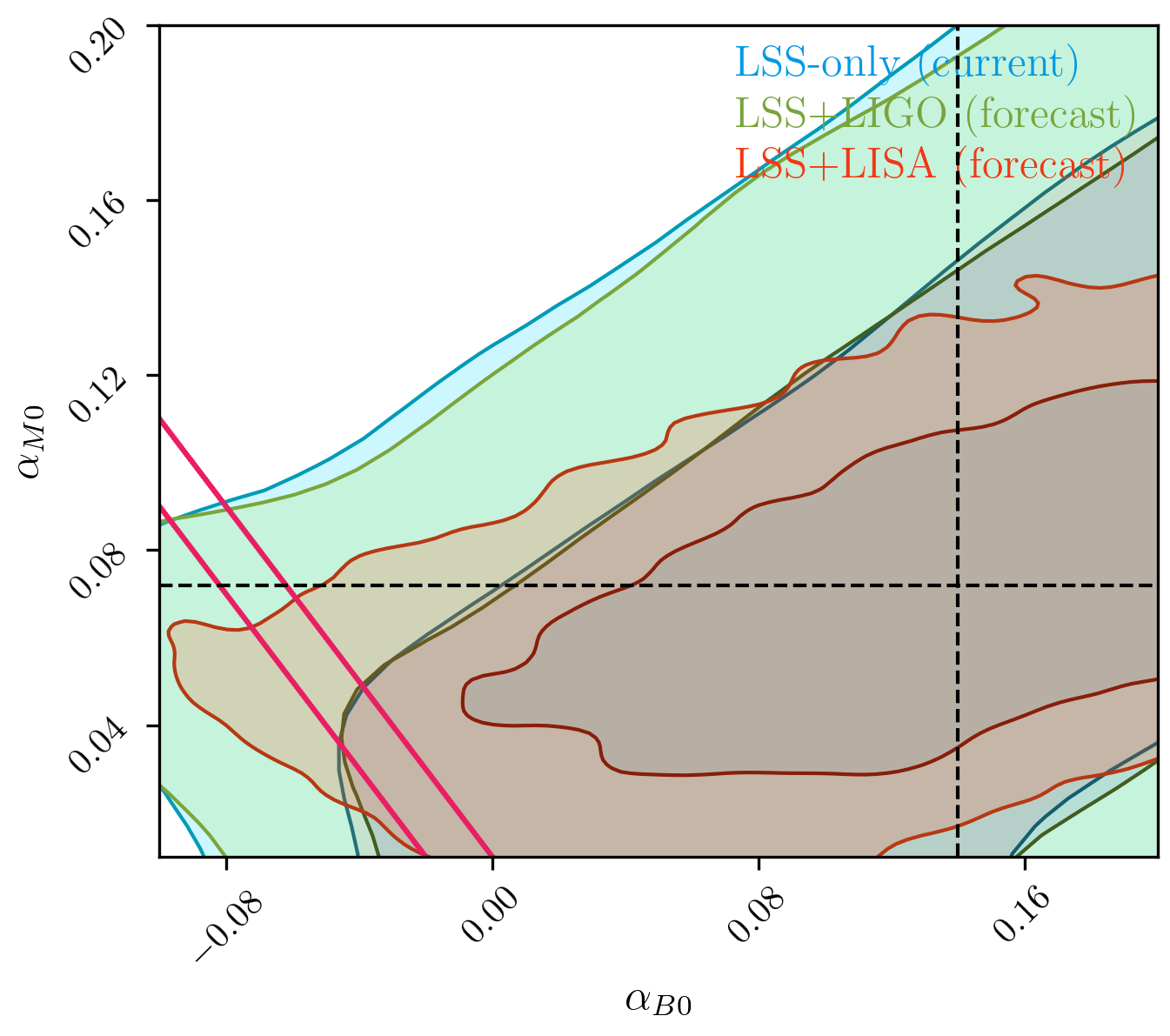}
  \end{centering}
  \caption{Constraints on Horndeski parameters in both the \proptoomega (\emph{left}) and \proptoscale (\emph{right}) cases, with the addition of the theoretical bound on Horndeski parameters \cref{eqn:creminelli} from \cite{Creminelli2017} shown as the magenta lines. The best fitting cosmology for the LSS-only data is shown as the dashed line; this is also the cosmology at which the two forecast data sets were generated at.}
  \label{fig:crem_bound}
\end{figure}

In \cref{fig:crem_bound} we show the similar effect of this prior on our own constraints, where it dominates over the information available from any of the combined data sets. Note that LSS-only constraints are compatible with this bound at the 68\% confidence level, but the best-fitting cosmology (indicated by the dashed lines) is not. In both \proptoomega and \proptoscale cases, deviations from this bound appear marginally detectable with the LSS+LISA forecast constraints at the best-fitting LSS-only cosmology. If GR were correct and the values of $\aM$ and $\aB$ are both zero, the size of the two forecast constraints will remain the same, but they would expect to be centred around the origin (and hence more compatible with the theory bound).

\subsection{Connection to the $\{\Xi_0, n\}$ Parameterisation}
\label{sub:alt_param}
As discussed in the introduction, an alternative parameterisation of modified GW luminosity distances was presented in \cite{Belgacem_2018_modified}. This parameterisation is more general than the one we used here, in that it parameterises the phenomenology of deviations between luminosity distances, and can hence be applied to gravity theories beyond Horndeski gravity. However, this means it cannot be directly and generally connected to EM-only constraints (such as those from LSS). Constraints on this parameterisation from LISA standard sirens were forecast in \cite{Belgacem2019}. Here we will connect our forecast bounds on $\aM$ to this alternative set of variables.

In the framework of \cite{Belgacem_2018_modified}, the ratio of GW and EM luminosity distances is parameterised by two constants $\Xi_0$ and $n$ in an expansion, as follows:
\begin{eqnarray}
  \label{eqn:xi}
  \Xi(z)&=&\frac{d_{GW}(z)}{d_L(z)}=\Xi_0+\frac{1-\Xi_0}{(1+z)^n}
\end{eqnarray}
Note the limits $\Xi(z\ll1)\sim 1$ and $\Xi(z\gg 1)\sim \Xi_0$, which matches the behaviour shown in the lower left panel of our \cref{fig:test_models}. The mapping between these expressions and our $\aM$ function is given (for any $\aM$ ansatz form) by:
\begin{eqnarray}
\Xi_0 &=& \exp \left[ \frac{1}{2} \int_0^\infty dz \, \frac{\alpha_M(z)}{(1+z)}\right] \\
n &=& \frac{\aM(z=0)}{2\left(\Xi_0-1\right)}
\end{eqnarray}
Table 1 of \cite{Belgacem_2018_modified} presents a dictionary for converting between the phenomenological parameterisation \cref{eqn:xi} and parameters in a number of modified gravity theories. For the two Horndeski theories considered here, in the case of \proptoomega:
\begin{eqnarray}
\Xi_0&=& 1-\frac{{\aM}_0}{6}\ln \Omega_{M0} \qquad\qquad n=-\frac{3}{\ln\Omega_{M0}},
\end{eqnarray}
and \proptoscale:
\begin{eqnarray}
\Xi_0&=& 1+\frac{{\aM}_0}{2} \qquad\qquad n=1.
\end{eqnarray}
For our tightest constraints (the case of the pop. III model for LISA) we present the constraints on $\Xi_0$ in \cref{tab:model_params_propto_scale,tab:model_params_propto_omega}. We find them to be comparable to those from \cite{Belgacem_2018_modified}, who used a different combination of data sets and a $w$CDM background model to find in the range $\sigma_{\Xi_0} = \Omega_{\Lambda 0} \times 0.022$ to $\Omega_{\Lambda 0} \times 0.044$.

\section{Conclusions}
\label{sec:conclusions}
In just five years, direct detection of gravitational waves has yielded a succession of discoveries about the gravitational universe. For cosmology, perhaps the most significant of these come from a single data point, that of GW170817. In this work we have anticipated the forthcoming improvements from the inevitable detection of further standard sirens.

In many ways, our analysis has been relatively conservative. The constraints in \cref{fig:exps} make no assumptions about the future increased accuracy of electromagnetic cosmological observables. At present it is difficult to be certain about the galaxy and weak lensing surveys that will be coincident with LISA; at the very least, they will share the significant challenges of modelling the nonlinear regime faced by the present and next generations of EM experiments. A natural extension of the current work is to include in our forecast the large-scale structure data provided by DESI, Euclid and the Vera Rubin Observatory, following \cite{Alonso_2017, Fert__2019}. Our current forecast did not include weak lensing data precisely due to the importance of modelling non-linear scales in modified gravity theories, but we note that other authors have formed constraints using approximate prescriptions \cite{Spurio_Mancini_2019}, finding improvement by factors $\sim 1.5$ over the LSS-only constraints (without weak lensing) found here.

Furthermore, in this analysis we have focussed solely on modifications of the GW amplitude via the GW luminosity distance. This is a very generic signature shared broadly by modified gravity theories well beyond the Horndeski class. However, changes to the GW phase and post-Newtonian corrections are also possible in other gravity models, and could provide equally valuable constraints if they can be disentangled from the astrophysics of the source itself (such as spins of the MBHs, which we have neglected here) \cite{Niu2020}. Whilst Horndeski gravity is increasingly under fire on both theoretical and observational fronts \cite{Creminelli_2018,2019PhRvD..99j3502N,Creminelli_2019, Noller2020}, it serves as a useful test case to demonstrate the complementarity of EM and GW data. Here we found that LISA standard sirens were able to enhance the bounds on \emph{some} MG parameters by a factor of $\sim 7$ over low-redshift data, but are only weakly sensitive to other key MG parameters, such as $\alpha_B$. The order of magnitude of these bounds depends only mildly on the details of the MG modelling (see \cref{sec:ansatz}); the shape of the parameter contours is more significantly affected, however.

Though we have said that the detection of further standard siren events is inevitable, this is a weak statement --- the prevalence of electromagnetic counterparts to GW events remains highly uncertain. This is particularly true for LISA sources, which necessarily depend on the gas environments around massive binary black holes and the mechanisms underlying quasar/flare events in galactic nuclei. Having two distance proxies to compare --- one from the EM counterpart redshift --- was crucial for the forecasts we performed here. So what if EM counterparts of binary MBH mergers are only ever identifiable for very nearby sources?

Fortunately, a swathe of new techniques for GW cosmology even in the absence of the EM counterparts is emerging.  Based on the original idea of Schutz \cite{1986Natur.323..310S} and subsequently built upon in \cite{PhysRevD.86.043011, Pozzo:it,2019ApJ...876L...7S,2020arXiv200614961P}, recent works have investigated the possibility of statistically inferring cosmological parameters \textit{without} unique source redshifts, by effectively marginalising over the redshifts of all detected galaxies in the source localisation volume. Other tools under consideration are using the clustering scale of host galaxies \cite{mukherjee2020accurate}, or the presence of a gap in the mass range of compact objects \cite{Farr_2019, ezquiaga2020jumping}, features which can potentially break the famous degeneracy of chirp mass and redshift in GW signals. The detection of higher-order harmonic modes of GWs could help to break other parameter degeneracies, such as that at between source distance and inclination, or mass ratio and black hole spins \cite{borhanian2020dark,mills2020measuring}.

Obtaining an independent measurement of the Hubble constant from GWs has been the primary driver for many of these new techniques. We have demonstrated here that the burgeoning field of GW cosmology has valuable contributions to make on other driving questions of the field, such as the nature of dark energy (and others \cite{Pardo_2018,corman2020constraining,hogg2020, martinelli2020,Bachega2020}). Collaboration between EM and GW experimental consortia will be key to obtaining optimum scientific results in this area in the future \cite{Maggiore2020JCAP}.

\acknowledgments{
The preparation of this manuscript was made possible by a number of software packages: \texttt{NumPy}, \texttt{SciPy} \citep{2020SciPy-NMeth}, \texttt{Matplotlib} \citep{2007CSE.....9...90H}, \cosmosis \cite{Zuntz_2015}, \hiclass \cite{2017JCAP...08..019Z,Bellini_2020} and \texttt{IPython}/\texttt{Jupyter} \citep{2007CSE.....9c..21P}.
We use \texttt{emcee} \citep{2013PASP..125..306F} for sampling posterior distributions, and \texttt{ChainConsumer} \citep{2016JOSS....1...45H} for analysis and plotting.

IH acknowledges support from the European Research Council in the form of a Consolidator Grant with number 681431 and from the Beecroft Trust. TB is supported by the Royal Society grant number URF\textbackslash R1\textbackslash 180009. We are pleased to thank Johannes Noller, Andrina Nicola, Daniel B. Thomas, Macarena Lagos, Andrew R. Williamson, Emilio Bellini and Michele Maggiore for useful discussions.
}

\bibliographystyle{apsrev4-1}
\bibliography{muggle_refs}

\begin{thebibliography}{136}%
\makeatletter
\providecommand \@ifxundefined [1]{%
 \@ifx{#1\undefined}
}%
\providecommand \@ifnum [1]{%
 \ifnum #1\expandafter \@firstoftwo
 \else \expandafter \@secondoftwo
 \fi
}%
\providecommand \@ifx [1]{%
 \ifx #1\expandafter \@firstoftwo
 \else \expandafter \@secondoftwo
 \fi
}%
\providecommand \natexlab [1]{#1}%
\providecommand \enquote  [1]{``#1''}%
\providecommand \bibnamefont  [1]{#1}%
\providecommand \bibfnamefont [1]{#1}%
\providecommand \citenamefont [1]{#1}%
\providecommand \href@noop [0]{\@secondoftwo}%
\providecommand \href [0]{\begingroup \@sanitize@url \@href}%
\providecommand \@href[1]{\@@startlink{#1}\@@href}%
\providecommand \@@href[1]{\endgroup#1\@@endlink}%
\providecommand \@sanitize@url [0]{\catcode `\\12\catcode `\$12\catcode
  `\&12\catcode `\#12\catcode `\^12\catcode `\_12\catcode `\%12\relax}%
\providecommand \@@startlink[1]{}%
\providecommand \@@endlink[0]{}%
\providecommand \url  [0]{\begingroup\@sanitize@url \@url }%
\providecommand \@url [1]{\endgroup\@href {#1}{\urlprefix }}%
\providecommand \urlprefix  [0]{URL }%
\providecommand \Eprint [0]{\href }%
\providecommand \doibase [0]{http://dx.doi.org/}%
\providecommand \selectlanguage [0]{\@gobble}%
\providecommand \bibinfo  [0]{\@secondoftwo}%
\providecommand \bibfield  [0]{\@secondoftwo}%
\providecommand \translation [1]{[#1]}%
\providecommand \BibitemOpen [0]{}%
\providecommand \bibitemStop [0]{}%
\providecommand \bibitemNoStop [0]{.\EOS\space}%
\providecommand \EOS [0]{\spacefactor3000\relax}%
\providecommand \BibitemShut  [1]{\csname bibitem#1\endcsname}%
\let\auto@bib@innerbib\@empty
\bibitem [{\citenamefont {{Abbott}}\ \emph {et~al.}(2017)\citenamefont
  {{Abbott}} \emph {et~al.}}]{LIGO2017}%
  \BibitemOpen
  \bibfield  {author} {\bibinfo {author} {\bibfnamefont {B.~P.}\ \bibnamefont
  {{Abbott}}} \emph {et~al.},\ }\href {\doibase 10.3847/2041-8213/aa91c9}
  {\bibfield  {journal} {\bibinfo  {journal} {\apjl}\ }\textbf {\bibinfo
  {volume} {848}},\ \bibinfo {eid} {L12} (\bibinfo {year} {2017})},\ \Eprint
  {http://arxiv.org/abs/1710.05833} {arXiv:1710.05833 [astro-ph.HE]}
  \BibitemShut {NoStop}%
\bibitem [{\citenamefont {Abbott}\ \emph {et~al.}(2017)\citenamefont {Abbott}
  \emph {et~al.}}]{PhysRevLett.119.161101}%
  \BibitemOpen
  \bibfield  {author} {\bibinfo {author} {\bibfnamefont {B.~P.}\ \bibnamefont
  {Abbott}} \emph {et~al.} (\bibinfo {collaboration} {LIGO Scientific
  Collaboration and Virgo Collaboration}),\ }\href {\doibase
  10.1103/PhysRevLett.119.161101} {\bibfield  {journal} {\bibinfo  {journal}
  {Phys. Rev. Lett.}\ }\textbf {\bibinfo {volume} {119}},\ \bibinfo {pages}
  {161101} (\bibinfo {year} {2017})}\BibitemShut {NoStop}%
\bibitem [{\citenamefont {{Baker}}\ \emph {et~al.}(2017)\citenamefont
  {{Baker}}, \citenamefont {{Bellini}}, \citenamefont {{Ferreira}},
  \citenamefont {{Lagos}}, \citenamefont {{Noller}},\ and\ \citenamefont
  {{Sawicki}}}]{Baker2017}%
  \BibitemOpen
  \bibfield  {author} {\bibinfo {author} {\bibfnamefont {T.}~\bibnamefont
  {{Baker}}}, \bibinfo {author} {\bibfnamefont {E.}~\bibnamefont {{Bellini}}},
  \bibinfo {author} {\bibfnamefont {P.~G.}\ \bibnamefont {{Ferreira}}},
  \bibinfo {author} {\bibfnamefont {M.}~\bibnamefont {{Lagos}}}, \bibinfo
  {author} {\bibfnamefont {J.}~\bibnamefont {{Noller}}}, \ and\ \bibinfo
  {author} {\bibfnamefont {I.}~\bibnamefont {{Sawicki}}},\ }\href {\doibase
  10.1103/PhysRevLett.119.251301} {\bibfield  {journal} {\bibinfo  {journal}
  {Physical Review Letters}\ }\textbf {\bibinfo {volume} {119}},\ \bibinfo
  {eid} {251301} (\bibinfo {year} {2017})},\ \Eprint
  {http://arxiv.org/abs/1710.06394} {arXiv:1710.06394} \BibitemShut {NoStop}%
\bibitem [{\citenamefont {{Mar{\'{\i}}a Ezquiaga}}\ and\ \citenamefont
  {{Zumalac{\'a}rregui}}(2017)}]{Zuma2017}%
  \BibitemOpen
  \bibfield  {author} {\bibinfo {author} {\bibfnamefont {J.}~\bibnamefont
  {{Mar{\'{\i}}a Ezquiaga}}}\ and\ \bibinfo {author} {\bibfnamefont
  {M.}~\bibnamefont {{Zumalac{\'a}rregui}}},\ }\href@noop {} {\bibfield
  {journal} {\bibinfo  {journal} {ArXiv e-prints}\ } (\bibinfo {year}
  {2017})},\ \Eprint {http://arxiv.org/abs/1710.05901} {arXiv:1710.05901}
  \BibitemShut {NoStop}%
\bibitem [{\citenamefont {{Creminelli}}\ and\ \citenamefont
  {{Vernizzi}}(2017)}]{Creminelli2017}%
  \BibitemOpen
  \bibfield  {author} {\bibinfo {author} {\bibfnamefont {P.}~\bibnamefont
  {{Creminelli}}}\ and\ \bibinfo {author} {\bibfnamefont {F.}~\bibnamefont
  {{Vernizzi}}},\ }\href {\doibase 10.1103/PhysRevLett.119.251302} {\bibfield
  {journal} {\bibinfo  {journal} {Physical Review Letters}\ }\textbf {\bibinfo
  {volume} {119}},\ \bibinfo {eid} {251302} (\bibinfo {year} {2017})},\ \Eprint
  {http://arxiv.org/abs/1710.05877} {arXiv:1710.05877} \BibitemShut {NoStop}%
\bibitem [{\citenamefont {{Sakstein}}\ and\ \citenamefont
  {{Jain}}(2017)}]{Sakstein2017}%
  \BibitemOpen
  \bibfield  {author} {\bibinfo {author} {\bibfnamefont {J.}~\bibnamefont
  {{Sakstein}}}\ and\ \bibinfo {author} {\bibfnamefont {B.}~\bibnamefont
  {{Jain}}},\ }\href {\doibase 10.1103/PhysRevLett.119.251303} {\bibfield
  {journal} {\bibinfo  {journal} {Physical Review Letters}\ }\textbf {\bibinfo
  {volume} {119}},\ \bibinfo {eid} {251303} (\bibinfo {year} {2017})},\ \Eprint
  {http://arxiv.org/abs/1710.05893} {arXiv:1710.05893} \BibitemShut {NoStop}%
\bibitem [{\citenamefont {Boran}\ \emph {et~al.}(2018)\citenamefont {Boran},
  \citenamefont {Desai}, \citenamefont {Kahya},\ and\ \citenamefont
  {Woodard}}]{Boran_2018}%
  \BibitemOpen
  \bibfield  {author} {\bibinfo {author} {\bibfnamefont {S.}~\bibnamefont
  {Boran}}, \bibinfo {author} {\bibfnamefont {S.}~\bibnamefont {Desai}},
  \bibinfo {author} {\bibfnamefont {E.}~\bibnamefont {Kahya}}, \ and\ \bibinfo
  {author} {\bibfnamefont {R.}~\bibnamefont {Woodard}},\ }\href {\doibase
  10.1103/physrevd.97.041501} {\bibfield  {journal} {\bibinfo  {journal}
  {Physical Review D}\ }\textbf {\bibinfo {volume} {97}} (\bibinfo {year}
  {2018}),\ 10.1103/physrevd.97.041501}\BibitemShut {NoStop}%
\bibitem [{\citenamefont {{Mastrogiovanni}}\ \emph {et~al.}(2020)\citenamefont
  {{Mastrogiovanni}}, \citenamefont {{Steer}},\ and\ \citenamefont
  {{Barsuglia}}}]{Mastrogiovanni2020_jointtests}%
  \BibitemOpen
  \bibfield  {author} {\bibinfo {author} {\bibfnamefont {S.}~\bibnamefont
  {{Mastrogiovanni}}}, \bibinfo {author} {\bibfnamefont {D.~A.}\ \bibnamefont
  {{Steer}}}, \ and\ \bibinfo {author} {\bibfnamefont {M.}~\bibnamefont
  {{Barsuglia}}},\ }\href@noop {} {\bibfield  {journal} {\bibinfo  {journal}
  {arXiv e-prints}\ ,\ \bibinfo {eid} {arXiv:2004.06102}} (\bibinfo {year}
  {2020})},\ \Eprint {http://arxiv.org/abs/2004.06102} {arXiv:2004.06102
  [gr-qc]} \BibitemShut {NoStop}%
\bibitem [{\citenamefont {{Ishak}}\ \emph {et~al.}(2019)\citenamefont
  {{Ishak}}, \citenamefont {{Baker}}, \citenamefont {{Bull}}, \citenamefont
  {{Pedersen}}, \citenamefont {{Blazek}}, \citenamefont {{Ferreira}},
  \citenamefont {{Leonard}}, \citenamefont {{Lin}}, \citenamefont {{Linder}},
  \citenamefont {{Pardo}},\ and\ \citenamefont
  {{Valogiannis}}}]{2019arXiv190509687I}%
  \BibitemOpen
  \bibfield  {author} {\bibinfo {author} {\bibfnamefont {M.}~\bibnamefont
  {{Ishak}}}, \bibinfo {author} {\bibfnamefont {T.}~\bibnamefont {{Baker}}},
  \bibinfo {author} {\bibfnamefont {P.}~\bibnamefont {{Bull}}}, \bibinfo
  {author} {\bibfnamefont {E.~M.}\ \bibnamefont {{Pedersen}}}, \bibinfo
  {author} {\bibfnamefont {J.}~\bibnamefont {{Blazek}}}, \bibinfo {author}
  {\bibfnamefont {P.~G.}\ \bibnamefont {{Ferreira}}}, \bibinfo {author}
  {\bibfnamefont {C.~D.}\ \bibnamefont {{Leonard}}}, \bibinfo {author}
  {\bibfnamefont {W.}~\bibnamefont {{Lin}}}, \bibinfo {author} {\bibfnamefont
  {E.}~\bibnamefont {{Linder}}}, \bibinfo {author} {\bibfnamefont
  {K.}~\bibnamefont {{Pardo}}}, \ and\ \bibinfo {author} {\bibfnamefont
  {G.}~\bibnamefont {{Valogiannis}}},\ }\href@noop {} {\bibfield  {journal}
  {\bibinfo  {journal} {arXiv e-prints}\ ,\ \bibinfo {eid} {arXiv:1905.09687}}
  (\bibinfo {year} {2019})},\ \Eprint {http://arxiv.org/abs/1905.09687}
  {arXiv:1905.09687 [astro-ph.CO]} \BibitemShut {NoStop}%
\bibitem [{\citenamefont {Tasinato}(2014)}]{Tasinato_2014}%
  \BibitemOpen
  \bibfield  {author} {\bibinfo {author} {\bibfnamefont {G.}~\bibnamefont
  {Tasinato}},\ }\href {\doibase 10.1007/jhep04(2014)067} {\bibfield  {journal}
  {\bibinfo  {journal} {Journal of High Energy Physics}\ }\textbf {\bibinfo
  {volume} {2014}} (\bibinfo {year} {2014}),\
  10.1007/jhep04(2014)067}\BibitemShut {NoStop}%
\bibitem [{\citenamefont {{Heisenberg}}(2014)}]{2014JCAP...05..015H}%
  \BibitemOpen
  \bibfield  {author} {\bibinfo {author} {\bibfnamefont {L.}~\bibnamefont
  {{Heisenberg}}},\ }\href {\doibase 10.1088/1475-7516/2014/05/015} {\bibfield
  {journal} {\bibinfo  {journal} {\jcap}\ }\textbf {\bibinfo {volume} {2014}},\
  \bibinfo {eid} {015} (\bibinfo {year} {2014})},\ \Eprint
  {http://arxiv.org/abs/1402.7026} {arXiv:1402.7026 [hep-th]} \BibitemShut
  {NoStop}%
\bibitem [{\citenamefont {{Skordis}}\ and\ \citenamefont
  {{Z{\l}o{\'s}nik}}(2019)}]{2019PhRvD.100j4013S}%
  \BibitemOpen
  \bibfield  {author} {\bibinfo {author} {\bibfnamefont {C.}~\bibnamefont
  {{Skordis}}}\ and\ \bibinfo {author} {\bibfnamefont {T.}~\bibnamefont
  {{Z{\l}o{\'s}nik}}},\ }\href {\doibase 10.1103/PhysRevD.100.104013}
  {\bibfield  {journal} {\bibinfo  {journal} {\prd}\ }\textbf {\bibinfo
  {volume} {100}},\ \bibinfo {eid} {104013} (\bibinfo {year} {2019})},\ \Eprint
  {http://arxiv.org/abs/1905.09465} {arXiv:1905.09465 [gr-qc]} \BibitemShut
  {NoStop}%
\bibitem [{\citenamefont {Schutz}(1986)}]{Schutz:1986gp}%
  \BibitemOpen
  \bibfield  {author} {\bibinfo {author} {\bibfnamefont {B.~F.}\ \bibnamefont
  {Schutz}},\ }\href {\doibase 10.1038/323310a0} {\bibfield  {journal}
  {\bibinfo  {journal} {Nature}\ }\textbf {\bibinfo {volume} {323}},\ \bibinfo
  {pages} {310} (\bibinfo {year} {1986})}\BibitemShut {NoStop}%
\bibitem [{\citenamefont {{Holz}}\ and\ \citenamefont
  {{Hughes}}(2005)}]{2005ApJ...629...15H}%
  \BibitemOpen
  \bibfield  {author} {\bibinfo {author} {\bibfnamefont {D.~E.}\ \bibnamefont
  {{Holz}}}\ and\ \bibinfo {author} {\bibfnamefont {S.~A.}\ \bibnamefont
  {{Hughes}}},\ }\href {\doibase 10.1086/431341} {\bibfield  {journal}
  {\bibinfo  {journal} {\apj}\ }\textbf {\bibinfo {volume} {629}},\ \bibinfo
  {pages} {15} (\bibinfo {year} {2005})},\ \Eprint
  {http://arxiv.org/abs/astro-ph/0504616} {arXiv:astro-ph/0504616 [astro-ph]}
  \BibitemShut {NoStop}%
\bibitem [{\citenamefont {Chen}\ \emph {et~al.}(2018)\citenamefont {Chen},
  \citenamefont {Fishbach},\ and\ \citenamefont {Holz}}]{Chen_2018}%
  \BibitemOpen
  \bibfield  {author} {\bibinfo {author} {\bibfnamefont {H.-Y.}\ \bibnamefont
  {Chen}}, \bibinfo {author} {\bibfnamefont {M.}~\bibnamefont {Fishbach}}, \
  and\ \bibinfo {author} {\bibfnamefont {D.~E.}\ \bibnamefont {Holz}},\ }\href
  {\doibase 10.1038/s41586-018-0606-0} {\bibfield  {journal} {\bibinfo
  {journal} {Nature}\ }\textbf {\bibinfo {volume} {562}},\ \bibinfo {pages}
  {545–547} (\bibinfo {year} {2018})}\BibitemShut {NoStop}%
\bibitem [{\citenamefont {Feeney}\ \emph {et~al.}(2018)\citenamefont {Feeney},
  \citenamefont {Mortlock},\ and\ \citenamefont {Dalmasso}}]{Feeney_2018}%
  \BibitemOpen
  \bibfield  {author} {\bibinfo {author} {\bibfnamefont {S.~M.}\ \bibnamefont
  {Feeney}}, \bibinfo {author} {\bibfnamefont {D.~J.}\ \bibnamefont
  {Mortlock}}, \ and\ \bibinfo {author} {\bibfnamefont {N.}~\bibnamefont
  {Dalmasso}},\ }\href {\doibase 10.1093/mnras/sty418} {\bibfield  {journal}
  {\bibinfo  {journal} {Monthly Notices of the Royal Astronomical Society}\
  }\textbf {\bibinfo {volume} {476}},\ \bibinfo {pages} {3861–3882} (\bibinfo
  {year} {2018})}\BibitemShut {NoStop}%
\bibitem [{\citenamefont {Feeney}\ \emph {et~al.}(2019)\citenamefont {Feeney},
  \citenamefont {Peiris}, \citenamefont {Williamson}, \citenamefont {Nissanke},
  \citenamefont {Mortlock}, \citenamefont {Alsing},\ and\ \citenamefont
  {Scolnic}}]{Feeney_2019}%
  \BibitemOpen
  \bibfield  {author} {\bibinfo {author} {\bibfnamefont {S.~M.}\ \bibnamefont
  {Feeney}}, \bibinfo {author} {\bibfnamefont {H.~V.}\ \bibnamefont {Peiris}},
  \bibinfo {author} {\bibfnamefont {A.~R.}\ \bibnamefont {Williamson}},
  \bibinfo {author} {\bibfnamefont {S.~M.}\ \bibnamefont {Nissanke}}, \bibinfo
  {author} {\bibfnamefont {D.~J.}\ \bibnamefont {Mortlock}}, \bibinfo {author}
  {\bibfnamefont {J.}~\bibnamefont {Alsing}}, \ and\ \bibinfo {author}
  {\bibfnamefont {D.}~\bibnamefont {Scolnic}},\ }\href {\doibase
  10.1103/physrevlett.122.061105} {\bibfield  {journal} {\bibinfo  {journal}
  {Physical Review Letters}\ }\textbf {\bibinfo {volume} {122}} (\bibinfo
  {year} {2019}),\ 10.1103/physrevlett.122.061105}\BibitemShut {NoStop}%
\bibitem [{\citenamefont {Lagos}\ \emph {et~al.}(2019)\citenamefont {Lagos},
  \citenamefont {Fishbach}, \citenamefont {Landry},\ and\ \citenamefont
  {Holz}}]{Lagos_2019}%
  \BibitemOpen
  \bibfield  {author} {\bibinfo {author} {\bibfnamefont {M.}~\bibnamefont
  {Lagos}}, \bibinfo {author} {\bibfnamefont {M.}~\bibnamefont {Fishbach}},
  \bibinfo {author} {\bibfnamefont {P.}~\bibnamefont {Landry}}, \ and\ \bibinfo
  {author} {\bibfnamefont {D.~E.}\ \bibnamefont {Holz}},\ }\href {\doibase
  10.1103/PhysRevD.99.083504} {\bibfield  {journal} {\bibinfo  {journal} {Phys.
  Rev. D}\ }\textbf {\bibinfo {volume} {99}},\ \bibinfo {pages} {083504}
  (\bibinfo {year} {2019})}\BibitemShut {NoStop}%
\bibitem [{\citenamefont {{Gray}}\ \emph {et~al.}(2019)\citenamefont {{Gray}},
  \citenamefont {{Maga{\~n}a Hernandez}}, \citenamefont {{Qi}}, \citenamefont
  {{Sur}}, \citenamefont {{Brady}}, \citenamefont {{Chen}}, \citenamefont
  {{Farr}}, \citenamefont {{Fishbach}}, \citenamefont {{Gair}}, \citenamefont
  {{Ghosh}}, \citenamefont {{Holz}}, \citenamefont {{Mastrogiovanni}},
  \citenamefont {{Messenger}}, \citenamefont {{Steer}},\ and\ \citenamefont
  {{Veitch}}}]{2019arXiv190806050G}%
  \BibitemOpen
  \bibfield  {author} {\bibinfo {author} {\bibfnamefont {R.}~\bibnamefont
  {{Gray}}}, \bibinfo {author} {\bibfnamefont {I.}~\bibnamefont {{Maga{\~n}a
  Hernandez}}}, \bibinfo {author} {\bibfnamefont {H.}~\bibnamefont {{Qi}}},
  \bibinfo {author} {\bibfnamefont {A.}~\bibnamefont {{Sur}}}, \bibinfo
  {author} {\bibfnamefont {P.~R.}\ \bibnamefont {{Brady}}}, \bibinfo {author}
  {\bibfnamefont {H.-Y.}\ \bibnamefont {{Chen}}}, \bibinfo {author}
  {\bibfnamefont {W.~M.}\ \bibnamefont {{Farr}}}, \bibinfo {author}
  {\bibfnamefont {M.}~\bibnamefont {{Fishbach}}}, \bibinfo {author}
  {\bibfnamefont {J.~R.}\ \bibnamefont {{Gair}}}, \bibinfo {author}
  {\bibfnamefont {A.}~\bibnamefont {{Ghosh}}}, \bibinfo {author} {\bibfnamefont
  {D.~E.}\ \bibnamefont {{Holz}}}, \bibinfo {author} {\bibfnamefont
  {S.}~\bibnamefont {{Mastrogiovanni}}}, \bibinfo {author} {\bibfnamefont
  {C.}~\bibnamefont {{Messenger}}}, \bibinfo {author} {\bibfnamefont {D.~A.}\
  \bibnamefont {{Steer}}}, \ and\ \bibinfo {author} {\bibfnamefont
  {J.}~\bibnamefont {{Veitch}}},\ }\href@noop {} {\bibfield  {journal}
  {\bibinfo  {journal} {arXiv e-prints}\ ,\ \bibinfo {eid} {arXiv:1908.06050}}
  (\bibinfo {year} {2019})},\ \Eprint {http://arxiv.org/abs/1908.06050}
  {arXiv:1908.06050 [gr-qc]} \BibitemShut {NoStop}%
\bibitem [{\citenamefont {{Mortlock}}\ \emph {et~al.}(2019)\citenamefont
  {{Mortlock}}, \citenamefont {{Feeney}}, \citenamefont {{Peiris}},
  \citenamefont {{Williamson}},\ and\ \citenamefont
  {{Nissanke}}}]{2019PhRvD.100j3523M}%
  \BibitemOpen
  \bibfield  {author} {\bibinfo {author} {\bibfnamefont {D.~J.}\ \bibnamefont
  {{Mortlock}}}, \bibinfo {author} {\bibfnamefont {S.~M.}\ \bibnamefont
  {{Feeney}}}, \bibinfo {author} {\bibfnamefont {H.~V.}\ \bibnamefont
  {{Peiris}}}, \bibinfo {author} {\bibfnamefont {A.~R.}\ \bibnamefont
  {{Williamson}}}, \ and\ \bibinfo {author} {\bibfnamefont {S.~M.}\
  \bibnamefont {{Nissanke}}},\ }\href {\doibase 10.1103/PhysRevD.100.103523}
  {\bibfield  {journal} {\bibinfo  {journal} {\prd}\ }\textbf {\bibinfo
  {volume} {100}},\ \bibinfo {eid} {103523} (\bibinfo {year} {2019})},\ \Eprint
  {http://arxiv.org/abs/1811.11723} {arXiv:1811.11723 [astro-ph.CO]}
  \BibitemShut {NoStop}%
\bibitem [{\citenamefont {Farr}\ \emph {et~al.}(2019)\citenamefont {Farr},
  \citenamefont {Fishbach}, \citenamefont {Ye},\ and\ \citenamefont
  {Holz}}]{Farr_2019}%
  \BibitemOpen
  \bibfield  {author} {\bibinfo {author} {\bibfnamefont {W.~M.}\ \bibnamefont
  {Farr}}, \bibinfo {author} {\bibfnamefont {M.}~\bibnamefont {Fishbach}},
  \bibinfo {author} {\bibfnamefont {J.}~\bibnamefont {Ye}}, \ and\ \bibinfo
  {author} {\bibfnamefont {D.~E.}\ \bibnamefont {Holz}},\ }\href {\doibase
  10.3847/2041-8213/ab4284} {\bibfield  {journal} {\bibinfo  {journal} {The
  Astrophysical Journal}\ }\textbf {\bibinfo {volume} {883}},\ \bibinfo {pages}
  {L42} (\bibinfo {year} {2019})}\BibitemShut {NoStop}%
\bibitem [{\citenamefont {{Soares-Santos}}\ \emph {et~al.}(2019)\citenamefont
  {{Soares-Santos}}, \citenamefont {{Palmese}}, \citenamefont {{Hartley}},
  \citenamefont {{Annis}}, \citenamefont {{Garcia-Bellido}}, \citenamefont
  {{Lahav}}, \citenamefont {{Doctor}}, \citenamefont {{Fishbach}},
  \citenamefont {{Holz}}, \citenamefont {{Lin}}, \citenamefont {{Pereira}},
  \citenamefont {{Garcia}}, \citenamefont {{Herner}}, \citenamefont
  {{Kessler}}, \citenamefont {{Peiris}}, \citenamefont {{Sako}}, \citenamefont
  {{Allam}}, \citenamefont {{Brout}}, \citenamefont {{Carnero Rosell}},
  \citenamefont {{Chen}}, \citenamefont {{Conselice}}, \citenamefont
  {{deRose}}, \citenamefont {{deVicente}}, \citenamefont {{Diehl}},
  \citenamefont {{Gill}}, \citenamefont {{Gschwend}}, \citenamefont
  {{Sevilla-Noarbe}}, \citenamefont {{Tucker}}, \citenamefont {{Wechsler}},
  \citenamefont {{Berger}}, \citenamefont {{Cowperthwaite}}, \citenamefont
  {{Metzger}}, \citenamefont {{Williams}}, \citenamefont {{Abbott}} \emph
  {et~al.}}]{2019ApJ...876L...7S}%
  \BibitemOpen
  \bibfield  {author} {\bibinfo {author} {\bibfnamefont {M.}~\bibnamefont
  {{Soares-Santos}}}, \bibinfo {author} {\bibfnamefont {A.}~\bibnamefont
  {{Palmese}}}, \bibinfo {author} {\bibfnamefont {W.}~\bibnamefont
  {{Hartley}}}, \bibinfo {author} {\bibfnamefont {J.}~\bibnamefont {{Annis}}},
  \bibinfo {author} {\bibfnamefont {J.}~\bibnamefont {{Garcia-Bellido}}},
  \bibinfo {author} {\bibfnamefont {O.}~\bibnamefont {{Lahav}}}, \bibinfo
  {author} {\bibfnamefont {Z.}~\bibnamefont {{Doctor}}}, \bibinfo {author}
  {\bibfnamefont {M.}~\bibnamefont {{Fishbach}}}, \bibinfo {author}
  {\bibfnamefont {D.~E.}\ \bibnamefont {{Holz}}}, \bibinfo {author}
  {\bibfnamefont {H.}~\bibnamefont {{Lin}}}, \bibinfo {author} {\bibfnamefont
  {M.~E.~S.}\ \bibnamefont {{Pereira}}}, \bibinfo {author} {\bibfnamefont
  {A.}~\bibnamefont {{Garcia}}}, \bibinfo {author} {\bibfnamefont
  {K.}~\bibnamefont {{Herner}}}, \bibinfo {author} {\bibfnamefont
  {R.}~\bibnamefont {{Kessler}}}, \bibinfo {author} {\bibfnamefont {H.~V.}\
  \bibnamefont {{Peiris}}}, \bibinfo {author} {\bibfnamefont {M.}~\bibnamefont
  {{Sako}}}, \bibinfo {author} {\bibfnamefont {S.}~\bibnamefont {{Allam}}},
  \bibinfo {author} {\bibfnamefont {D.}~\bibnamefont {{Brout}}}, \bibinfo
  {author} {\bibfnamefont {A.}~\bibnamefont {{Carnero Rosell}}}, \bibinfo
  {author} {\bibfnamefont {H.~Y.}\ \bibnamefont {{Chen}}}, \bibinfo {author}
  {\bibfnamefont {C.}~\bibnamefont {{Conselice}}}, \bibinfo {author}
  {\bibfnamefont {J.}~\bibnamefont {{deRose}}}, \bibinfo {author}
  {\bibfnamefont {J.}~\bibnamefont {{deVicente}}}, \bibinfo {author}
  {\bibfnamefont {H.~T.}\ \bibnamefont {{Diehl}}}, \bibinfo {author}
  {\bibfnamefont {M.~S.~S.}\ \bibnamefont {{Gill}}}, \bibinfo {author}
  {\bibfnamefont {J.}~\bibnamefont {{Gschwend}}}, \bibinfo {author}
  {\bibfnamefont {I.}~\bibnamefont {{Sevilla-Noarbe}}}, \bibinfo {author}
  {\bibfnamefont {D.~L.}\ \bibnamefont {{Tucker}}}, \bibinfo {author}
  {\bibfnamefont {R.}~\bibnamefont {{Wechsler}}}, \bibinfo {author}
  {\bibfnamefont {E.}~\bibnamefont {{Berger}}}, \bibinfo {author}
  {\bibfnamefont {P.~S.}\ \bibnamefont {{Cowperthwaite}}}, \bibinfo {author}
  {\bibfnamefont {B.~D.}\ \bibnamefont {{Metzger}}}, \bibinfo {author}
  {\bibfnamefont {P.~K.~G.}\ \bibnamefont {{Williams}}}, \bibinfo {author}
  {\bibfnamefont {T.~M.~C.}\ \bibnamefont {{Abbott}}},  \emph {et~al.},\ }\href
  {\doibase 10.3847/2041-8213/ab14f1} {\bibfield  {journal} {\bibinfo
  {journal} {\apjl}\ }\textbf {\bibinfo {volume} {876}},\ \bibinfo {eid} {L7}
  (\bibinfo {year} {2019})},\ \Eprint {http://arxiv.org/abs/1901.01540}
  {arXiv:1901.01540 [astro-ph.CO]} \BibitemShut {NoStop}%
\bibitem [{\citenamefont {{Palmese}}\ \emph {et~al.}(2020)\citenamefont
  {{Palmese}}, \citenamefont {{deVicente}}, \citenamefont {{Pereira}},
  \citenamefont {{Annis}}, \citenamefont {{Hartley}}, \citenamefont {{Herner}},
  \citenamefont {{Soares-Santos}}, \citenamefont {{Crocce}}, \citenamefont
  {{Huterer}}, \citenamefont {{Magana Hernandez}}, \citenamefont {{Davis}},
  \citenamefont {{Garcia}}, \citenamefont {{Garcia-Bellido}}, \citenamefont
  {{Gschwend}}, \citenamefont {{Holz}}, \citenamefont {{Kessler}},
  \citenamefont {{Lahav}}, \citenamefont {{Morgan}}, \citenamefont
  {{Nicolaou}}, \citenamefont {{Conselice}}, \citenamefont {{Foley}},
  \citenamefont {{Gill}}, \citenamefont {{Abbott}} \emph
  {et~al.}}]{2020arXiv200614961P}%
  \BibitemOpen
  \bibfield  {author} {\bibinfo {author} {\bibfnamefont {A.}~\bibnamefont
  {{Palmese}}}, \bibinfo {author} {\bibfnamefont {J.}~\bibnamefont
  {{deVicente}}}, \bibinfo {author} {\bibfnamefont {M.~E.~S.}\ \bibnamefont
  {{Pereira}}}, \bibinfo {author} {\bibfnamefont {J.}~\bibnamefont {{Annis}}},
  \bibinfo {author} {\bibfnamefont {W.}~\bibnamefont {{Hartley}}}, \bibinfo
  {author} {\bibfnamefont {K.}~\bibnamefont {{Herner}}}, \bibinfo {author}
  {\bibfnamefont {M.}~\bibnamefont {{Soares-Santos}}}, \bibinfo {author}
  {\bibfnamefont {M.}~\bibnamefont {{Crocce}}}, \bibinfo {author}
  {\bibfnamefont {D.}~\bibnamefont {{Huterer}}}, \bibinfo {author}
  {\bibfnamefont {I.}~\bibnamefont {{Magana Hernandez}}}, \bibinfo {author}
  {\bibfnamefont {T.~M.}\ \bibnamefont {{Davis}}}, \bibinfo {author}
  {\bibfnamefont {A.}~\bibnamefont {{Garcia}}}, \bibinfo {author}
  {\bibfnamefont {J.}~\bibnamefont {{Garcia-Bellido}}}, \bibinfo {author}
  {\bibfnamefont {J.}~\bibnamefont {{Gschwend}}}, \bibinfo {author}
  {\bibfnamefont {D.~E.}\ \bibnamefont {{Holz}}}, \bibinfo {author}
  {\bibfnamefont {R.}~\bibnamefont {{Kessler}}}, \bibinfo {author}
  {\bibfnamefont {O.}~\bibnamefont {{Lahav}}}, \bibinfo {author} {\bibfnamefont
  {R.}~\bibnamefont {{Morgan}}}, \bibinfo {author} {\bibfnamefont
  {C.}~\bibnamefont {{Nicolaou}}}, \bibinfo {author} {\bibfnamefont
  {C.}~\bibnamefont {{Conselice}}}, \bibinfo {author} {\bibfnamefont {R.~J.}\
  \bibnamefont {{Foley}}}, \bibinfo {author} {\bibfnamefont {M.~S.~S.}\
  \bibnamefont {{Gill}}}, \bibinfo {author} {\bibfnamefont {T.~M.~C.}\
  \bibnamefont {{Abbott}}},  \emph {et~al.},\ }\href@noop {} {\bibfield
  {journal} {\bibinfo  {journal} {arXiv e-prints}\ ,\ \bibinfo {eid}
  {arXiv:2006.14961}} (\bibinfo {year} {2020})},\ \Eprint
  {http://arxiv.org/abs/2006.14961} {arXiv:2006.14961 [astro-ph.CO]}
  \BibitemShut {NoStop}%
\bibitem [{\citenamefont {Freedman}(2017)}]{freedman2017cosmology}%
  \BibitemOpen
  \bibfield  {author} {\bibinfo {author} {\bibfnamefont {W.~L.}\ \bibnamefont
  {Freedman}},\ }\href@noop {} {\enquote {\bibinfo {title} {Cosmology at at
  crossroads: Tension with the hubble constant},}\ } (\bibinfo {year} {2017}),\
  \Eprint {http://arxiv.org/abs/1706.02739} {arXiv:1706.02739 [astro-ph.CO]}
  \BibitemShut {NoStop}%
\bibitem [{\citenamefont {Verde}\ \emph {et~al.}(2019)\citenamefont {Verde},
  \citenamefont {Treu},\ and\ \citenamefont {Riess}}]{Verde_2019}%
  \BibitemOpen
  \bibfield  {author} {\bibinfo {author} {\bibfnamefont {L.}~\bibnamefont
  {Verde}}, \bibinfo {author} {\bibfnamefont {T.}~\bibnamefont {Treu}}, \ and\
  \bibinfo {author} {\bibfnamefont {A.~G.}\ \bibnamefont {Riess}},\ }\href
  {\doibase 10.1038/s41550-019-0902-0} {\bibfield  {journal} {\bibinfo
  {journal} {Nature Astronomy}\ }\textbf {\bibinfo {volume} {3}},\ \bibinfo
  {pages} {891–895} (\bibinfo {year} {2019})}\BibitemShut {NoStop}%
\bibitem [{\citenamefont {Belgacem}\ \emph
  {et~al.}(2018{\natexlab{a}})\citenamefont {Belgacem}, \citenamefont {Dirian},
  \citenamefont {Foffa},\ and\ \citenamefont {Maggiore}}]{Belgacem_2018_gwlum}%
  \BibitemOpen
  \bibfield  {author} {\bibinfo {author} {\bibfnamefont {E.}~\bibnamefont
  {Belgacem}}, \bibinfo {author} {\bibfnamefont {Y.}~\bibnamefont {Dirian}},
  \bibinfo {author} {\bibfnamefont {S.}~\bibnamefont {Foffa}}, \ and\ \bibinfo
  {author} {\bibfnamefont {M.}~\bibnamefont {Maggiore}},\ }\href {\doibase
  10.1103/physrevd.97.104066} {\bibfield  {journal} {\bibinfo  {journal}
  {Physical Review D}\ }\textbf {\bibinfo {volume} {97}} (\bibinfo {year}
  {2018}{\natexlab{a}}),\ 10.1103/physrevd.97.104066}\BibitemShut {NoStop}%
\bibitem [{\citenamefont {Belgacem}\ \emph
  {et~al.}(2018{\natexlab{b}})\citenamefont {Belgacem}, \citenamefont {Dirian},
  \citenamefont {Foffa},\ and\ \citenamefont
  {Maggiore}}]{Belgacem_2018_modified}%
  \BibitemOpen
  \bibfield  {author} {\bibinfo {author} {\bibfnamefont {E.}~\bibnamefont
  {Belgacem}}, \bibinfo {author} {\bibfnamefont {Y.}~\bibnamefont {Dirian}},
  \bibinfo {author} {\bibfnamefont {S.}~\bibnamefont {Foffa}}, \ and\ \bibinfo
  {author} {\bibfnamefont {M.}~\bibnamefont {Maggiore}},\ }\href {\doibase
  10.1103/physrevd.98.023510} {\bibfield  {journal} {\bibinfo  {journal}
  {Physical Review D}\ }\textbf {\bibinfo {volume} {98}} (\bibinfo {year}
  {2018}{\natexlab{b}}),\ 10.1103/physrevd.98.023510}\BibitemShut {NoStop}%
\bibitem [{\citenamefont {{Belgacem}}\ \emph {et~al.}(2019)\citenamefont
  {{Belgacem}}, \citenamefont {{Calcagni}}, \citenamefont {{Crisostomi}} \emph
  {et~al.}}]{Belgacem2019}%
  \BibitemOpen
  \bibfield  {author} {\bibinfo {author} {\bibfnamefont {E.}~\bibnamefont
  {{Belgacem}}}, \bibinfo {author} {\bibfnamefont {G.}~\bibnamefont
  {{Calcagni}}}, \bibinfo {author} {\bibfnamefont {M.}~\bibnamefont
  {{Crisostomi}}},  \emph {et~al.},\ }\href {\doibase
  10.1088/1475-7516/2019/07/024} {\bibfield  {journal} {\bibinfo  {journal}
  {\jcap}\ }\textbf {\bibinfo {volume} {2019}},\ \bibinfo {eid} {024} (\bibinfo
  {year} {2019})},\ \Eprint {http://arxiv.org/abs/1906.01593} {arXiv:1906.01593
  [astro-ph.CO]} \BibitemShut {NoStop}%
\bibitem [{\citenamefont {{Barausse}}\ \emph {et~al.}(2020)\citenamefont
  {{Barausse}}, \citenamefont {{Berti}}, \citenamefont {{Hertog}} \emph
  {et~al.}}]{Barausse2020}%
  \BibitemOpen
  \bibfield  {author} {\bibinfo {author} {\bibfnamefont {E.}~\bibnamefont
  {{Barausse}}}, \bibinfo {author} {\bibfnamefont {E.}~\bibnamefont {{Berti}}},
  \bibinfo {author} {\bibfnamefont {T.}~\bibnamefont {{Hertog}}},  \emph
  {et~al.},\ }\href@noop {} {\bibfield  {journal} {\bibinfo  {journal} {arXiv
  e-prints}\ ,\ \bibinfo {eid} {arXiv:2001.09793}} (\bibinfo {year} {2020})},\
  \Eprint {http://arxiv.org/abs/2001.09793} {arXiv:2001.09793 [gr-qc]}
  \BibitemShut {NoStop}%
\bibitem [{\citenamefont {Crisostomi}\ \emph {et~al.}(2016)\citenamefont
  {Crisostomi}, \citenamefont {Koyama},\ and\ \citenamefont
  {Tasinato}}]{Crisostomi_2016}%
  \BibitemOpen
  \bibfield  {author} {\bibinfo {author} {\bibfnamefont {M.}~\bibnamefont
  {Crisostomi}}, \bibinfo {author} {\bibfnamefont {K.}~\bibnamefont {Koyama}},
  \ and\ \bibinfo {author} {\bibfnamefont {G.}~\bibnamefont {Tasinato}},\
  }\href {\doibase 10.1088/1475-7516/2016/04/044} {\bibfield  {journal}
  {\bibinfo  {journal} {Journal of Cosmology and Astroparticle Physics}\
  }\textbf {\bibinfo {volume} {2016}},\ \bibinfo {pages} {044–044} (\bibinfo
  {year} {2016})}\BibitemShut {NoStop}%
\bibitem [{\citenamefont {Langlois}\ and\ \citenamefont
  {Noui}(2016)}]{Langlois_2016}%
  \BibitemOpen
  \bibfield  {author} {\bibinfo {author} {\bibfnamefont {D.}~\bibnamefont
  {Langlois}}\ and\ \bibinfo {author} {\bibfnamefont {K.}~\bibnamefont
  {Noui}},\ }\href {\doibase 10.1088/1475-7516/2016/02/034} {\bibfield
  {journal} {\bibinfo  {journal} {Journal of Cosmology and Astroparticle
  Physics}\ }\textbf {\bibinfo {volume} {2016}},\ \bibinfo {pages} {034–034}
  (\bibinfo {year} {2016})}\BibitemShut {NoStop}%
\bibitem [{\citenamefont {Ben~Achour}\ \emph {et~al.}(2016)\citenamefont
  {Ben~Achour}, \citenamefont {Langlois},\ and\ \citenamefont
  {Noui}}]{Ben_Achour_2016}%
  \BibitemOpen
  \bibfield  {author} {\bibinfo {author} {\bibfnamefont {J.}~\bibnamefont
  {Ben~Achour}}, \bibinfo {author} {\bibfnamefont {D.}~\bibnamefont
  {Langlois}}, \ and\ \bibinfo {author} {\bibfnamefont {K.}~\bibnamefont
  {Noui}},\ }\href {\doibase 10.1103/physrevd.93.124005} {\bibfield  {journal}
  {\bibinfo  {journal} {Physical Review D}\ }\textbf {\bibinfo {volume} {93}}
  (\bibinfo {year} {2016}),\ 10.1103/physrevd.93.124005}\BibitemShut {NoStop}%
\bibitem [{\citenamefont {Achour}\ \emph {et~al.}(2016)\citenamefont {Achour},
  \citenamefont {Crisostomi}, \citenamefont {Koyama}, \citenamefont {Langlois},
  \citenamefont {Noui},\ and\ \citenamefont {Tasinato}}]{Achour_2016}%
  \BibitemOpen
  \bibfield  {author} {\bibinfo {author} {\bibfnamefont {J.~B.}\ \bibnamefont
  {Achour}}, \bibinfo {author} {\bibfnamefont {M.}~\bibnamefont {Crisostomi}},
  \bibinfo {author} {\bibfnamefont {K.}~\bibnamefont {Koyama}}, \bibinfo
  {author} {\bibfnamefont {D.}~\bibnamefont {Langlois}}, \bibinfo {author}
  {\bibfnamefont {K.}~\bibnamefont {Noui}}, \ and\ \bibinfo {author}
  {\bibfnamefont {G.}~\bibnamefont {Tasinato}},\ }\href {\doibase
  10.1007/jhep12(2016)100} {\bibfield  {journal} {\bibinfo  {journal} {Journal
  of High Energy Physics}\ }\textbf {\bibinfo {volume} {2016}} (\bibinfo {year}
  {2016}),\ 10.1007/jhep12(2016)100}\BibitemShut {NoStop}%
\bibitem [{\citenamefont {Maggiore}\ and\ \citenamefont
  {Mancarella}(2014)}]{Maggiore_2014}%
  \BibitemOpen
  \bibfield  {author} {\bibinfo {author} {\bibfnamefont {M.}~\bibnamefont
  {Maggiore}}\ and\ \bibinfo {author} {\bibfnamefont {M.}~\bibnamefont
  {Mancarella}},\ }\href {\doibase 10.1103/physrevd.90.023005} {\bibfield
  {journal} {\bibinfo  {journal} {Physical Review D}\ }\textbf {\bibinfo
  {volume} {90}} (\bibinfo {year} {2014}),\
  10.1103/physrevd.90.023005}\BibitemShut {NoStop}%
\bibitem [{\citenamefont {{Belgacem}}\ \emph {et~al.}(2018)\citenamefont
  {{Belgacem}}, \citenamefont {{Dirian}}, \citenamefont {{Foffa}},\ and\
  \citenamefont {{Maggiore}}}]{2018JCAP...03..002B}%
  \BibitemOpen
  \bibfield  {author} {\bibinfo {author} {\bibfnamefont {E.}~\bibnamefont
  {{Belgacem}}}, \bibinfo {author} {\bibfnamefont {Y.}~\bibnamefont
  {{Dirian}}}, \bibinfo {author} {\bibfnamefont {S.}~\bibnamefont {{Foffa}}}, \
  and\ \bibinfo {author} {\bibfnamefont {M.}~\bibnamefont {{Maggiore}}},\
  }\href {\doibase 10.1088/1475-7516/2018/03/002} {\bibfield  {journal}
  {\bibinfo  {journal} {\jcap}\ }\textbf {\bibinfo {volume} {2018}},\ \bibinfo
  {eid} {002} (\bibinfo {year} {2018})},\ \Eprint
  {http://arxiv.org/abs/1712.07066} {arXiv:1712.07066 [hep-th]} \BibitemShut
  {NoStop}%
\bibitem [{\citenamefont {Max}\ \emph {et~al.}(2017)\citenamefont {Max},
  \citenamefont {Platscher},\ and\ \citenamefont {Smirnov}}]{Max_2017}%
  \BibitemOpen
  \bibfield  {author} {\bibinfo {author} {\bibfnamefont {K.}~\bibnamefont
  {Max}}, \bibinfo {author} {\bibfnamefont {M.}~\bibnamefont {Platscher}}, \
  and\ \bibinfo {author} {\bibfnamefont {J.}~\bibnamefont {Smirnov}},\ }\href
  {\doibase 10.1103/physrevlett.119.111101} {\bibfield  {journal} {\bibinfo
  {journal} {Physical Review Letters}\ }\textbf {\bibinfo {volume} {119}}
  (\bibinfo {year} {2017}),\ 10.1103/physrevlett.119.111101}\BibitemShut
  {NoStop}%
\bibitem [{\citenamefont {Jiménez}\ \emph {et~al.}(2020)\citenamefont
  {Jiménez}, \citenamefont {Ezquiaga},\ and\ \citenamefont
  {Heisenberg}}]{Jim_nez_2020}%
  \BibitemOpen
  \bibfield  {author} {\bibinfo {author} {\bibfnamefont {J.~B.}\ \bibnamefont
  {Jiménez}}, \bibinfo {author} {\bibfnamefont {J.~M.}\ \bibnamefont
  {Ezquiaga}}, \ and\ \bibinfo {author} {\bibfnamefont {L.}~\bibnamefont
  {Heisenberg}},\ }\href {\doibase 10.1088/1475-7516/2020/04/027} {\bibfield
  {journal} {\bibinfo  {journal} {Journal of Cosmology and Astroparticle
  Physics}\ }\textbf {\bibinfo {volume} {2020}},\ \bibinfo {pages} {027–027}
  (\bibinfo {year} {2020})}\BibitemShut {NoStop}%
\bibitem [{\citenamefont {Palmese}\ and\ \citenamefont
  {Kim}(2020)}]{palmese2020probing}%
  \BibitemOpen
  \bibfield  {author} {\bibinfo {author} {\bibfnamefont {A.}~\bibnamefont
  {Palmese}}\ and\ \bibinfo {author} {\bibfnamefont {A.~G.}\ \bibnamefont
  {Kim}},\ }\href@noop {} {\enquote {\bibinfo {title} {Probing gravity and
  growth of structure with gravitational waves and galaxies' peculiar
  velocity},}\ } (\bibinfo {year} {2020}),\ \Eprint
  {http://arxiv.org/abs/2005.04325} {arXiv:2005.04325 [astro-ph.CO]}
  \BibitemShut {NoStop}%
\bibitem [{\citenamefont {{Horndeski}}(1974)}]{1974IJTP...10..363H}%
  \BibitemOpen
  \bibfield  {author} {\bibinfo {author} {\bibfnamefont {G.~W.}\ \bibnamefont
  {{Horndeski}}},\ }\href {\doibase 10.1007/BF01807638} {\bibfield  {journal}
  {\bibinfo  {journal} {International Journal of Theoretical Physics}\ }\textbf
  {\bibinfo {volume} {10}},\ \bibinfo {pages} {363} (\bibinfo {year}
  {1974})}\BibitemShut {NoStop}%
\bibitem [{\citenamefont {{Nicolis}}\ \emph {et~al.}(2009)\citenamefont
  {{Nicolis}}, \citenamefont {{Rattazzi}},\ and\ \citenamefont
  {{Trincherini}}}]{2009PhRvD..79f4036N}%
  \BibitemOpen
  \bibfield  {author} {\bibinfo {author} {\bibfnamefont {A.}~\bibnamefont
  {{Nicolis}}}, \bibinfo {author} {\bibfnamefont {R.}~\bibnamefont
  {{Rattazzi}}}, \ and\ \bibinfo {author} {\bibfnamefont {E.}~\bibnamefont
  {{Trincherini}}},\ }\href {\doibase 10.1103/PhysRevD.79.064036} {\bibfield
  {journal} {\bibinfo  {journal} {\prd}\ }\textbf {\bibinfo {volume} {79}},\
  \bibinfo {eid} {064036} (\bibinfo {year} {2009})},\ \Eprint
  {http://arxiv.org/abs/0811.2197} {arXiv:0811.2197 [hep-th]} \BibitemShut
  {NoStop}%
\bibitem [{\citenamefont {Deffayet}\ \emph {et~al.}(2011)\citenamefont
  {Deffayet}, \citenamefont {Gao}, \citenamefont {Steer},\ and\ \citenamefont
  {Zahariade}}]{PhysRevD.84.064039}%
  \BibitemOpen
  \bibfield  {author} {\bibinfo {author} {\bibfnamefont {C.}~\bibnamefont
  {Deffayet}}, \bibinfo {author} {\bibfnamefont {X.}~\bibnamefont {Gao}},
  \bibinfo {author} {\bibfnamefont {D.~A.}\ \bibnamefont {Steer}}, \ and\
  \bibinfo {author} {\bibfnamefont {G.}~\bibnamefont {Zahariade}},\ }\href
  {\doibase 10.1103/PhysRevD.84.064039} {\bibfield  {journal} {\bibinfo
  {journal} {Phys. Rev. D}\ }\textbf {\bibinfo {volume} {84}},\ \bibinfo
  {pages} {064039} (\bibinfo {year} {2011})}\BibitemShut {NoStop}%
\bibitem [{\citenamefont {Ostrogradsky}(1850)}]{Ostrogradsky:1850fid}%
  \BibitemOpen
  \bibfield  {author} {\bibinfo {author} {\bibfnamefont {M.}~\bibnamefont
  {Ostrogradsky}},\ }\href@noop {} {\bibfield  {journal} {\bibinfo  {journal}
  {Mem. Acad. St. Petersbourg}\ }\textbf {\bibinfo {volume} {6}},\ \bibinfo
  {pages} {385} (\bibinfo {year} {1850})}\BibitemShut {NoStop}%
\bibitem [{\citenamefont {Woodard}(2007)}]{Woodard2007}%
  \BibitemOpen
  \bibfield  {author} {\bibinfo {author} {\bibfnamefont {R.}~\bibnamefont
  {Woodard}},\ }\enquote {\bibinfo {title} {Avoiding dark energy with 1/r
  modifications of gravity},}\ in\ \href {\doibase
  10.1007/978-3-540-71013-4_14} {\emph {\bibinfo {booktitle} {The Invisible
  Universe: Dark Matter and Dark Energy}}},\ \bibinfo {editor} {edited by\
  \bibinfo {editor} {\bibfnamefont {L.}~\bibnamefont {Papantonopoulos}}}\
  (\bibinfo  {publisher} {Springer Berlin Heidelberg},\ \bibinfo {address}
  {Berlin, Heidelberg},\ \bibinfo {year} {2007})\ pp.\ \bibinfo {pages}
  {403--433}\BibitemShut {NoStop}%
\bibitem [{\citenamefont {Frusciante}\ and\ \citenamefont
  {Perenon}(2020)}]{Frusciante_2020}%
  \BibitemOpen
  \bibfield  {author} {\bibinfo {author} {\bibfnamefont {N.}~\bibnamefont
  {Frusciante}}\ and\ \bibinfo {author} {\bibfnamefont {L.}~\bibnamefont
  {Perenon}},\ }\href {\doibase 10.1016/j.physrep.2020.02.004} {\bibfield
  {journal} {\bibinfo  {journal} {Physics Reports}\ }\textbf {\bibinfo {volume}
  {857}},\ \bibinfo {pages} {1–63} (\bibinfo {year} {2020})}\BibitemShut
  {NoStop}%
\bibitem [{\citenamefont {Bellini}\ \emph {et~al.}(2016)\citenamefont
  {Bellini}, \citenamefont {Cuesta}, \citenamefont {Jimenez},\ and\
  \citenamefont {Verde}}]{Bellini_2016}%
  \BibitemOpen
  \bibfield  {author} {\bibinfo {author} {\bibfnamefont {E.}~\bibnamefont
  {Bellini}}, \bibinfo {author} {\bibfnamefont {A.~J.}\ \bibnamefont {Cuesta}},
  \bibinfo {author} {\bibfnamefont {R.}~\bibnamefont {Jimenez}}, \ and\
  \bibinfo {author} {\bibfnamefont {L.}~\bibnamefont {Verde}},\ }\href
  {\doibase 10.1088/1475-7516/2016/02/053} {\bibfield  {journal} {\bibinfo
  {journal} {Journal of Cosmology and Astroparticle Physics}\ }\textbf
  {\bibinfo {volume} {2016}},\ \bibinfo {pages} {053–053} (\bibinfo {year}
  {2016})}\BibitemShut {NoStop}%
\bibitem [{\citenamefont {Kreisch}\ and\ \citenamefont
  {Komatsu}(2018)}]{Kreisch_2018}%
  \BibitemOpen
  \bibfield  {author} {\bibinfo {author} {\bibfnamefont {C.}~\bibnamefont
  {Kreisch}}\ and\ \bibinfo {author} {\bibfnamefont {E.}~\bibnamefont
  {Komatsu}},\ }\href {\doibase 10.1088/1475-7516/2018/12/030} {\bibfield
  {journal} {\bibinfo  {journal} {Journal of Cosmology and Astroparticle
  Physics}\ }\textbf {\bibinfo {volume} {2018}},\ \bibinfo {pages} {030–030}
  (\bibinfo {year} {2018})}\BibitemShut {NoStop}%
\bibitem [{\citenamefont {{Noller}}\ and\ \citenamefont
  {{Nicola}}(2019)}]{2019PhRvD..99j3502N}%
  \BibitemOpen
  \bibfield  {author} {\bibinfo {author} {\bibfnamefont {J.}~\bibnamefont
  {{Noller}}}\ and\ \bibinfo {author} {\bibfnamefont {A.}~\bibnamefont
  {{Nicola}}},\ }\href {\doibase 10.1103/PhysRevD.99.103502} {\bibfield
  {journal} {\bibinfo  {journal} {\prd}\ }\textbf {\bibinfo {volume} {99}},\
  \bibinfo {eid} {103502} (\bibinfo {year} {2019})},\ \Eprint
  {http://arxiv.org/abs/1811.12928} {arXiv:1811.12928 [astro-ph.CO]}
  \BibitemShut {NoStop}%
\bibitem [{\citenamefont {Spurio Mancini}\ \emph {et~al.}(2019)\citenamefont
  {Spurio Mancini}, \citenamefont {Köhlinger}, \citenamefont {Joachimi},
  \citenamefont {Pettorino}, \citenamefont {Schäfer}, \citenamefont
  {Reischke}, \citenamefont {van Uitert}, \citenamefont {Brieden},
  \citenamefont {Archidiacono},\ and\ \citenamefont
  {Lesgourgues}}]{Spurio_Mancini_2019}%
  \BibitemOpen
  \bibfield  {author} {\bibinfo {author} {\bibfnamefont {A.}~\bibnamefont
  {Spurio Mancini}}, \bibinfo {author} {\bibfnamefont {F.}~\bibnamefont
  {Köhlinger}}, \bibinfo {author} {\bibfnamefont {B.}~\bibnamefont
  {Joachimi}}, \bibinfo {author} {\bibfnamefont {V.}~\bibnamefont {Pettorino}},
  \bibinfo {author} {\bibfnamefont {B.~M.}\ \bibnamefont {Schäfer}}, \bibinfo
  {author} {\bibfnamefont {R.}~\bibnamefont {Reischke}}, \bibinfo {author}
  {\bibfnamefont {E.}~\bibnamefont {van Uitert}}, \bibinfo {author}
  {\bibfnamefont {S.}~\bibnamefont {Brieden}}, \bibinfo {author} {\bibfnamefont
  {M.}~\bibnamefont {Archidiacono}}, \ and\ \bibinfo {author} {\bibfnamefont
  {J.}~\bibnamefont {Lesgourgues}},\ }\href {\doibase 10.1093/mnras/stz2581}
  {\bibfield  {journal} {\bibinfo  {journal} {Monthly Notices of the Royal
  Astronomical Society}\ }\textbf {\bibinfo {volume} {490}},\ \bibinfo {pages}
  {2155–2177} (\bibinfo {year} {2019})}\BibitemShut {NoStop}%
\bibitem [{\citenamefont {Alonso}\ \emph {et~al.}(2017)\citenamefont {Alonso},
  \citenamefont {Bellini}, \citenamefont {Ferreira},\ and\ \citenamefont
  {Zumalacárregui}}]{Alonso_2017}%
  \BibitemOpen
  \bibfield  {author} {\bibinfo {author} {\bibfnamefont {D.}~\bibnamefont
  {Alonso}}, \bibinfo {author} {\bibfnamefont {E.}~\bibnamefont {Bellini}},
  \bibinfo {author} {\bibfnamefont {P.}~\bibnamefont {Ferreira}}, \ and\
  \bibinfo {author} {\bibfnamefont {M.}~\bibnamefont {Zumalacárregui}},\
  }\href {\doibase 10.1103/physrevd.95.063502} {\bibfield  {journal} {\bibinfo
  {journal} {Physical Review D}\ }\textbf {\bibinfo {volume} {95}} (\bibinfo
  {year} {2017}),\ 10.1103/physrevd.95.063502}\BibitemShut {NoStop}%
\bibitem [{\citenamefont {Reischke}\ \emph {et~al.}(2018)\citenamefont
  {Reischke}, \citenamefont {Mancini}, \citenamefont {Schäfer},\ and\
  \citenamefont {Merkel}}]{Reischke_2018}%
  \BibitemOpen
  \bibfield  {author} {\bibinfo {author} {\bibfnamefont {R.}~\bibnamefont
  {Reischke}}, \bibinfo {author} {\bibfnamefont {A.~S.}\ \bibnamefont
  {Mancini}}, \bibinfo {author} {\bibfnamefont {B.~M.}\ \bibnamefont
  {Schäfer}}, \ and\ \bibinfo {author} {\bibfnamefont {P.~M.}\ \bibnamefont
  {Merkel}},\ }\href {\doibase 10.1093/mnras/sty2919} {\bibfield  {journal}
  {\bibinfo  {journal} {Monthly Notices of the Royal Astronomical Society}\ }
  (\bibinfo {year} {2018}),\ 10.1093/mnras/sty2919}\BibitemShut {NoStop}%
\bibitem [{\citenamefont {Spurio Mancini}\ \emph {et~al.}(2018)\citenamefont
  {Spurio Mancini}, \citenamefont {Reischke}, \citenamefont {Pettorino},
  \citenamefont {Schäfer},\ and\ \citenamefont
  {Zumalacárregui}}]{Spurio_Mancini_2018}%
  \BibitemOpen
  \bibfield  {author} {\bibinfo {author} {\bibfnamefont {A.}~\bibnamefont
  {Spurio Mancini}}, \bibinfo {author} {\bibfnamefont {R.}~\bibnamefont
  {Reischke}}, \bibinfo {author} {\bibfnamefont {V.}~\bibnamefont {Pettorino}},
  \bibinfo {author} {\bibfnamefont {B.~M.}\ \bibnamefont {Schäfer}}, \ and\
  \bibinfo {author} {\bibfnamefont {M.}~\bibnamefont {Zumalacárregui}},\
  }\href {\doibase 10.1093/mnras/sty2092} {\bibfield  {journal} {\bibinfo
  {journal} {Monthly Notices of the Royal Astronomical Society}\ }\textbf
  {\bibinfo {volume} {480}},\ \bibinfo {pages} {3725–3738} (\bibinfo {year}
  {2018})}\BibitemShut {NoStop}%
\bibitem [{\citenamefont {Lagos}\ \emph {et~al.}(2018)\citenamefont {Lagos},
  \citenamefont {Bellini}, \citenamefont {Noller}, \citenamefont {Ferreira},\
  and\ \citenamefont {Baker}}]{Lagos_2018}%
  \BibitemOpen
  \bibfield  {author} {\bibinfo {author} {\bibfnamefont {M.}~\bibnamefont
  {Lagos}}, \bibinfo {author} {\bibfnamefont {E.}~\bibnamefont {Bellini}},
  \bibinfo {author} {\bibfnamefont {J.}~\bibnamefont {Noller}}, \bibinfo
  {author} {\bibfnamefont {P.~G.}\ \bibnamefont {Ferreira}}, \ and\ \bibinfo
  {author} {\bibfnamefont {T.}~\bibnamefont {Baker}},\ }\href {\doibase
  10.1088/1475-7516/2018/03/021} {\bibfield  {journal} {\bibinfo  {journal}
  {Journal of Cosmology and Astroparticle Physics}\ }\textbf {\bibinfo {volume}
  {2018}},\ \bibinfo {pages} {021} (\bibinfo {year} {2018})}\BibitemShut
  {NoStop}%
\bibitem [{\citenamefont {{Baker}}\ \emph {et~al.}(2019)\citenamefont
  {{Baker}}, \citenamefont {{Barreira}}, \citenamefont {{Desmond}},
  \citenamefont {{Ferreira}}, \citenamefont {{Jain}}, \citenamefont {{Koyama}},
  \citenamefont {{Li}}, \citenamefont {{Lombriser}}, \citenamefont {{Nicola}},
  \citenamefont {{Sakstein}},\ and\ \citenamefont
  {{Schmidt}}}]{2019arXiv190803430B}%
  \BibitemOpen
  \bibfield  {author} {\bibinfo {author} {\bibfnamefont {T.}~\bibnamefont
  {{Baker}}}, \bibinfo {author} {\bibfnamefont {A.}~\bibnamefont {{Barreira}}},
  \bibinfo {author} {\bibfnamefont {H.}~\bibnamefont {{Desmond}}}, \bibinfo
  {author} {\bibfnamefont {P.}~\bibnamefont {{Ferreira}}}, \bibinfo {author}
  {\bibfnamefont {B.}~\bibnamefont {{Jain}}}, \bibinfo {author} {\bibfnamefont
  {K.}~\bibnamefont {{Koyama}}}, \bibinfo {author} {\bibfnamefont
  {B.}~\bibnamefont {{Li}}}, \bibinfo {author} {\bibfnamefont {L.}~\bibnamefont
  {{Lombriser}}}, \bibinfo {author} {\bibfnamefont {A.}~\bibnamefont
  {{Nicola}}}, \bibinfo {author} {\bibfnamefont {J.}~\bibnamefont
  {{Sakstein}}}, \ and\ \bibinfo {author} {\bibfnamefont {F.}~\bibnamefont
  {{Schmidt}}},\ }\href@noop {} {\bibfield  {journal} {\bibinfo  {journal}
  {arXiv e-prints}\ ,\ \bibinfo {eid} {arXiv:1908.03430}} (\bibinfo {year}
  {2019})},\ \Eprint {http://arxiv.org/abs/1908.03430} {arXiv:1908.03430
  [astro-ph.CO]} \BibitemShut {NoStop}%
\bibitem [{\citenamefont {{Bonilla}}\ \emph {et~al.}(2020)\citenamefont
  {{Bonilla}}, \citenamefont {{D'Agostino}}, \citenamefont {{Nunes}},\ and\
  \citenamefont {{de Araujo}}}]{Bonilla2020}%
  \BibitemOpen
  \bibfield  {author} {\bibinfo {author} {\bibfnamefont {A.}~\bibnamefont
  {{Bonilla}}}, \bibinfo {author} {\bibfnamefont {R.}~\bibnamefont
  {{D'Agostino}}}, \bibinfo {author} {\bibfnamefont {R.~C.}\ \bibnamefont
  {{Nunes}}}, \ and\ \bibinfo {author} {\bibfnamefont {J.~C.~N.}\ \bibnamefont
  {{de Araujo}}},\ }\href {\doibase 10.1088/1475-7516/2020/03/015} {\bibfield
  {journal} {\bibinfo  {journal} {\jcap}\ }\textbf {\bibinfo {volume} {2020}},\
  \bibinfo {eid} {015} (\bibinfo {year} {2020})},\ \Eprint
  {http://arxiv.org/abs/1910.05631} {arXiv:1910.05631 [gr-qc]} \BibitemShut
  {NoStop}%
\bibitem [{\citenamefont {Heisenberg}\ \emph {et~al.}(2020)\citenamefont
  {Heisenberg}, \citenamefont {Noller},\ and\ \citenamefont
  {Zosso}}]{heisenberg2020horndeski}%
  \BibitemOpen
  \bibfield  {author} {\bibinfo {author} {\bibfnamefont {L.}~\bibnamefont
  {Heisenberg}}, \bibinfo {author} {\bibfnamefont {J.}~\bibnamefont {Noller}},
  \ and\ \bibinfo {author} {\bibfnamefont {J.}~\bibnamefont {Zosso}},\
  }\href@noop {} {\enquote {\bibinfo {title} {Horndeski under the quantum
  loupe},}\ } (\bibinfo {year} {2020}),\ \Eprint
  {http://arxiv.org/abs/2004.11655} {arXiv:2004.11655 [hep-th]} \BibitemShut
  {NoStop}%
\bibitem [{\citenamefont {Deffayet}\ \emph {et~al.}(2010)\citenamefont
  {Deffayet}, \citenamefont {Pujolas}, \citenamefont {Sawicki},\ and\
  \citenamefont {Vikman}}]{Deffayet:2010qz}%
  \BibitemOpen
  \bibfield  {author} {\bibinfo {author} {\bibfnamefont {C.}~\bibnamefont
  {Deffayet}}, \bibinfo {author} {\bibfnamefont {O.}~\bibnamefont {Pujolas}},
  \bibinfo {author} {\bibfnamefont {I.}~\bibnamefont {Sawicki}}, \ and\
  \bibinfo {author} {\bibfnamefont {A.}~\bibnamefont {Vikman}},\ }\href
  {\doibase 10.1088/1475-7516/2010/10/026} {\bibfield  {journal} {\bibinfo
  {journal} {JCAP}\ }\textbf {\bibinfo {volume} {10}},\ \bibinfo {pages} {026}
  (\bibinfo {year} {2010})},\ \Eprint {http://arxiv.org/abs/1008.0048}
  {arXiv:1008.0048 [hep-th]} \BibitemShut {NoStop}%
\bibitem [{\citenamefont {Bellini}\ and\ \citenamefont
  {Sawicki}(2014)}]{Bellini:2014fua}%
  \BibitemOpen
  \bibfield  {author} {\bibinfo {author} {\bibfnamefont {E.}~\bibnamefont
  {Bellini}}\ and\ \bibinfo {author} {\bibfnamefont {I.}~\bibnamefont
  {Sawicki}},\ }\href {\doibase 10.1088/1475-7516/2014/07/050} {\bibfield
  {journal} {\bibinfo  {journal} {JCAP}\ }\textbf {\bibinfo {volume} {1407}},\
  \bibinfo {pages} {050} (\bibinfo {year} {2014})},\ \Eprint
  {http://arxiv.org/abs/1404.3713} {arXiv:1404.3713 [astro-ph.CO]} \BibitemShut
  {NoStop}%
\bibitem [{\citenamefont {Creminelli}\ \emph {et~al.}(2018)\citenamefont
  {Creminelli}, \citenamefont {Lewandowski}, \citenamefont {Tambalo},\ and\
  \citenamefont {Vernizzi}}]{Creminelli_2018}%
  \BibitemOpen
  \bibfield  {author} {\bibinfo {author} {\bibfnamefont {P.}~\bibnamefont
  {Creminelli}}, \bibinfo {author} {\bibfnamefont {M.}~\bibnamefont
  {Lewandowski}}, \bibinfo {author} {\bibfnamefont {G.}~\bibnamefont
  {Tambalo}}, \ and\ \bibinfo {author} {\bibfnamefont {F.}~\bibnamefont
  {Vernizzi}},\ }\href {\doibase 10.1088/1475-7516/2018/12/025} {\bibfield
  {journal} {\bibinfo  {journal} {Journal of Cosmology and Astroparticle
  Physics}\ }\textbf {\bibinfo {volume} {2018}},\ \bibinfo {pages} {025–025}
  (\bibinfo {year} {2018})}\BibitemShut {NoStop}%
\bibitem [{\citenamefont {Creminelli}\ \emph {et~al.}(2019)\citenamefont
  {Creminelli}, \citenamefont {Tambalo}, \citenamefont {Vernizzi},\ and\
  \citenamefont {Yingcharoenrat}}]{Creminelli_2019}%
  \BibitemOpen
  \bibfield  {author} {\bibinfo {author} {\bibfnamefont {P.}~\bibnamefont
  {Creminelli}}, \bibinfo {author} {\bibfnamefont {G.}~\bibnamefont {Tambalo}},
  \bibinfo {author} {\bibfnamefont {F.}~\bibnamefont {Vernizzi}}, \ and\
  \bibinfo {author} {\bibfnamefont {V.}~\bibnamefont {Yingcharoenrat}},\ }\href
  {\doibase 10.1088/1475-7516/2019/10/072} {\bibfield  {journal} {\bibinfo
  {journal} {Journal of Cosmology and Astroparticle Physics}\ }\textbf
  {\bibinfo {volume} {2019}},\ \bibinfo {pages} {072–072} (\bibinfo {year}
  {2019})}\BibitemShut {NoStop}%
\bibitem [{\citenamefont {{Noller}}(2020)}]{Noller2020}%
  \BibitemOpen
  \bibfield  {author} {\bibinfo {author} {\bibfnamefont {J.}~\bibnamefont
  {{Noller}}},\ }\href {\doibase 10.1103/PhysRevD.101.063524} {\bibfield
  {journal} {\bibinfo  {journal} {\prd}\ }\textbf {\bibinfo {volume} {101}},\
  \bibinfo {eid} {063524} (\bibinfo {year} {2020})},\ \Eprint
  {http://arxiv.org/abs/2001.05469} {arXiv:2001.05469 [astro-ph.CO]}
  \BibitemShut {NoStop}%
\bibitem [{\citenamefont {de~Rham}\ and\ \citenamefont
  {Melville}(2018)}]{PhysRevLett.121.221101}%
  \BibitemOpen
  \bibfield  {author} {\bibinfo {author} {\bibfnamefont {C.}~\bibnamefont
  {de~Rham}}\ and\ \bibinfo {author} {\bibfnamefont {S.}~\bibnamefont
  {Melville}},\ }\href {\doibase 10.1103/PhysRevLett.121.221101} {\bibfield
  {journal} {\bibinfo  {journal} {Phys. Rev. Lett.}\ }\textbf {\bibinfo
  {volume} {121}},\ \bibinfo {pages} {221101} (\bibinfo {year}
  {2018})}\BibitemShut {NoStop}%
\bibitem [{\citenamefont {{Nishizawa}}(2018)}]{2018PhRvD..97j4037N}%
  \BibitemOpen
  \bibfield  {author} {\bibinfo {author} {\bibfnamefont {A.}~\bibnamefont
  {{Nishizawa}}},\ }\href {\doibase 10.1103/PhysRevD.97.104037} {\bibfield
  {journal} {\bibinfo  {journal} {\prd}\ }\textbf {\bibinfo {volume} {97}},\
  \bibinfo {eid} {104037} (\bibinfo {year} {2018})},\ \Eprint
  {http://arxiv.org/abs/1710.04825} {arXiv:1710.04825 [gr-qc]} \BibitemShut
  {NoStop}%
\bibitem [{\citenamefont {{Amendola}}\ \emph {et~al.}(2018)\citenamefont
  {{Amendola}}, \citenamefont {{Sawicki}}, \citenamefont {{Kunz}},\ and\
  \citenamefont {{Saltas}}}]{2018JCAP...08..030A}%
  \BibitemOpen
  \bibfield  {author} {\bibinfo {author} {\bibfnamefont {L.}~\bibnamefont
  {{Amendola}}}, \bibinfo {author} {\bibfnamefont {I.}~\bibnamefont
  {{Sawicki}}}, \bibinfo {author} {\bibfnamefont {M.}~\bibnamefont {{Kunz}}}, \
  and\ \bibinfo {author} {\bibfnamefont {I.~D.}\ \bibnamefont {{Saltas}}},\
  }\href {\doibase 10.1088/1475-7516/2018/08/030} {\bibfield  {journal}
  {\bibinfo  {journal} {\jcap}\ }\textbf {\bibinfo {volume} {2018}},\ \bibinfo
  {eid} {030} (\bibinfo {year} {2018})},\ \Eprint
  {http://arxiv.org/abs/1712.08623} {arXiv:1712.08623 [astro-ph.CO]}
  \BibitemShut {NoStop}%
\bibitem [{\citenamefont {Joyce}\ \emph {et~al.}(2015)\citenamefont {Joyce},
  \citenamefont {Jain}, \citenamefont {Khoury},\ and\ \citenamefont
  {Trodden}}]{Joyce_2015}%
  \BibitemOpen
  \bibfield  {author} {\bibinfo {author} {\bibfnamefont {A.}~\bibnamefont
  {Joyce}}, \bibinfo {author} {\bibfnamefont {B.}~\bibnamefont {Jain}},
  \bibinfo {author} {\bibfnamefont {J.}~\bibnamefont {Khoury}}, \ and\ \bibinfo
  {author} {\bibfnamefont {M.}~\bibnamefont {Trodden}},\ }\href {\doibase
  10.1016/j.physrep.2014.12.002} {\bibfield  {journal} {\bibinfo  {journal}
  {Physics Reports}\ }\textbf {\bibinfo {volume} {568}},\ \bibinfo {pages}
  {1–98} (\bibinfo {year} {2015})}\BibitemShut {NoStop}%
\bibitem [{\citenamefont {Will}(2014)}]{Will_2014}%
  \BibitemOpen
  \bibfield  {author} {\bibinfo {author} {\bibfnamefont {C.~M.}\ \bibnamefont
  {Will}},\ }\href {\doibase 10.12942/lrr-2014-4} {\bibfield  {journal}
  {\bibinfo  {journal} {Living Reviews in Relativity}\ }\textbf {\bibinfo
  {volume} {17}} (\bibinfo {year} {2014}),\ 10.12942/lrr-2014-4}\BibitemShut
  {NoStop}%
\bibitem [{\citenamefont {{Nissanke}}\ \emph {et~al.}(2010)\citenamefont
  {{Nissanke}}, \citenamefont {{Holz}}, \citenamefont {{Hughes}}, \citenamefont
  {{Dalal}},\ and\ \citenamefont {{Sievers}}}]{Nissanke_2009}%
  \BibitemOpen
  \bibfield  {author} {\bibinfo {author} {\bibfnamefont {S.}~\bibnamefont
  {{Nissanke}}}, \bibinfo {author} {\bibfnamefont {D.~E.}\ \bibnamefont
  {{Holz}}}, \bibinfo {author} {\bibfnamefont {S.~A.}\ \bibnamefont
  {{Hughes}}}, \bibinfo {author} {\bibfnamefont {N.}~\bibnamefont {{Dalal}}}, \
  and\ \bibinfo {author} {\bibfnamefont {J.~L.}\ \bibnamefont {{Sievers}}},\
  }\href {\doibase 10.1088/0004-637X/725/1/496} {\bibfield  {journal} {\bibinfo
   {journal} {\apj}\ }\textbf {\bibinfo {volume} {725}},\ \bibinfo {pages}
  {496} (\bibinfo {year} {2010})},\ \Eprint {http://arxiv.org/abs/0904.1017}
  {arXiv:0904.1017 [astro-ph.CO]} \BibitemShut {NoStop}%
\bibitem [{\citenamefont {{Dalang}}\ \emph {et~al.}(2019)\citenamefont
  {{Dalang}}, \citenamefont {{Fleury}},\ and\ \citenamefont
  {{Lombriser}}}]{Dalang_paper2}%
  \BibitemOpen
  \bibfield  {author} {\bibinfo {author} {\bibfnamefont {C.}~\bibnamefont
  {{Dalang}}}, \bibinfo {author} {\bibfnamefont {P.}~\bibnamefont {{Fleury}}},
  \ and\ \bibinfo {author} {\bibfnamefont {L.}~\bibnamefont {{Lombriser}}},\
  }\href@noop {} {\bibfield  {journal} {\bibinfo  {journal} {arXiv e-prints}\
  ,\ \bibinfo {eid} {arXiv:1912.06117}} (\bibinfo {year} {2019})},\ \Eprint
  {http://arxiv.org/abs/1912.06117} {arXiv:1912.06117 [gr-qc]} \BibitemShut
  {NoStop}%
\bibitem [{\citenamefont {Mastrogiovanni}\ \emph {et~al.}(2020)\citenamefont
  {Mastrogiovanni}, \citenamefont {Steer},\ and\ \citenamefont
  {Barsuglia}}]{mastrogiovanni2020probing}%
  \BibitemOpen
  \bibfield  {author} {\bibinfo {author} {\bibfnamefont {S.}~\bibnamefont
  {Mastrogiovanni}}, \bibinfo {author} {\bibfnamefont {D.}~\bibnamefont
  {Steer}}, \ and\ \bibinfo {author} {\bibfnamefont {M.}~\bibnamefont
  {Barsuglia}},\ }\href@noop {} {\enquote {\bibinfo {title} {Probing modified
  gravity theories and cosmology using gravitational-waves and associated
  electromagnetic counterparts},}\ } (\bibinfo {year} {2020}),\ \Eprint
  {http://arxiv.org/abs/2004.01632} {arXiv:2004.01632 [gr-qc]} \BibitemShut
  {NoStop}%
\bibitem [{\citenamefont {{D'Agostino}}\ and\ \citenamefont
  {{Nunes}}(2019)}]{dagostino2019}%
  \BibitemOpen
  \bibfield  {author} {\bibinfo {author} {\bibfnamefont {R.}~\bibnamefont
  {{D'Agostino}}}\ and\ \bibinfo {author} {\bibfnamefont {R.~C.}\ \bibnamefont
  {{Nunes}}},\ }\href {\doibase 10.1103/PhysRevD.100.044041} {\bibfield
  {journal} {\bibinfo  {journal} {\prd}\ }\textbf {\bibinfo {volume} {100}},\
  \bibinfo {eid} {044041} (\bibinfo {year} {2019})},\ \Eprint
  {http://arxiv.org/abs/1907.05516} {arXiv:1907.05516 [gr-qc]} \BibitemShut
  {NoStop}%
\bibitem [{\citenamefont {Dalang}\ and\ \citenamefont
  {Lombriser}(2019)}]{Dalang_2019}%
  \BibitemOpen
  \bibfield  {author} {\bibinfo {author} {\bibfnamefont {C.}~\bibnamefont
  {Dalang}}\ and\ \bibinfo {author} {\bibfnamefont {L.}~\bibnamefont
  {Lombriser}},\ }\href {\doibase 10.1088/1475-7516/2019/10/013} {\bibfield
  {journal} {\bibinfo  {journal} {Journal of Cosmology and Astroparticle
  Physics}\ }\textbf {\bibinfo {volume} {2019}},\ \bibinfo {pages} {013}
  (\bibinfo {year} {2019})}\BibitemShut {NoStop}%
\bibitem [{\citenamefont {Blas}\ \emph {et~al.}(2011)\citenamefont {Blas},
  \citenamefont {Lesgourgues},\ and\ \citenamefont {Tram}}]{Blas_2011}%
  \BibitemOpen
  \bibfield  {author} {\bibinfo {author} {\bibfnamefont {D.}~\bibnamefont
  {Blas}}, \bibinfo {author} {\bibfnamefont {J.}~\bibnamefont {Lesgourgues}}, \
  and\ \bibinfo {author} {\bibfnamefont {T.}~\bibnamefont {Tram}},\ }\href
  {\doibase 10.1088/1475-7516/2011/07/034} {\bibfield  {journal} {\bibinfo
  {journal} {Journal of Cosmology and Astroparticle Physics}\ }\textbf
  {\bibinfo {volume} {2011}},\ \bibinfo {pages} {034–034} (\bibinfo {year}
  {2011})}\BibitemShut {NoStop}%
\bibitem [{\citenamefont {{Zumalac{\'a}rregui}}\ \emph
  {et~al.}(2017)\citenamefont {{Zumalac{\'a}rregui}}, \citenamefont
  {{Bellini}}, \citenamefont {{Sawicki}}, \citenamefont {{Lesgourgues}},\ and\
  \citenamefont {{Ferreira}}}]{2017JCAP...08..019Z}%
  \BibitemOpen
  \bibfield  {author} {\bibinfo {author} {\bibfnamefont {M.}~\bibnamefont
  {{Zumalac{\'a}rregui}}}, \bibinfo {author} {\bibfnamefont {E.}~\bibnamefont
  {{Bellini}}}, \bibinfo {author} {\bibfnamefont {I.}~\bibnamefont
  {{Sawicki}}}, \bibinfo {author} {\bibfnamefont {J.}~\bibnamefont
  {{Lesgourgues}}}, \ and\ \bibinfo {author} {\bibfnamefont {P.~G.}\
  \bibnamefont {{Ferreira}}},\ }\href {\doibase 10.1088/1475-7516/2017/08/019}
  {\bibfield  {journal} {\bibinfo  {journal} {\jcap}\ }\textbf {\bibinfo
  {volume} {2017}},\ \bibinfo {eid} {019} (\bibinfo {year} {2017})},\ \Eprint
  {http://arxiv.org/abs/1605.06102} {arXiv:1605.06102 [astro-ph.CO]}
  \BibitemShut {NoStop}%
\bibitem [{\citenamefont {Bellini}\ \emph {et~al.}(2020)\citenamefont
  {Bellini}, \citenamefont {Sawicki},\ and\ \citenamefont
  {Zumalacárregui}}]{Bellini_2020}%
  \BibitemOpen
  \bibfield  {author} {\bibinfo {author} {\bibfnamefont {E.}~\bibnamefont
  {Bellini}}, \bibinfo {author} {\bibfnamefont {I.}~\bibnamefont {Sawicki}}, \
  and\ \bibinfo {author} {\bibfnamefont {M.}~\bibnamefont {Zumalacárregui}},\
  }\href {\doibase 10.1088/1475-7516/2020/02/008} {\bibfield  {journal}
  {\bibinfo  {journal} {Journal of Cosmology and Astroparticle Physics}\
  }\textbf {\bibinfo {volume} {2020}},\ \bibinfo {pages} {008–008} (\bibinfo
  {year} {2020})}\BibitemShut {NoStop}%
\bibitem [{\citenamefont {Hu}\ \emph {et~al.}(2014)\citenamefont {Hu},
  \citenamefont {Raveri}, \citenamefont {Frusciante},\ and\ \citenamefont
  {Silvestri}}]{Hu_2014}%
  \BibitemOpen
  \bibfield  {author} {\bibinfo {author} {\bibfnamefont {B.}~\bibnamefont
  {Hu}}, \bibinfo {author} {\bibfnamefont {M.}~\bibnamefont {Raveri}}, \bibinfo
  {author} {\bibfnamefont {N.}~\bibnamefont {Frusciante}}, \ and\ \bibinfo
  {author} {\bibfnamefont {A.}~\bibnamefont {Silvestri}},\ }\href {\doibase
  10.1103/physrevd.89.103530} {\bibfield  {journal} {\bibinfo  {journal}
  {Physical Review D}\ }\textbf {\bibinfo {volume} {89}} (\bibinfo {year}
  {2014}),\ 10.1103/physrevd.89.103530}\BibitemShut {NoStop}%
\bibitem [{\citenamefont {Raveri}\ \emph {et~al.}(2014)\citenamefont {Raveri},
  \citenamefont {Hu}, \citenamefont {Frusciante},\ and\ \citenamefont
  {Silvestri}}]{Raveri_2014}%
  \BibitemOpen
  \bibfield  {author} {\bibinfo {author} {\bibfnamefont {M.}~\bibnamefont
  {Raveri}}, \bibinfo {author} {\bibfnamefont {B.}~\bibnamefont {Hu}}, \bibinfo
  {author} {\bibfnamefont {N.}~\bibnamefont {Frusciante}}, \ and\ \bibinfo
  {author} {\bibfnamefont {A.}~\bibnamefont {Silvestri}},\ }\href {\doibase
  10.1103/physrevd.90.043513} {\bibfield  {journal} {\bibinfo  {journal}
  {Physical Review D}\ }\textbf {\bibinfo {volume} {90}} (\bibinfo {year}
  {2014}),\ 10.1103/physrevd.90.043513}\BibitemShut {NoStop}%
\bibitem [{\citenamefont {Bellini}\ \emph {et~al.}(2018)\citenamefont
  {Bellini}, \citenamefont {Barreira}, \citenamefont {Frusciante},
  \citenamefont {Hu}, \citenamefont {Peirone}, \citenamefont {Raveri},
  \citenamefont {Zumalacárregui}, \citenamefont {Avilez-Lopez}, \citenamefont
  {Ballardini}, \citenamefont {Battye},\ and\ \citenamefont
  {et~al.}}]{Bellini_2018}%
  \BibitemOpen
  \bibfield  {author} {\bibinfo {author} {\bibfnamefont {E.}~\bibnamefont
  {Bellini}}, \bibinfo {author} {\bibfnamefont {A.}~\bibnamefont {Barreira}},
  \bibinfo {author} {\bibfnamefont {N.}~\bibnamefont {Frusciante}}, \bibinfo
  {author} {\bibfnamefont {B.}~\bibnamefont {Hu}}, \bibinfo {author}
  {\bibfnamefont {S.}~\bibnamefont {Peirone}}, \bibinfo {author} {\bibfnamefont
  {M.}~\bibnamefont {Raveri}}, \bibinfo {author} {\bibfnamefont
  {M.}~\bibnamefont {Zumalacárregui}}, \bibinfo {author} {\bibfnamefont
  {A.}~\bibnamefont {Avilez-Lopez}}, \bibinfo {author} {\bibfnamefont
  {M.}~\bibnamefont {Ballardini}}, \bibinfo {author} {\bibfnamefont
  {R.}~\bibnamefont {Battye}}, \ and\ \bibinfo {author} {\bibnamefont
  {et~al.}},\ }\href {\doibase 10.1103/physrevd.97.023520} {\bibfield
  {journal} {\bibinfo  {journal} {Physical Review D}\ }\textbf {\bibinfo
  {volume} {97}} (\bibinfo {year} {2018}),\
  10.1103/physrevd.97.023520}\BibitemShut {NoStop}%
\bibitem [{\citenamefont {Sawicki}\ and\ \citenamefont
  {Bellini}(2015)}]{Sawicki_2015}%
  \BibitemOpen
  \bibfield  {author} {\bibinfo {author} {\bibfnamefont {I.}~\bibnamefont
  {Sawicki}}\ and\ \bibinfo {author} {\bibfnamefont {E.}~\bibnamefont
  {Bellini}},\ }\href {\doibase 10.1103/physrevd.92.084061} {\bibfield
  {journal} {\bibinfo  {journal} {Physical Review D}\ }\textbf {\bibinfo
  {volume} {92}} (\bibinfo {year} {2015}),\
  10.1103/physrevd.92.084061}\BibitemShut {NoStop}%
\bibitem [{\citenamefont {Noller}\ \emph {et~al.}(2014)\citenamefont {Noller},
  \citenamefont {von Braun-Bates},\ and\ \citenamefont
  {Ferreira}}]{PhysRevD.89.023521}%
  \BibitemOpen
  \bibfield  {author} {\bibinfo {author} {\bibfnamefont {J.}~\bibnamefont
  {Noller}}, \bibinfo {author} {\bibfnamefont {F.}~\bibnamefont {von
  Braun-Bates}}, \ and\ \bibinfo {author} {\bibfnamefont {P.~G.}\ \bibnamefont
  {Ferreira}},\ }\href {\doibase 10.1103/PhysRevD.89.023521} {\bibfield
  {journal} {\bibinfo  {journal} {Phys. Rev. D}\ }\textbf {\bibinfo {volume}
  {89}},\ \bibinfo {pages} {023521} (\bibinfo {year} {2014})}\BibitemShut
  {NoStop}%
\bibitem [{\citenamefont {Silvestri}\ \emph {et~al.}(2013)\citenamefont
  {Silvestri}, \citenamefont {Pogosian},\ and\ \citenamefont
  {Buniy}}]{PhysRevD.87.104015}%
  \BibitemOpen
  \bibfield  {author} {\bibinfo {author} {\bibfnamefont {A.}~\bibnamefont
  {Silvestri}}, \bibinfo {author} {\bibfnamefont {L.}~\bibnamefont {Pogosian}},
  \ and\ \bibinfo {author} {\bibfnamefont {R.~V.}\ \bibnamefont {Buniy}},\
  }\href {\doibase 10.1103/PhysRevD.87.104015} {\bibfield  {journal} {\bibinfo
  {journal} {Phys. Rev. D}\ }\textbf {\bibinfo {volume} {87}},\ \bibinfo
  {pages} {104015} (\bibinfo {year} {2013})}\BibitemShut {NoStop}%
\bibitem [{\citenamefont {Perenon}\ \emph {et~al.}(2020)\citenamefont
  {Perenon}, \citenamefont {Ilić}, \citenamefont {Maartens},\ and\
  \citenamefont {de~la Cruz-Dombriz}}]{perenon2020improvements}%
  \BibitemOpen
  \bibfield  {author} {\bibinfo {author} {\bibfnamefont {L.}~\bibnamefont
  {Perenon}}, \bibinfo {author} {\bibfnamefont {S.}~\bibnamefont {Ilić}},
  \bibinfo {author} {\bibfnamefont {R.}~\bibnamefont {Maartens}}, \ and\
  \bibinfo {author} {\bibfnamefont {A.}~\bibnamefont {de~la Cruz-Dombriz}},\
  }\href@noop {} {\enquote {\bibinfo {title} {Improvements in cosmological
  constraints from breaking growth degeneracy},}\ } (\bibinfo {year} {2020}),\
  \Eprint {http://arxiv.org/abs/2005.00418} {arXiv:2005.00418 [astro-ph.CO]}
  \BibitemShut {NoStop}%
\bibitem [{\citenamefont {Linder}(2020)}]{linder2020limited}%
  \BibitemOpen
  \bibfield  {author} {\bibinfo {author} {\bibfnamefont {E.~V.}\ \bibnamefont
  {Linder}},\ }\href@noop {} {\enquote {\bibinfo {title} {Limited modified
  gravity},}\ } (\bibinfo {year} {2020}),\ \Eprint
  {http://arxiv.org/abs/2003.10453} {arXiv:2003.10453 [astro-ph.CO]}
  \BibitemShut {NoStop}%
\bibitem [{\citenamefont {Linder}\ \emph {et~al.}(2016)\citenamefont {Linder},
  \citenamefont {Sengör},\ and\ \citenamefont {Watson}}]{Linder_2016}%
  \BibitemOpen
  \bibfield  {author} {\bibinfo {author} {\bibfnamefont {E.~V.}\ \bibnamefont
  {Linder}}, \bibinfo {author} {\bibfnamefont {G.}~\bibnamefont {Sengör}}, \
  and\ \bibinfo {author} {\bibfnamefont {S.}~\bibnamefont {Watson}},\ }\href
  {\doibase 10.1088/1475-7516/2016/05/053} {\bibfield  {journal} {\bibinfo
  {journal} {Journal of Cosmology and Astroparticle Physics}\ }\textbf
  {\bibinfo {volume} {2016}},\ \bibinfo {pages} {053–053} (\bibinfo {year}
  {2016})}\BibitemShut {NoStop}%
\bibitem [{\citenamefont {Linder}(2017)}]{Linder_2017}%
  \BibitemOpen
  \bibfield  {author} {\bibinfo {author} {\bibfnamefont {E.~V.}\ \bibnamefont
  {Linder}},\ }\href {\doibase 10.1103/physrevd.95.023518} {\bibfield
  {journal} {\bibinfo  {journal} {Physical Review D}\ }\textbf {\bibinfo
  {volume} {95}} (\bibinfo {year} {2017}),\
  10.1103/physrevd.95.023518}\BibitemShut {NoStop}%
\bibitem [{\citenamefont {{Gleyzes}}(2017)}]{2017PhRvD..96f3516G}%
  \BibitemOpen
  \bibfield  {author} {\bibinfo {author} {\bibfnamefont {J.}~\bibnamefont
  {{Gleyzes}}},\ }\href {\doibase 10.1103/PhysRevD.96.063516} {\bibfield
  {journal} {\bibinfo  {journal} {\prd}\ }\textbf {\bibinfo {volume} {96}},\
  \bibinfo {eid} {063516} (\bibinfo {year} {2017})},\ \Eprint
  {http://arxiv.org/abs/1705.04714} {arXiv:1705.04714 [astro-ph.CO]}
  \BibitemShut {NoStop}%
\bibitem [{\citenamefont {{Burrage}}\ and\ \citenamefont
  {{Dombrowski}}(2020)}]{Burrage2020}%
  \BibitemOpen
  \bibfield  {author} {\bibinfo {author} {\bibfnamefont {C.}~\bibnamefont
  {{Burrage}}}\ and\ \bibinfo {author} {\bibfnamefont {J.}~\bibnamefont
  {{Dombrowski}}},\ }\href@noop {} {\bibfield  {journal} {\bibinfo  {journal}
  {arXiv e-prints}\ ,\ \bibinfo {eid} {arXiv:2004.14260}} (\bibinfo {year}
  {2020})},\ \Eprint {http://arxiv.org/abs/2004.14260} {arXiv:2004.14260
  [astro-ph.CO]} \BibitemShut {NoStop}%
\bibitem [{\citenamefont {{Wolf}}\ and\ \citenamefont
  {{Lagos}}(2019)}]{2019arXiv191010580W}%
  \BibitemOpen
  \bibfield  {author} {\bibinfo {author} {\bibfnamefont {W.~J.}\ \bibnamefont
  {{Wolf}}}\ and\ \bibinfo {author} {\bibfnamefont {M.}~\bibnamefont
  {{Lagos}}},\ }\href@noop {} {\bibfield  {journal} {\bibinfo  {journal} {arXiv
  e-prints}\ ,\ \bibinfo {eid} {arXiv:1910.10580}} (\bibinfo {year} {2019})},\
  \Eprint {http://arxiv.org/abs/1910.10580} {arXiv:1910.10580 [gr-qc]}
  \BibitemShut {NoStop}%
\bibitem [{\citenamefont {Wang}\ \emph {et~al.}(2012)\citenamefont {Wang},
  \citenamefont {Hui},\ and\ \citenamefont {Khoury}}]{PhysRevLett.109.241301}%
  \BibitemOpen
  \bibfield  {author} {\bibinfo {author} {\bibfnamefont {J.}~\bibnamefont
  {Wang}}, \bibinfo {author} {\bibfnamefont {L.}~\bibnamefont {Hui}}, \ and\
  \bibinfo {author} {\bibfnamefont {J.}~\bibnamefont {Khoury}},\ }\href
  {\doibase 10.1103/PhysRevLett.109.241301} {\bibfield  {journal} {\bibinfo
  {journal} {Phys. Rev. Lett.}\ }\textbf {\bibinfo {volume} {109}},\ \bibinfo
  {pages} {241301} (\bibinfo {year} {2012})}\BibitemShut {NoStop}%
\bibitem [{\citenamefont {Koyama}(2020)}]{koyama2020testing}%
  \BibitemOpen
  \bibfield  {author} {\bibinfo {author} {\bibfnamefont {K.}~\bibnamefont
  {Koyama}},\ }\href@noop {} {\enquote {\bibinfo {title} {Testing brans-dicke
  gravity with screening by scalar gravitational wave memory},}\ } (\bibinfo
  {year} {2020}),\ \Eprint {http://arxiv.org/abs/2006.15914} {arXiv:2006.15914
  [gr-qc]} \BibitemShut {NoStop}%
\bibitem [{\citenamefont {{Tamanini}}\ \emph {et~al.}(2016)\citenamefont
  {{Tamanini}}, \citenamefont {{Caprini}}, \citenamefont {{Barausse}},
  \citenamefont {{Sesana}}, \citenamefont {{Klein}},\ and\ \citenamefont
  {{Petiteau}}}]{Tamanini2016}%
  \BibitemOpen
  \bibfield  {author} {\bibinfo {author} {\bibfnamefont {N.}~\bibnamefont
  {{Tamanini}}}, \bibinfo {author} {\bibfnamefont {C.}~\bibnamefont
  {{Caprini}}}, \bibinfo {author} {\bibfnamefont {E.}~\bibnamefont
  {{Barausse}}}, \bibinfo {author} {\bibfnamefont {A.}~\bibnamefont
  {{Sesana}}}, \bibinfo {author} {\bibfnamefont {A.}~\bibnamefont {{Klein}}}, \
  and\ \bibinfo {author} {\bibfnamefont {A.}~\bibnamefont {{Petiteau}}},\
  }\href {\doibase 10.1088/1475-7516/2016/04/002} {\bibfield  {journal}
  {\bibinfo  {journal} {\jcap}\ }\textbf {\bibinfo {volume} {2016}},\ \bibinfo
  {eid} {002} (\bibinfo {year} {2016})},\ \Eprint
  {http://arxiv.org/abs/1601.07112} {arXiv:1601.07112 [astro-ph.CO]}
  \BibitemShut {NoStop}%
\bibitem [{\citenamefont {{Wang}}\ \emph {et~al.}(2019)\citenamefont {{Wang}},
  \citenamefont {{Zhao}}, \citenamefont {{Zhang}},\ and\ \citenamefont
  {{Zhang}}}]{wang2019}%
  \BibitemOpen
  \bibfield  {author} {\bibinfo {author} {\bibfnamefont {L.-F.}\ \bibnamefont
  {{Wang}}}, \bibinfo {author} {\bibfnamefont {Z.-W.}\ \bibnamefont {{Zhao}}},
  \bibinfo {author} {\bibfnamefont {J.-F.}\ \bibnamefont {{Zhang}}}, \ and\
  \bibinfo {author} {\bibfnamefont {X.}~\bibnamefont {{Zhang}}},\ }\href@noop
  {} {\bibfield  {journal} {\bibinfo  {journal} {arXiv e-prints}\ ,\ \bibinfo
  {eid} {arXiv:1907.01838}} (\bibinfo {year} {2019})},\ \Eprint
  {http://arxiv.org/abs/1907.01838} {arXiv:1907.01838 [astro-ph.CO]}
  \BibitemShut {NoStop}%
\bibitem [{\citenamefont {Begelman}\ \emph {et~al.}(1980)\citenamefont
  {Begelman}, \citenamefont {Blandford},\ and\ \citenamefont
  {Rees}}]{Begelman:1980vb}%
  \BibitemOpen
  \bibfield  {author} {\bibinfo {author} {\bibfnamefont {M.}~\bibnamefont
  {Begelman}}, \bibinfo {author} {\bibfnamefont {R.}~\bibnamefont {Blandford}},
  \ and\ \bibinfo {author} {\bibfnamefont {M.}~\bibnamefont {Rees}},\ }\href
  {\doibase 10.1038/287307a0} {\bibfield  {journal} {\bibinfo  {journal}
  {Nature}\ }\textbf {\bibinfo {volume} {287}},\ \bibinfo {pages} {307}
  (\bibinfo {year} {1980})}\BibitemShut {NoStop}%
\bibitem [{\citenamefont {{Milosavljevi{\'c}}}\ and\ \citenamefont
  {{Merritt}}(2003)}]{2003AIPC..686..201M}%
  \BibitemOpen
  \bibfield  {author} {\bibinfo {author} {\bibfnamefont {M.}~\bibnamefont
  {{Milosavljevi{\'c}}}}\ and\ \bibinfo {author} {\bibfnamefont
  {D.}~\bibnamefont {{Merritt}}},\ }in\ \href {\doibase 10.1063/1.1629432}
  {\emph {\bibinfo {booktitle} {The Astrophysics of Gravitational Wave
  Sources}}},\ \bibinfo {series} {American Institute of Physics Conference
  Series}, Vol.\ \bibinfo {volume} {686},\ \bibinfo {editor} {edited by\
  \bibinfo {editor} {\bibfnamefont {J.~M.}\ \bibnamefont {{Centrella}}}}\
  (\bibinfo {year} {2003})\ pp.\ \bibinfo {pages} {201--210},\ \Eprint
  {http://arxiv.org/abs/astro-ph/0212270} {arXiv:astro-ph/0212270 [astro-ph]}
  \BibitemShut {NoStop}%
\bibitem [{\citenamefont {Vasiliev}\ \emph {et~al.}(2015)\citenamefont
  {Vasiliev}, \citenamefont {Antonini},\ and\ \citenamefont
  {Merritt}}]{Vasiliev_2015}%
  \BibitemOpen
  \bibfield  {author} {\bibinfo {author} {\bibfnamefont {E.}~\bibnamefont
  {Vasiliev}}, \bibinfo {author} {\bibfnamefont {F.}~\bibnamefont {Antonini}},
  \ and\ \bibinfo {author} {\bibfnamefont {D.}~\bibnamefont {Merritt}},\ }\href
  {\doibase 10.1088/0004-637x/810/1/49} {\bibfield  {journal} {\bibinfo
  {journal} {The Astrophysical Journal}\ }\textbf {\bibinfo {volume} {810}},\
  \bibinfo {pages} {49} (\bibinfo {year} {2015})}\BibitemShut {NoStop}%
\bibitem [{\citenamefont {Barausse}(2012)}]{Barausse_2012}%
  \BibitemOpen
  \bibfield  {author} {\bibinfo {author} {\bibfnamefont {E.}~\bibnamefont
  {Barausse}},\ }\href {\doibase 10.1111/j.1365-2966.2012.21057.x} {\bibfield
  {journal} {\bibinfo  {journal} {Monthly Notices of the Royal Astronomical
  Society}\ }\textbf {\bibinfo {volume} {423}},\ \bibinfo {pages} {2533–2557}
  (\bibinfo {year} {2012})}\BibitemShut {NoStop}%
\bibitem [{\citenamefont {{Klein}}\ \emph {et~al.}(2016)\citenamefont
  {{Klein}}, \citenamefont {{Barausse}}, \citenamefont {{Sesana}},
  \citenamefont {{Petiteau}}, \citenamefont {{Berti}}, \citenamefont {{Babak}},
  \citenamefont {{Gair}}, \citenamefont {{Aoudia}}, \citenamefont {{Hinder}},
  \citenamefont {{Ohme}},\ and\ \citenamefont
  {{Wardell}}}]{2016PhRvD..93b4003K}%
  \BibitemOpen
  \bibfield  {author} {\bibinfo {author} {\bibfnamefont {A.}~\bibnamefont
  {{Klein}}}, \bibinfo {author} {\bibfnamefont {E.}~\bibnamefont {{Barausse}}},
  \bibinfo {author} {\bibfnamefont {A.}~\bibnamefont {{Sesana}}}, \bibinfo
  {author} {\bibfnamefont {A.}~\bibnamefont {{Petiteau}}}, \bibinfo {author}
  {\bibfnamefont {E.}~\bibnamefont {{Berti}}}, \bibinfo {author} {\bibfnamefont
  {S.}~\bibnamefont {{Babak}}}, \bibinfo {author} {\bibfnamefont
  {J.}~\bibnamefont {{Gair}}}, \bibinfo {author} {\bibfnamefont
  {S.}~\bibnamefont {{Aoudia}}}, \bibinfo {author} {\bibfnamefont
  {I.}~\bibnamefont {{Hinder}}}, \bibinfo {author} {\bibfnamefont
  {F.}~\bibnamefont {{Ohme}}}, \ and\ \bibinfo {author} {\bibfnamefont
  {B.}~\bibnamefont {{Wardell}}},\ }\href {\doibase 10.1103/PhysRevD.93.024003}
  {\bibfield  {journal} {\bibinfo  {journal} {\prd}\ }\textbf {\bibinfo
  {volume} {93}},\ \bibinfo {eid} {024003} (\bibinfo {year} {2016})},\ \Eprint
  {http://arxiv.org/abs/1511.05581} {arXiv:1511.05581 [gr-qc]} \BibitemShut
  {NoStop}%
\bibitem [{\citenamefont {Baibhav}\ \emph {et~al.}(2019)\citenamefont
  {Baibhav}, \citenamefont {Berti}, \citenamefont {Gerosa}, \citenamefont
  {Mapelli}, \citenamefont {Giacobbo}, \citenamefont {Bouffanais},\ and\
  \citenamefont {Di~Carlo}}]{Baibhav_2019}%
  \BibitemOpen
  \bibfield  {author} {\bibinfo {author} {\bibfnamefont {V.}~\bibnamefont
  {Baibhav}}, \bibinfo {author} {\bibfnamefont {E.}~\bibnamefont {Berti}},
  \bibinfo {author} {\bibfnamefont {D.}~\bibnamefont {Gerosa}}, \bibinfo
  {author} {\bibfnamefont {M.}~\bibnamefont {Mapelli}}, \bibinfo {author}
  {\bibfnamefont {N.}~\bibnamefont {Giacobbo}}, \bibinfo {author}
  {\bibfnamefont {Y.}~\bibnamefont {Bouffanais}}, \ and\ \bibinfo {author}
  {\bibfnamefont {U.~N.}\ \bibnamefont {Di~Carlo}},\ }\href {\doibase
  10.1103/physrevd.100.064060} {\bibfield  {journal} {\bibinfo  {journal}
  {Physical Review D}\ }\textbf {\bibinfo {volume} {100}} (\bibinfo {year}
  {2019}),\ 10.1103/physrevd.100.064060}\BibitemShut {NoStop}%
\bibitem [{\citenamefont {{Gergely}}\ and\ \citenamefont
  {{Biermann}}(2012)}]{2012arXiv1208.5251G}%
  \BibitemOpen
  \bibfield  {author} {\bibinfo {author} {\bibfnamefont {L.~{\'A}.}\
  \bibnamefont {{Gergely}}}\ and\ \bibinfo {author} {\bibfnamefont {P.~L.}\
  \bibnamefont {{Biermann}}},\ }\href@noop {} {\bibfield  {journal} {\bibinfo
  {journal} {arXiv e-prints}\ ,\ \bibinfo {eid} {arXiv:1208.5251}} (\bibinfo
  {year} {2012})},\ \Eprint {http://arxiv.org/abs/1208.5251} {arXiv:1208.5251
  [gr-qc]} \BibitemShut {NoStop}%
\bibitem [{\citenamefont {Petiteau}\ \emph {et~al.}(2016)\citenamefont
  {Petiteau}, \citenamefont {Hewitson}, \citenamefont {Heinzel}, \citenamefont
  {Fitzsimons},\ and\ \citenamefont {Halloin}}]{Noisecurve}%
  \BibitemOpen
  \bibfield  {author} {\bibinfo {author} {\bibfnamefont {A.}~\bibnamefont
  {Petiteau}}, \bibinfo {author} {\bibfnamefont {M.}~\bibnamefont {Hewitson}},
  \bibinfo {author} {\bibfnamefont {G.}~\bibnamefont {Heinzel}}, \bibinfo
  {author} {\bibfnamefont {E.}~\bibnamefont {Fitzsimons}}, \ and\ \bibinfo
  {author} {\bibfnamefont {H.}~\bibnamefont {Halloin}},\ }\href@noop {}
  {\bibfield  {journal} {\bibinfo  {journal} {Tech. rep. (LISA Consortium),
  lISA-CST-TN-0001}\ } (\bibinfo {year} {2016})}\BibitemShut {NoStop}%
\bibitem [{\citenamefont {{Sesana}}(2016)}]{2016PhRvL.116w1102S}%
  \BibitemOpen
  \bibfield  {author} {\bibinfo {author} {\bibfnamefont {A.}~\bibnamefont
  {{Sesana}}},\ }\href {\doibase 10.1103/PhysRevLett.116.231102} {\bibfield
  {journal} {\bibinfo  {journal} {\prl}\ }\textbf {\bibinfo {volume} {116}},\
  \bibinfo {eid} {231102} (\bibinfo {year} {2016})},\ \Eprint
  {http://arxiv.org/abs/1602.06951} {arXiv:1602.06951 [gr-qc]} \BibitemShut
  {NoStop}%
\bibitem [{\citenamefont {{Cutler}}\ and\ \citenamefont
  {{Flanagan}}(1994)}]{1994PhRvD..49.2658C}%
  \BibitemOpen
  \bibfield  {author} {\bibinfo {author} {\bibfnamefont {C.}~\bibnamefont
  {{Cutler}}}\ and\ \bibinfo {author} {\bibfnamefont {{\'E}.~E.}\ \bibnamefont
  {{Flanagan}}},\ }\href {\doibase 10.1103/PhysRevD.49.2658} {\bibfield
  {journal} {\bibinfo  {journal} {\prd}\ }\textbf {\bibinfo {volume} {49}},\
  \bibinfo {pages} {2658} (\bibinfo {year} {1994})},\ \Eprint
  {http://arxiv.org/abs/gr-qc/9402014} {arXiv:gr-qc/9402014 [gr-qc]}
  \BibitemShut {NoStop}%
\bibitem [{\citenamefont {Moore}\ \emph {et~al.}(2014)\citenamefont {Moore},
  \citenamefont {Cole},\ and\ \citenamefont {Berry}}]{Moore_2014}%
  \BibitemOpen
  \bibfield  {author} {\bibinfo {author} {\bibfnamefont {C.~J.}\ \bibnamefont
  {Moore}}, \bibinfo {author} {\bibfnamefont {R.~H.}\ \bibnamefont {Cole}}, \
  and\ \bibinfo {author} {\bibfnamefont {C.~P.~L.}\ \bibnamefont {Berry}},\
  }\href {\doibase 10.1088/0264-9381/32/1/015014} {\bibfield  {journal}
  {\bibinfo  {journal} {Classical and Quantum Gravity}\ }\textbf {\bibinfo
  {volume} {32}},\ \bibinfo {pages} {015014} (\bibinfo {year}
  {2014})}\BibitemShut {NoStop}%
\bibitem [{\citenamefont {{Amaro-Seoane}}\ \emph {et~al.}(2017)\citenamefont
  {{Amaro-Seoane}}, \citenamefont {{Audley}}, \citenamefont {{Babak}} \emph
  {et~al.}}]{2017arXiv170200786A}%
  \BibitemOpen
  \bibfield  {author} {\bibinfo {author} {\bibfnamefont {P.}~\bibnamefont
  {{Amaro-Seoane}}}, \bibinfo {author} {\bibfnamefont {H.}~\bibnamefont
  {{Audley}}}, \bibinfo {author} {\bibfnamefont {S.}~\bibnamefont {{Babak}}},
  \emph {et~al.},\ }\href@noop {} {\bibfield  {journal} {\bibinfo  {journal}
  {arXiv e-prints}\ ,\ \bibinfo {eid} {arXiv:1702.00786}} (\bibinfo {year}
  {2017})},\ \Eprint {http://arxiv.org/abs/1702.00786} {arXiv:1702.00786
  [astro-ph.IM]} \BibitemShut {NoStop}%
\bibitem [{\citenamefont {Ravi}(2018)}]{Ravi:2018jyk}%
  \BibitemOpen
  \bibfield  {author} {\bibinfo {author} {\bibfnamefont {V.}~\bibnamefont
  {Ravi}},\ }\href@noop {} {\bibfield  {journal} {\bibinfo  {journal} {ASP
  Conf. Ser.}\ }\textbf {\bibinfo {volume} {517}},\ \bibinfo {pages} {781}
  (\bibinfo {year} {2018})},\ \Eprint {http://arxiv.org/abs/1806.08446}
  {arXiv:1806.08446 [astro-ph.HE]} \BibitemShut {NoStop}%
\bibitem [{\citenamefont {Ezquiaga}\ \emph {et~al.}(2020)\citenamefont
  {Ezquiaga}, \citenamefont {Hu},\ and\ \citenamefont
  {Lagos}}]{ezquiaga2020apparent}%
  \BibitemOpen
  \bibfield  {author} {\bibinfo {author} {\bibfnamefont {J.~M.}\ \bibnamefont
  {Ezquiaga}}, \bibinfo {author} {\bibfnamefont {W.}~\bibnamefont {Hu}}, \ and\
  \bibinfo {author} {\bibfnamefont {M.}~\bibnamefont {Lagos}},\ }\href@noop {}
  {\enquote {\bibinfo {title} {Apparent superluminality of lensed gravitational
  waves},}\ } (\bibinfo {year} {2020}),\ \Eprint
  {http://arxiv.org/abs/2005.10702} {arXiv:2005.10702 [astro-ph.CO]}
  \BibitemShut {NoStop}%
\bibitem [{\citenamefont {Mukherjee}\ \emph
  {et~al.}(2020{\natexlab{a}})\citenamefont {Mukherjee}, \citenamefont
  {Wandelt},\ and\ \citenamefont {Silk}}]{Mukherjee_2020_multimessenger}%
  \BibitemOpen
  \bibfield  {author} {\bibinfo {author} {\bibfnamefont {S.}~\bibnamefont
  {Mukherjee}}, \bibinfo {author} {\bibfnamefont {B.~D.}\ \bibnamefont
  {Wandelt}}, \ and\ \bibinfo {author} {\bibfnamefont {J.}~\bibnamefont
  {Silk}},\ }\href {\doibase 10.1103/physrevd.101.103509} {\bibfield  {journal}
  {\bibinfo  {journal} {Physical Review D}\ }\textbf {\bibinfo {volume} {101}}
  (\bibinfo {year} {2020}{\natexlab{a}}),\
  10.1103/physrevd.101.103509}\BibitemShut {NoStop}%
\bibitem [{\citenamefont {Mukherjee}\ \emph
  {et~al.}(2020{\natexlab{b}})\citenamefont {Mukherjee}, \citenamefont
  {Wandelt},\ and\ \citenamefont {Silk}}]{Mukherjee_2020_probing}%
  \BibitemOpen
  \bibfield  {author} {\bibinfo {author} {\bibfnamefont {S.}~\bibnamefont
  {Mukherjee}}, \bibinfo {author} {\bibfnamefont {B.~D.}\ \bibnamefont
  {Wandelt}}, \ and\ \bibinfo {author} {\bibfnamefont {J.}~\bibnamefont
  {Silk}},\ }\href {\doibase 10.1093/mnras/staa827} {\bibfield  {journal}
  {\bibinfo  {journal} {Monthly Notices of the Royal Astronomical Society}\
  }\textbf {\bibinfo {volume} {494}},\ \bibinfo {pages} {1956–1970} (\bibinfo
  {year} {2020}{\natexlab{b}})}\BibitemShut {NoStop}%
\bibitem [{\citenamefont {{Hirata}}\ \emph {et~al.}(2010)\citenamefont
  {{Hirata}}, \citenamefont {{Holz}},\ and\ \citenamefont
  {{Cutler}}}]{2010PhRvD..81l4046H}%
  \BibitemOpen
  \bibfield  {author} {\bibinfo {author} {\bibfnamefont {C.~M.}\ \bibnamefont
  {{Hirata}}}, \bibinfo {author} {\bibfnamefont {D.~E.}\ \bibnamefont
  {{Holz}}}, \ and\ \bibinfo {author} {\bibfnamefont {C.}~\bibnamefont
  {{Cutler}}},\ }\href {\doibase 10.1103/PhysRevD.81.124046} {\bibfield
  {journal} {\bibinfo  {journal} {\prd}\ }\textbf {\bibinfo {volume} {81}},\
  \bibinfo {eid} {124046} (\bibinfo {year} {2010})},\ \Eprint
  {http://arxiv.org/abs/1004.3988} {arXiv:1004.3988 [astro-ph.CO]} \BibitemShut
  {NoStop}%
\bibitem [{\citenamefont {{Mukherjee}}\ \emph {et~al.}(2019)\citenamefont
  {{Mukherjee}}, \citenamefont {{Lavaux}}, \citenamefont {{Bouchet}},
  \citenamefont {{Jasche}}, \citenamefont {{Wand elt}}, \citenamefont
  {{Nissanke}}, \citenamefont {{Leclercq}},\ and\ \citenamefont
  {{Hotokezaka}}}]{2019arXiv190908627M}%
  \BibitemOpen
  \bibfield  {author} {\bibinfo {author} {\bibfnamefont {S.}~\bibnamefont
  {{Mukherjee}}}, \bibinfo {author} {\bibfnamefont {G.}~\bibnamefont
  {{Lavaux}}}, \bibinfo {author} {\bibfnamefont {F.~R.}\ \bibnamefont
  {{Bouchet}}}, \bibinfo {author} {\bibfnamefont {J.}~\bibnamefont {{Jasche}}},
  \bibinfo {author} {\bibfnamefont {B.~D.}\ \bibnamefont {{Wand elt}}},
  \bibinfo {author} {\bibfnamefont {S.~M.}\ \bibnamefont {{Nissanke}}},
  \bibinfo {author} {\bibfnamefont {F.}~\bibnamefont {{Leclercq}}}, \ and\
  \bibinfo {author} {\bibfnamefont {K.}~\bibnamefont {{Hotokezaka}}},\
  }\href@noop {} {\bibfield  {journal} {\bibinfo  {journal} {arXiv e-prints}\
  ,\ \bibinfo {eid} {arXiv:1909.08627}} (\bibinfo {year} {2019})},\ \Eprint
  {http://arxiv.org/abs/1909.08627} {arXiv:1909.08627 [astro-ph.CO]}
  \BibitemShut {NoStop}%
\bibitem [{\citenamefont {{Kocsis}}\ \emph {et~al.}(2006)\citenamefont
  {{Kocsis}}, \citenamefont {{Frei}}, \citenamefont {{Haiman}},\ and\
  \citenamefont {{Menou}}}]{2006ApJ...637...27K}%
  \BibitemOpen
  \bibfield  {author} {\bibinfo {author} {\bibfnamefont {B.}~\bibnamefont
  {{Kocsis}}}, \bibinfo {author} {\bibfnamefont {Z.}~\bibnamefont {{Frei}}},
  \bibinfo {author} {\bibfnamefont {Z.}~\bibnamefont {{Haiman}}}, \ and\
  \bibinfo {author} {\bibfnamefont {K.}~\bibnamefont {{Menou}}},\ }\href
  {\doibase 10.1086/498236} {\bibfield  {journal} {\bibinfo  {journal} {\apj}\
  }\textbf {\bibinfo {volume} {637}},\ \bibinfo {pages} {27} (\bibinfo {year}
  {2006})},\ \Eprint {http://arxiv.org/abs/astro-ph/0505394}
  {arXiv:astro-ph/0505394 [astro-ph]} \BibitemShut {NoStop}%
\bibitem [{\citenamefont {{Planck Collaboration}}\ \emph
  {et~al.}(2014)\citenamefont {{Planck Collaboration}}, \citenamefont {{Ade}},
  \citenamefont {{Aghanim}} \emph {et~al.}}]{2014A&A...571A..15P}%
  \BibitemOpen
  \bibfield  {author} {\bibinfo {author} {\bibnamefont {{Planck
  Collaboration}}}, \bibinfo {author} {\bibfnamefont {P.~A.~R.}\ \bibnamefont
  {{Ade}}}, \bibinfo {author} {\bibfnamefont {N.}~\bibnamefont {{Aghanim}}},
  \emph {et~al.},\ }\href {\doibase 10.1051/0004-6361/201321573} {\bibfield
  {journal} {\bibinfo  {journal} {\aap}\ }\textbf {\bibinfo {volume} {571}},\
  \bibinfo {eid} {A15} (\bibinfo {year} {2014})},\ \Eprint
  {http://arxiv.org/abs/1303.5075} {arXiv:1303.5075 [astro-ph.CO]} \BibitemShut
  {NoStop}%
\bibitem [{\citenamefont {{Alam}}\ \emph {et~al.}(2017)\citenamefont {{Alam}},
  \citenamefont {{Ata}} \emph {et~al.}}]{2017MNRAS.470.2617A}%
  \BibitemOpen
  \bibfield  {author} {\bibinfo {author} {\bibfnamefont {S.}~\bibnamefont
  {{Alam}}}, \bibinfo {author} {\bibfnamefont {M.}~\bibnamefont {{Ata}}},
  \emph {et~al.},\ }\href {\doibase 10.1093/mnras/stx721} {\bibfield  {journal}
  {\bibinfo  {journal} {\mnras}\ }\textbf {\bibinfo {volume} {470}},\ \bibinfo
  {pages} {2617} (\bibinfo {year} {2017})},\ \Eprint
  {http://arxiv.org/abs/1607.03155} {arXiv:1607.03155 [astro-ph.CO]}
  \BibitemShut {NoStop}%
\bibitem [{\citenamefont {{Beutler}}\ \emph {et~al.}(2012)\citenamefont
  {{Beutler}}, \citenamefont {{Blake}}, \citenamefont {{Colless}},
  \citenamefont {{Jones}}, \citenamefont {{Staveley-Smith}}, \citenamefont
  {{Poole}}, \citenamefont {{Campbell}}, \citenamefont {{Parker}},
  \citenamefont {{Saunders}},\ and\ \citenamefont
  {{Watson}}}]{2012MNRAS.423.3430B}%
  \BibitemOpen
  \bibfield  {author} {\bibinfo {author} {\bibfnamefont {F.}~\bibnamefont
  {{Beutler}}}, \bibinfo {author} {\bibfnamefont {C.}~\bibnamefont {{Blake}}},
  \bibinfo {author} {\bibfnamefont {M.}~\bibnamefont {{Colless}}}, \bibinfo
  {author} {\bibfnamefont {D.~H.}\ \bibnamefont {{Jones}}}, \bibinfo {author}
  {\bibfnamefont {L.}~\bibnamefont {{Staveley-Smith}}}, \bibinfo {author}
  {\bibfnamefont {G.~B.}\ \bibnamefont {{Poole}}}, \bibinfo {author}
  {\bibfnamefont {L.}~\bibnamefont {{Campbell}}}, \bibinfo {author}
  {\bibfnamefont {Q.}~\bibnamefont {{Parker}}}, \bibinfo {author}
  {\bibfnamefont {W.}~\bibnamefont {{Saunders}}}, \ and\ \bibinfo {author}
  {\bibfnamefont {F.}~\bibnamefont {{Watson}}},\ }\href {\doibase
  10.1111/j.1365-2966.2012.21136.x} {\bibfield  {journal} {\bibinfo  {journal}
  {\mnras}\ }\textbf {\bibinfo {volume} {423}},\ \bibinfo {pages} {3430}
  (\bibinfo {year} {2012})},\ \Eprint {http://arxiv.org/abs/1204.4725}
  {arXiv:1204.4725 [astro-ph.CO]} \BibitemShut {NoStop}%
\bibitem [{\citenamefont {{Kazin}}\ \emph {et~al.}(2014)\citenamefont
  {{Kazin}}, \citenamefont {{Koda}}, \citenamefont {{Blake}} \emph
  {et~al.}}]{2014MNRAS.441.3524K}%
  \BibitemOpen
  \bibfield  {author} {\bibinfo {author} {\bibfnamefont {E.~A.}\ \bibnamefont
  {{Kazin}}}, \bibinfo {author} {\bibfnamefont {J.}~\bibnamefont {{Koda}}},
  \bibinfo {author} {\bibfnamefont {C.}~\bibnamefont {{Blake}}},  \emph
  {et~al.},\ }\href {\doibase 10.1093/mnras/stu778} {\bibfield  {journal}
  {\bibinfo  {journal} {\mnras}\ }\textbf {\bibinfo {volume} {441}},\ \bibinfo
  {pages} {3524} (\bibinfo {year} {2014})},\ \Eprint
  {http://arxiv.org/abs/1401.0358} {arXiv:1401.0358 [astro-ph.CO]} \BibitemShut
  {NoStop}%
\bibitem [{\citenamefont {{Ross}}\ \emph {et~al.}(2015)\citenamefont {{Ross}},
  \citenamefont {{Samushia}}, \citenamefont {{Howlett}}, \citenamefont
  {{Percival}}, \citenamefont {{Burden}},\ and\ \citenamefont
  {{Manera}}}]{2015MNRAS.449..835R}%
  \BibitemOpen
  \bibfield  {author} {\bibinfo {author} {\bibfnamefont {A.~J.}\ \bibnamefont
  {{Ross}}}, \bibinfo {author} {\bibfnamefont {L.}~\bibnamefont {{Samushia}}},
  \bibinfo {author} {\bibfnamefont {C.}~\bibnamefont {{Howlett}}}, \bibinfo
  {author} {\bibfnamefont {W.~J.}\ \bibnamefont {{Percival}}}, \bibinfo
  {author} {\bibfnamefont {A.}~\bibnamefont {{Burden}}}, \ and\ \bibinfo
  {author} {\bibfnamefont {M.}~\bibnamefont {{Manera}}},\ }\href {\doibase
  10.1093/mnras/stv154} {\bibfield  {journal} {\bibinfo  {journal} {\mnras}\
  }\textbf {\bibinfo {volume} {449}},\ \bibinfo {pages} {835} (\bibinfo {year}
  {2015})},\ \Eprint {http://arxiv.org/abs/1409.3242} {arXiv:1409.3242
  [astro-ph.CO]} \BibitemShut {NoStop}%
\bibitem [{\citenamefont {Zuntz}\ \emph {et~al.}(2015)\citenamefont {Zuntz},
  \citenamefont {Paterno}, \citenamefont {Jennings}, \citenamefont {Rudd},
  \citenamefont {Manzotti}, \citenamefont {Dodelson}, \citenamefont {Bridle},
  \citenamefont {Sehrish},\ and\ \citenamefont {Kowalkowski}}]{Zuntz_2015}%
  \BibitemOpen
  \bibfield  {author} {\bibinfo {author} {\bibfnamefont {J.}~\bibnamefont
  {Zuntz}}, \bibinfo {author} {\bibfnamefont {M.}~\bibnamefont {Paterno}},
  \bibinfo {author} {\bibfnamefont {E.}~\bibnamefont {Jennings}}, \bibinfo
  {author} {\bibfnamefont {D.}~\bibnamefont {Rudd}}, \bibinfo {author}
  {\bibfnamefont {A.}~\bibnamefont {Manzotti}}, \bibinfo {author}
  {\bibfnamefont {S.}~\bibnamefont {Dodelson}}, \bibinfo {author}
  {\bibfnamefont {S.}~\bibnamefont {Bridle}}, \bibinfo {author} {\bibfnamefont
  {S.}~\bibnamefont {Sehrish}}, \ and\ \bibinfo {author} {\bibfnamefont
  {J.}~\bibnamefont {Kowalkowski}},\ }\href {\doibase
  10.1016/j.ascom.2015.05.005} {\bibfield  {journal} {\bibinfo  {journal}
  {Astronomy and Computing}\ }\textbf {\bibinfo {volume} {12}},\ \bibinfo
  {pages} {45–59} (\bibinfo {year} {2015})}\BibitemShut {NoStop}%
\bibitem [{\citenamefont {{Foreman-Mackey}}\ \emph {et~al.}(2013)\citenamefont
  {{Foreman-Mackey}}, \citenamefont {{Hogg}}, \citenamefont {{Lang}},\ and\
  \citenamefont {{Goodman}}}]{2013PASP..125..306F}%
  \BibitemOpen
  \bibfield  {author} {\bibinfo {author} {\bibfnamefont {D.}~\bibnamefont
  {{Foreman-Mackey}}}, \bibinfo {author} {\bibfnamefont {D.~W.}\ \bibnamefont
  {{Hogg}}}, \bibinfo {author} {\bibfnamefont {D.}~\bibnamefont {{Lang}}}, \
  and\ \bibinfo {author} {\bibfnamefont {J.}~\bibnamefont {{Goodman}}},\ }\href
  {\doibase 10.1086/670067} {\bibfield  {journal} {\bibinfo  {journal} {\pasp}\
  }\textbf {\bibinfo {volume} {125}},\ \bibinfo {pages} {306} (\bibinfo {year}
  {2013})},\ \Eprint {http://arxiv.org/abs/1202.3665} {arXiv:1202.3665
  [astro-ph.IM]} \BibitemShut {NoStop}%
\bibitem [{\citenamefont {Ferté}\ \emph {et~al.}(2019)\citenamefont {Ferté},
  \citenamefont {Kirk}, \citenamefont {Liddle},\ and\ \citenamefont
  {Zuntz}}]{Fert__2019}%
  \BibitemOpen
  \bibfield  {author} {\bibinfo {author} {\bibfnamefont {A.}~\bibnamefont
  {Ferté}}, \bibinfo {author} {\bibfnamefont {D.}~\bibnamefont {Kirk}},
  \bibinfo {author} {\bibfnamefont {A.~R.}\ \bibnamefont {Liddle}}, \ and\
  \bibinfo {author} {\bibfnamefont {J.}~\bibnamefont {Zuntz}},\ }\href
  {\doibase 10.1103/physrevd.99.083512} {\bibfield  {journal} {\bibinfo
  {journal} {Physical Review D}\ }\textbf {\bibinfo {volume} {99}} (\bibinfo
  {year} {2019}),\ 10.1103/physrevd.99.083512}\BibitemShut {NoStop}%
\bibitem [{\citenamefont {{Niu}}\ \emph {et~al.}(2020)\citenamefont {{Niu}},
  \citenamefont {{Zhang}}, \citenamefont {{Liu}}, \citenamefont {{Yu}},
  \citenamefont {{Wang}},\ and\ \citenamefont {{Zhao}}}]{Niu2020}%
  \BibitemOpen
  \bibfield  {author} {\bibinfo {author} {\bibfnamefont {R.}~\bibnamefont
  {{Niu}}}, \bibinfo {author} {\bibfnamefont {X.}~\bibnamefont {{Zhang}}},
  \bibinfo {author} {\bibfnamefont {T.}~\bibnamefont {{Liu}}}, \bibinfo
  {author} {\bibfnamefont {J.}~\bibnamefont {{Yu}}}, \bibinfo {author}
  {\bibfnamefont {B.}~\bibnamefont {{Wang}}}, \ and\ \bibinfo {author}
  {\bibfnamefont {W.}~\bibnamefont {{Zhao}}},\ }\href {\doibase
  10.3847/1538-4357/ab6d03} {\bibfield  {journal} {\bibinfo  {journal} {\apj}\
  }\textbf {\bibinfo {volume} {890}},\ \bibinfo {eid} {163} (\bibinfo {year}
  {2020})},\ \Eprint {http://arxiv.org/abs/1910.10592} {arXiv:1910.10592
  [gr-qc]} \BibitemShut {NoStop}%
\bibitem [{\citenamefont {{Schutz}}(1986)}]{1986Natur.323..310S}%
  \BibitemOpen
  \bibfield  {author} {\bibinfo {author} {\bibfnamefont {B.~F.}\ \bibnamefont
  {{Schutz}}},\ }\href {\doibase 10.1038/323310a0} {\bibfield  {journal}
  {\bibinfo  {journal} {\nat}\ }\textbf {\bibinfo {volume} {323}},\ \bibinfo
  {pages} {310} (\bibinfo {year} {1986})}\BibitemShut {NoStop}%
\bibitem [{\citenamefont {Del~Pozzo}(2012)}]{PhysRevD.86.043011}%
  \BibitemOpen
  \bibfield  {author} {\bibinfo {author} {\bibfnamefont {W.}~\bibnamefont
  {Del~Pozzo}},\ }\href {\doibase 10.1103/PhysRevD.86.043011} {\bibfield
  {journal} {\bibinfo  {journal} {Phys. Rev. D}\ }\textbf {\bibinfo {volume}
  {86}},\ \bibinfo {pages} {043011} (\bibinfo {year} {2012})}\BibitemShut
  {NoStop}%
\bibitem [{\citenamefont {{Del Pozzo}}\ \emph {et~al.}(2018)\citenamefont {{Del
  Pozzo}}, \citenamefont {{Sesana}},\ and\ \citenamefont {{Klein}}}]{Pozzo:it}%
  \BibitemOpen
  \bibfield  {author} {\bibinfo {author} {\bibfnamefont {W.}~\bibnamefont {{Del
  Pozzo}}}, \bibinfo {author} {\bibfnamefont {A.}~\bibnamefont {{Sesana}}}, \
  and\ \bibinfo {author} {\bibfnamefont {A.}~\bibnamefont {{Klein}}},\ }\href
  {\doibase 10.1093/mnras/sty057} {\bibfield  {journal} {\bibinfo  {journal}
  {\mnras}\ }\textbf {\bibinfo {volume} {475}},\ \bibinfo {pages} {3485}
  (\bibinfo {year} {2018})},\ \Eprint {http://arxiv.org/abs/1703.01300}
  {arXiv:1703.01300 [astro-ph.CO]} \BibitemShut {NoStop}%
\bibitem [{\citenamefont {Mukherjee}\ \emph
  {et~al.}(2020{\natexlab{c}})\citenamefont {Mukherjee}, \citenamefont
  {Wandelt}, \citenamefont {Nissanke},\ and\ \citenamefont
  {Silvestri}}]{mukherjee2020accurate}%
  \BibitemOpen
  \bibfield  {author} {\bibinfo {author} {\bibfnamefont {S.}~\bibnamefont
  {Mukherjee}}, \bibinfo {author} {\bibfnamefont {B.~D.}\ \bibnamefont
  {Wandelt}}, \bibinfo {author} {\bibfnamefont {S.~M.}\ \bibnamefont
  {Nissanke}}, \ and\ \bibinfo {author} {\bibfnamefont {A.}~\bibnamefont
  {Silvestri}},\ }\href@noop {} {\enquote {\bibinfo {title} {Accurate and
  precision cosmology with redshift unknown gravitational wave sources},}\ }
  (\bibinfo {year} {2020}{\natexlab{c}}),\ \Eprint
  {http://arxiv.org/abs/2007.02943} {arXiv:2007.02943 [astro-ph.CO]}
  \BibitemShut {NoStop}%
\bibitem [{\citenamefont {Ezquiaga}\ and\ \citenamefont
  {Holz}(2020)}]{ezquiaga2020jumping}%
  \BibitemOpen
  \bibfield  {author} {\bibinfo {author} {\bibfnamefont {J.~M.}\ \bibnamefont
  {Ezquiaga}}\ and\ \bibinfo {author} {\bibfnamefont {D.~E.}\ \bibnamefont
  {Holz}},\ }\href@noop {} {\enquote {\bibinfo {title} {Jumping the gap:
  searching for ligo's biggest black holes},}\ } (\bibinfo {year} {2020}),\
  \Eprint {http://arxiv.org/abs/2006.02211} {arXiv:2006.02211 [astro-ph.HE]}
  \BibitemShut {NoStop}%
\bibitem [{\citenamefont {Borhanian}\ \emph {et~al.}(2020)\citenamefont
  {Borhanian}, \citenamefont {Dhani}, \citenamefont {Gupta}, \citenamefont
  {Arun},\ and\ \citenamefont {Sathyaprakash}}]{borhanian2020dark}%
  \BibitemOpen
  \bibfield  {author} {\bibinfo {author} {\bibfnamefont {S.}~\bibnamefont
  {Borhanian}}, \bibinfo {author} {\bibfnamefont {A.}~\bibnamefont {Dhani}},
  \bibinfo {author} {\bibfnamefont {A.}~\bibnamefont {Gupta}}, \bibinfo
  {author} {\bibfnamefont {K.~G.}\ \bibnamefont {Arun}}, \ and\ \bibinfo
  {author} {\bibfnamefont {B.~S.}\ \bibnamefont {Sathyaprakash}},\ }\href@noop
  {} {\enquote {\bibinfo {title} {Dark sirens to resolve the hubble-lemaître
  tension},}\ } (\bibinfo {year} {2020}),\ \Eprint
  {http://arxiv.org/abs/2007.02883} {arXiv:2007.02883 [astro-ph.CO]}
  \BibitemShut {NoStop}%
\bibitem [{\citenamefont {Mills}\ and\ \citenamefont
  {Fairhurst}(2020)}]{mills2020measuring}%
  \BibitemOpen
  \bibfield  {author} {\bibinfo {author} {\bibfnamefont {C.}~\bibnamefont
  {Mills}}\ and\ \bibinfo {author} {\bibfnamefont {S.}~\bibnamefont
  {Fairhurst}},\ }\href@noop {} {\enquote {\bibinfo {title} {Measuring
  gravitational-wave higher-order modes},}\ } (\bibinfo {year} {2020}),\
  \Eprint {http://arxiv.org/abs/2007.04313} {arXiv:2007.04313 [gr-qc]}
  \BibitemShut {NoStop}%
\bibitem [{\citenamefont {Pardo}\ \emph {et~al.}(2018)\citenamefont {Pardo},
  \citenamefont {Fishbach}, \citenamefont {Holz},\ and\ \citenamefont
  {Spergel}}]{Pardo_2018}%
  \BibitemOpen
  \bibfield  {author} {\bibinfo {author} {\bibfnamefont {K.}~\bibnamefont
  {Pardo}}, \bibinfo {author} {\bibfnamefont {M.}~\bibnamefont {Fishbach}},
  \bibinfo {author} {\bibfnamefont {D.~E.}\ \bibnamefont {Holz}}, \ and\
  \bibinfo {author} {\bibfnamefont {D.~N.}\ \bibnamefont {Spergel}},\ }\href
  {\doibase 10.1088/1475-7516/2018/07/048} {\bibfield  {journal} {\bibinfo
  {journal} {Journal of Cosmology and Astroparticle Physics}\ }\textbf
  {\bibinfo {volume} {2018}},\ \bibinfo {pages} {048–048} (\bibinfo {year}
  {2018})}\BibitemShut {NoStop}%
\bibitem [{\citenamefont {Corman}\ \emph {et~al.}(2020)\citenamefont {Corman},
  \citenamefont {Escamilla-Rivera},\ and\ \citenamefont
  {Hendry}}]{corman2020constraining}%
  \BibitemOpen
  \bibfield  {author} {\bibinfo {author} {\bibfnamefont {M.}~\bibnamefont
  {Corman}}, \bibinfo {author} {\bibfnamefont {C.}~\bibnamefont
  {Escamilla-Rivera}}, \ and\ \bibinfo {author} {\bibfnamefont {M.~A.}\
  \bibnamefont {Hendry}},\ }\href@noop {} {\enquote {\bibinfo {title}
  {Constraining extra dimensions on cosmological scales with lisa future
  gravitational wave siren data},}\ } (\bibinfo {year} {2020}),\ \Eprint
  {http://arxiv.org/abs/2004.04009} {arXiv:2004.04009 [astro-ph.CO]}
  \BibitemShut {NoStop}%
\bibitem [{\citenamefont {Hogg}\ \emph {et~al.}(2020)\citenamefont {Hogg},
  \citenamefont {Martinelli},\ and\ \citenamefont {Nesseris}}]{hogg2020}%
  \BibitemOpen
  \bibfield  {author} {\bibinfo {author} {\bibfnamefont {N.}~\bibnamefont
  {Hogg}}, \bibinfo {author} {\bibfnamefont {M.}~\bibnamefont {Martinelli}}, \
  and\ \bibinfo {author} {\bibfnamefont {S.}~\bibnamefont {Nesseris}},\
  }\href@noop {} {\enquote {\bibinfo {title} {Constraints on the distance
  duality relation with standard sirens},}\ } (\bibinfo {year} {2020}),\
  \Eprint {http://arxiv.org/abs/in prep} {in prep} \BibitemShut {NoStop}%
\bibitem [{\citenamefont {Martinelli}\ \emph {et~al.}(2020)\citenamefont
  {Martinelli}, \citenamefont {Martins},\ and\ \citenamefont
  {Nesseris}}]{martinelli2020}%
  \BibitemOpen
  \bibfield  {author} {\bibinfo {author} {\bibfnamefont {M.}~\bibnamefont
  {Martinelli}}, \bibinfo {author} {\bibfnamefont {C.~J.}\ \bibnamefont
  {Martins}}, \ and\ \bibinfo {author} {\bibfnamefont {S.}~\bibnamefont
  {Nesseris}},\ }\href@noop {} {\enquote {\bibinfo {title} {Euclid: Forecast
  constraints on the cosmic distance duality relation with complementary
  external probes},}\ } (\bibinfo {year} {2020}),\ \Eprint
  {http://arxiv.org/abs/in prep} {in prep} \BibitemShut {NoStop}%
\bibitem [{\citenamefont {{Bachega}}\ \emph {et~al.}(2020)\citenamefont
  {{Bachega}}, \citenamefont {{Costa}}, \citenamefont {{Abdalla}},\ and\
  \citenamefont {{Fornazier}}}]{Bachega2020}%
  \BibitemOpen
  \bibfield  {author} {\bibinfo {author} {\bibfnamefont {R.~R.~A.}\
  \bibnamefont {{Bachega}}}, \bibinfo {author} {\bibfnamefont {A.~A.}\
  \bibnamefont {{Costa}}}, \bibinfo {author} {\bibfnamefont {E.}~\bibnamefont
  {{Abdalla}}}, \ and\ \bibinfo {author} {\bibfnamefont {K.~S.~F.}\
  \bibnamefont {{Fornazier}}},\ }\href {\doibase 10.1088/1475-7516/2020/05/021}
  {\bibfield  {journal} {\bibinfo  {journal} {\jcap}\ }\textbf {\bibinfo
  {volume} {2020}},\ \bibinfo {eid} {021} (\bibinfo {year} {2020})},\ \Eprint
  {http://arxiv.org/abs/1906.08909} {arXiv:1906.08909 [astro-ph.CO]}
  \BibitemShut {NoStop}%
\bibitem [{\citenamefont {{Maggiore}}\ \emph {et~al.}(2020)\citenamefont
  {{Maggiore}}, \citenamefont {{Van Den Broeck}}, \citenamefont {{Bartolo}},
  \citenamefont {{Belgacem}}, \citenamefont {{Bertacca}}, \citenamefont
  {{Bizouard}}, \citenamefont {{Branchesi}}, \citenamefont {{Clesse}},
  \citenamefont {{Foffa}}, \citenamefont {{Garc{\'\i}a-Bellido}}, \citenamefont
  {{Grimm}}, \citenamefont {{Harms}}, \citenamefont {{Hinderer}}, \citenamefont
  {{Matarrese}}, \citenamefont {{Palomba}}, \citenamefont {{Peloso}},
  \citenamefont {{Ricciardone}},\ and\ \citenamefont
  {{Sakellariadou}}}]{Maggiore2020JCAP}%
  \BibitemOpen
  \bibfield  {author} {\bibinfo {author} {\bibfnamefont {M.}~\bibnamefont
  {{Maggiore}}}, \bibinfo {author} {\bibfnamefont {C.}~\bibnamefont {{Van Den
  Broeck}}}, \bibinfo {author} {\bibfnamefont {N.}~\bibnamefont {{Bartolo}}},
  \bibinfo {author} {\bibfnamefont {E.}~\bibnamefont {{Belgacem}}}, \bibinfo
  {author} {\bibfnamefont {D.}~\bibnamefont {{Bertacca}}}, \bibinfo {author}
  {\bibfnamefont {M.~A.}\ \bibnamefont {{Bizouard}}}, \bibinfo {author}
  {\bibfnamefont {M.}~\bibnamefont {{Branchesi}}}, \bibinfo {author}
  {\bibfnamefont {S.}~\bibnamefont {{Clesse}}}, \bibinfo {author}
  {\bibfnamefont {S.}~\bibnamefont {{Foffa}}}, \bibinfo {author} {\bibfnamefont
  {J.}~\bibnamefont {{Garc{\'\i}a-Bellido}}}, \bibinfo {author} {\bibfnamefont
  {S.}~\bibnamefont {{Grimm}}}, \bibinfo {author} {\bibfnamefont
  {J.}~\bibnamefont {{Harms}}}, \bibinfo {author} {\bibfnamefont
  {T.}~\bibnamefont {{Hinderer}}}, \bibinfo {author} {\bibfnamefont
  {S.}~\bibnamefont {{Matarrese}}}, \bibinfo {author} {\bibfnamefont
  {C.}~\bibnamefont {{Palomba}}}, \bibinfo {author} {\bibfnamefont
  {M.}~\bibnamefont {{Peloso}}}, \bibinfo {author} {\bibfnamefont
  {A.}~\bibnamefont {{Ricciardone}}}, \ and\ \bibinfo {author} {\bibfnamefont
  {M.}~\bibnamefont {{Sakellariadou}}},\ }\href {\doibase
  10.1088/1475-7516/2020/03/050} {\bibfield  {journal} {\bibinfo  {journal}
  {\jcap}\ }\textbf {\bibinfo {volume} {2020}},\ \bibinfo {eid} {050} (\bibinfo
  {year} {2020})},\ \Eprint {http://arxiv.org/abs/1912.02622} {arXiv:1912.02622
  [astro-ph.CO]} \BibitemShut {NoStop}%
\bibitem [{\citenamefont {{Virtanen}}\ \emph {et~al.}(2020)\citenamefont
  {{Virtanen}}, \citenamefont {{Gommers}}, \citenamefont {{Oliphant}} \emph
  {et~al.}}]{2020SciPy-NMeth}%
  \BibitemOpen
  \bibfield  {author} {\bibinfo {author} {\bibfnamefont {P.}~\bibnamefont
  {{Virtanen}}}, \bibinfo {author} {\bibfnamefont {R.}~\bibnamefont
  {{Gommers}}}, \bibinfo {author} {\bibfnamefont {T.~E.}\ \bibnamefont
  {{Oliphant}}},  \emph {et~al.},\ }\href {\doibase
  https://doi.org/10.1038/s41592-019-0686-2} {\bibfield  {journal} {\bibinfo
  {journal} {Nature Methods}\ } (\bibinfo {year} {2020}),\
  https://doi.org/10.1038/s41592-019-0686-2}\BibitemShut {NoStop}%
\bibitem [{\citenamefont {{Hunter}}(2007)}]{2007CSE.....9...90H}%
  \BibitemOpen
  \bibfield  {author} {\bibinfo {author} {\bibfnamefont {J.~D.}\ \bibnamefont
  {{Hunter}}},\ }\href {\doibase 10.1109/MCSE.2007.55} {\bibfield  {journal}
  {\bibinfo  {journal} {Computing in Science and Engineering}\ }\textbf
  {\bibinfo {volume} {9}},\ \bibinfo {pages} {90} (\bibinfo {year}
  {2007})}\BibitemShut {NoStop}%
\bibitem [{\citenamefont {{Perez}}\ and\ \citenamefont
  {{Granger}}(2007)}]{2007CSE.....9c..21P}%
  \BibitemOpen
  \bibfield  {author} {\bibinfo {author} {\bibfnamefont {F.}~\bibnamefont
  {{Perez}}}\ and\ \bibinfo {author} {\bibfnamefont {B.~E.}\ \bibnamefont
  {{Granger}}},\ }\href {\doibase 10.1109/MCSE.2007.53} {\bibfield  {journal}
  {\bibinfo  {journal} {Computing in Science and Engineering}\ }\textbf
  {\bibinfo {volume} {9}},\ \bibinfo {pages} {21} (\bibinfo {year}
  {2007})}\BibitemShut {NoStop}%
\bibitem [{\citenamefont {{Hinton}}(2016)}]{2016JOSS....1...45H}%
  \BibitemOpen
  \bibfield  {author} {\bibinfo {author} {\bibfnamefont {S.~R.}\ \bibnamefont
  {{Hinton}}},\ }\href {\doibase 10.21105/joss.00045} {\bibfield  {journal}
  {\bibinfo  {journal} {The Journal of Open Source Software}\ }\textbf
  {\bibinfo {volume} {1}},\ \bibinfo {eid} {00045} (\bibinfo {year}
  {2016})}\BibitemShut {NoStop}%
\bibitem [{\citenamefont {Maggiore}(2007)}]{Maggiore:1900zz}%
  \BibitemOpen
  \bibfield  {author} {\bibinfo {author} {\bibfnamefont {M.}~\bibnamefont
  {Maggiore}},\ }\href {http://www.oup.com/uk/catalogue/?ci=9780198570745}
  {\emph {\bibinfo {title} {{Gravitational Waves. Vol. 1: Theory and
  Experiments}}}},\ Oxford Master Series in Physics\ (\bibinfo  {publisher}
  {Oxford University Press},\ \bibinfo {year} {2007})\BibitemShut {NoStop}%
\end{thebibliography}%

\appendix
\section{GW Amplitudes}
\label{app:waveform}
To lowest PN order, the GW amplitude in the time domain for a circular, inspiralling binary is \cite{Maggiore:1900zz}:
\begin{align}
h_0(t_{\rm obs})&=\frac{4}{d_L}\left(\frac{G\Mc}{c^2}\right)^{5/3}\left(\frac{\pi f_{gw} (t^{\rm obs})}{c}\right)^{2/3}    \label{hc}
\end{align}
Here $\Mc$ is the redshifted chirpmass, $\Mc = (1+z) M_c$, and $t_{\rm obs}$ is a time variable in the observer's frame. The amplitude multiplies the two polarisation and phase factors as per eqs.~(\ref{hamp2}) and (\ref{hamp1}).The evolution of $ f_{gw}(t{^{\rm obs})}$ is given by solving eq.~(\ref{eqn:fevolve}), which leads to:
\begin{align}
f_{gw}(\tauo)&=\frac{1}{\pi}\left(\frac{5}{256}\frac{1}{\tauo}\right)^{3/8}\left(\frac{G\Mc}{c^3}\right)^{-5/8}    \label{feq}
\end{align}
where we have replaced the general time $t_{\rm obs}$ with the time-to-coalescence interval in the observer's frame, $\tauo=t^c_{\rm obs}-t_{\rm obs}$ (where $t^c_{\rm obs}$ denotes the coalescence time). Substituting eq.~(\ref{feq}) into eq.~(\ref{hc}):
\begin{align}
h_0&=\frac{4c}{d_L}\left(\frac{G\Mc}{c^3}\right)^{5/4}\left(\frac{5}{256}\frac{1}{\tauo}\right)^{1/4} \label{hc2}
\end{align}
The authors of \cite{2019PhRvD.100j3523M} gives the characteristic coalescence observer-frame timescale as:
\begin{align}
\tau_c &\simeq \frac{5G\Mc}{c^3}    
\end{align}
We use this timescale to write the time to coalescence in the observer frame as:
\begin{align}
\tauo &= \tau_{c}\left(\frac{t^c_{\rm obs}}{\tau_{c}}-\frac{t_{\rm obs}}{\tau_{c}}\right)=\frac{5G\Mc}{c^3} \left(\hat{t}_c-\hat{t}\right)\label{fra}
\end{align}
where $\hat{t}$ and $\hat{t}_c$ are now dimensionless quantities, and we have dropped the subscript `obs' for ease of notation. $\hat{t}_c$ is of order unity, and $\hat{t}$ may be of order unity or smaller. Substituting eq.~(\ref{fra}) into in eq.~(\ref{hc2}) yields
\begin{align}
h_0\simeq \frac{1}{d_L}\frac{G\Mc}{c^2}\left(\hat{t}_c-\hat{t}\right)^{-1/4}    \label{ire}
\end{align}
For the portion of the inspiral where the approximation of eq.~(\ref{hc}) is valid (i.e. where a lowest-PN order description suffices), we expect the factor $\left(\hat{t}_c-\hat{t}\right)^{-1/4}$ to be of order unity and slowly evolving. Therefore it does not significantly affect the likelihood for the plus and cross amplitudes that we evaluate in section \ref{subsec:dl_error}, and so is dropped, leading to the Gaussian central values in eq.~(\ref{eqn:Aplus_Across_inclination}).
\vspace{5mm}

\section{LSS-only Constraints}
\label{app:LSSonly}
\begin{figure}
  \begin{centering}
    \includegraphics[width=\textwidth]{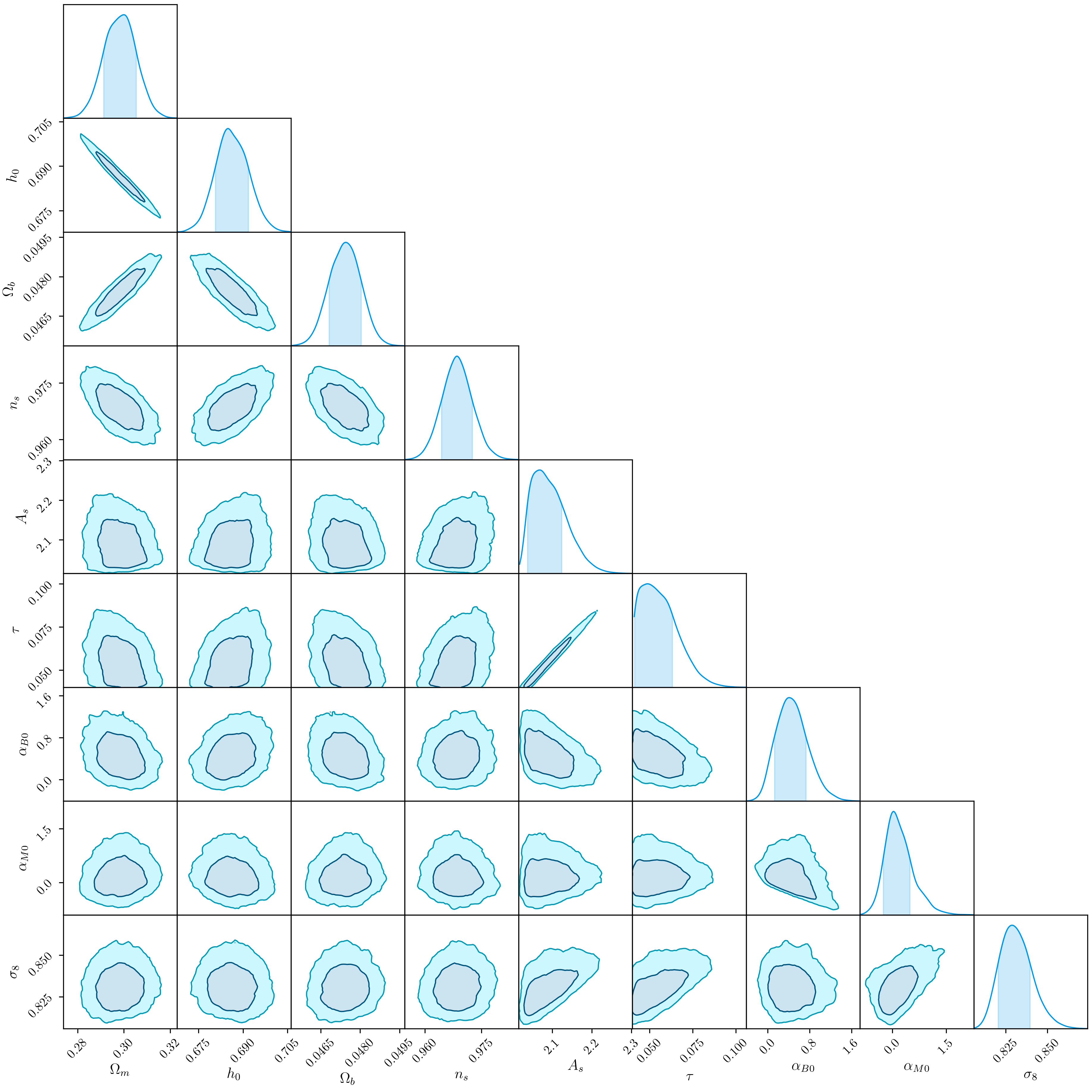}
  \end{centering}
  \caption{Constraints on cosmological and Horndeski parameters using the Large Scale Structure data sets described in the text, for the ansatz in which $\proptoomega$}
  \label{fig:LSS_omega}
\end{figure}

\begin{figure}
  \begin{centering}
    \includegraphics[width=\textwidth]{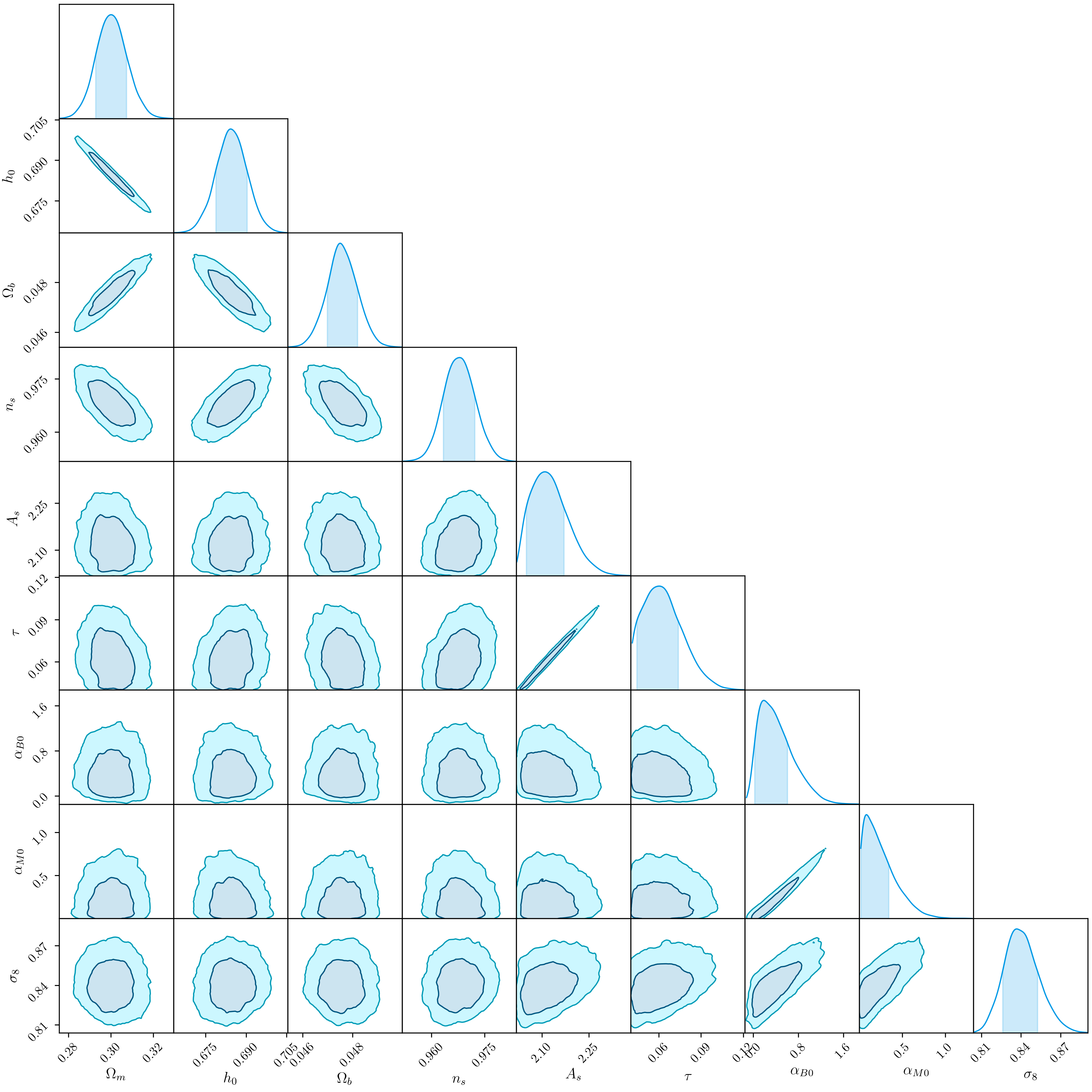}
  \end{centering}
  \caption{Constraints on cosmological and Horndeski parameters using the Large Scale Structure data sets described in the text, for the ansatz in which $\proptoscale$}
  \label{fig:LSS_scale}
\end{figure}

Here, we present the joint constraints on cosmological and Horndeski modified gravity parameters using current Large Scale Structure surveys. The data sets used in these constraints are described in \cref{sec:experiments}, the assumed likelihood and sampling are described in \cref{sub:likelihood}. The assumed priors on the cosmological parameters are shown in \cref{tab:model_params}. \Cref{tab:model_params} also shows the resulting parameter best fit and 68\% confidence intervals for the parameters, and contour plots of the posteriors are shown for the $\proptoomega$ model in \cref{fig:LSS_omega} and the $\proptoscale$ model in \cref{fig:LSS_scale}.
\begin{table}
\renewcommand{\arraystretch}{1.5}
    \centering
    \begin{tabular}{lccc}
        \hline
		Parameter & Prior & LSS-only $\propto a$ constraint & LSS-only $\propto \Omega_{\rm DE}$ constraint \\ 
		\hline
		$\Omega_m$ & $\mathcal{U}(0.1, 0.9)$ & $0.3002^{+0.0069}_{-0.0074}$ & $0.3007^{+0.0045}_{-0.0093}$ \\
		$h_0$ & $\mathcal{U}(0.55, 0.91)$ & $0.6840^{+0.0062}_{-0.0052}$ & $0.6850^{+0.0067}_{-0.0043}$ \\ 
		$\Omega_b$ & $\mathcal{U}(0.03, 0.12)$ & $0.0475^{+0.0007}_{-0.0005}$ & $0.0474^{+0.0006}_{-0.0006}$ \\ 
		$n_s$ & $\mathcal{U}(0.87, 1.07)$ & $0.9682^{+0.0039}_{-0.0048}$ & $0.9685^{+0.0041}_{-0.0040}$ \\ 
		$A_s$ & $\mathcal{U}(0.5, 5.0)$ & $2.106^{+0.060}_{-0.057}$ & $2.070^{+0.050}_{-0.034}$ \\ 
		$\tau$ & $\mathcal{U}(0.04, 0.125)$ & $0.061^{+0.013}_{-0.018}$ & $0.0487^{+0.0132}_{-0.0085}$ \\ 
		$\alpha_{B0}$ & $\mathcal{U}(-1.0, 3.0)$ & $0.18^{+0.41}_{-0.15}$ & $0.38^{+0.34}_{-0.25}$ \\ 
		$\alpha_{M0}$ & $\mathcal{U}(-1.0, 6.0)$ & $0.068^{+0.254}_{-0.063}$ & $0.02^{+0.46}_{-0.27}$ \\ 
		$\sigma_8$ & -- & $0.8359^{+0.0164}_{-0.0096}$ & $0.8293^{+0.0101}_{-0.0089}$ \\ 
		\hline
    \end{tabular}
    \caption{Priors on parameters used for the analyses described in the text and parameter constraints from current Large Scale Structure data sets, as described in \cref{sec:conclusions}. Priors of the form $\mathcal{U}(a,b)$ are uniform distributions on the range $[a,b]$. $\sigma_8$ is a derived parameter and so does not require a prior.}
    \label{tab:model_params}
\end{table}

\end{document}